\documentclass[11pt, a4paper]{article}
\usepackage[T1]{fontenc}
\usepackage{a4wide, url}
\usepackage[english]{babel}
\usepackage{graphicx}
\usepackage{natbib}
\usepackage[textfont=sc,labelfont=bf]{caption}
\usepackage{setspace}
\usepackage{multirow}
\usepackage{amsmath}
\usepackage{amsfonts}
\usepackage{amssymb}
\usepackage{pgf} 
\usepackage{bm}
\usepackage{epstopdf}
\usepackage{subfig}
\usepackage{longtable}
\usepackage{rotating}
\def\rot{\rotatebox}
\usepackage[utf8]{inputenc}
\usepackage{pdflscape}
\usepackage{adjustbox}
\usepackage{float}
\usepackage{tikz}
\usepackage{mathtools}
\usepackage{multirow}
\usepackage{booktabs}
\DeclarePairedDelimiter\abs{\lvert}{\rvert}

\doublespace
		
\title{\textsc{TailCoR}}
\author{{\normalsize Sla\dj ana  \textsc{Babi\' c}$^{1}$, Christophe \textsc{Ley}$^{1}$, Lorenzo \textsc{Ricci}$^{2}$ and \normalsize David \textsc{Veredas}$^{3}$} \\ \\
}

\date{}

\begin{document}
\maketitle

\vspace{-0.93cm}
\begin{abstract}
Economic and financial crises are characterised by unusually large events. These tail events co-move because of linear and/or nonlinear dependencies. We introduce TailCoR, a metric that combines (and disentangles) these linear and non-linear dependencies. TailCoR between two variables is based on the tail inter quantile range of a simple projection. It is dimension-free, it performs well in small samples, and no optimisations are needed.
\end{abstract}

\thispagestyle{empty}

\footnotetext[1]{Ghent University} 
\footnotetext[2]{European Stability Mechanism} 
\footnotetext[3]{Vlerick Business School and Ghent University\\
Corresponding address: David Veredas, Av. du Boulevard 21, B1210, Brussels, Belgium. Phone: +3222254121. Email: david.veredas@vlerick.com.}

\newpage

\section{Introduction}

\subsubsection*{Empirical motivation}
Two major global crises have stricken the economy and financial markets since 2000: The 2007-2008 Great Financial Crisis and the 2020 Great Lockdown due to COVID-19. These crises highlight the importance of tail -- or rare -- events that can have different natures:  corporate and Government defaults, stock market crashes, public health emergencies, or political decisions, to name a few.

When they occur, their shock wave spreads over the economic and financial systems through large co-movements. These unfrequent events  locate on the tails of the probability distributions. From a statistical point of view, joint tail events are caused by linear and/or nonlinear correlations. Indeed, correlations on the tails may happen because variables are linearly correlated (i.e., the Pearson correlations are different from zero) and/or nonlinearly correlated, in the sense that variables are dependent at the tails of their distributions only.\footnote{Other forms of nonlinearity may occur but we do not take them into account. They can however be considered within the theory of TailCoR.}


The analysis of tail correlations and their linear and nonlinear components has numerous financial applications. \cite{GGV15} document that while the linear dependence between short selling activity and asset returns is very weak, the correlation between extreme negative returns and extreme increases in short selling is highly significant. This finding sheds light on the contradiction between the widespread view among practitioners that short selling has mostly positive effects, and the bans of this activity by the competent authorities. \cite{PoonRockingerTawn04} argue that in the case of a portfolio the relationships among the constituents when they are on the tails give a better understanding on the tail behavior of the portfolio. The disentangling between the linear and nonlinear components allows not only to design portfolio allocation procedures that minimize the linear risk but also the tail risk of the portfolio (\cite{JKV00}).

Tail correlations also relate to the literature of tail risk spillovers using quantiles (namely VaR and CoVaR of \cite{CoVaR2016}). In this context, \cite{HautSchaSchie2015} introduce systemic risk betas as the total time-varying marginal effect of a firm's VaR on the system's VaR. These effects are computed as a function of the interdependence between firm's tail risk exposures. \cite{HardWangYu2016} propose a semiparametric measure to estimate systemic interconnectedness across financial institutions based on tail-driven spillover effects in a high dimensional network. \cite{CheSun2020} use this methodology for studying systemically important insurers.




\subsubsection*{Literature review}

Several statistical and econometric measures have been proposed to assess the degree of tail dependence. The tail dependence coefficients (also called extremal dependence structures and co-exceedance probabilities) stemming from extreme value theory (EVT henceforth) are the most used. These coefficients rely on two building blocks. The first is the assumption that tails of the distribution (either the joint and/or the marginal distributions) asymptotically decay according to a power law (see \cite{KiriRooWadsSeg2019} and \cite{RooSegWads2018} for references on the multivariate generalized Pareto distribution). The second building block is the extreme value copula (also called stable tail dependence function or STDF).\footnote{See e.g. \cite{Joe:1997}, \cite{Embrechts2000}, \cite{Beirlant:2004} and \cite{McNeilFreyEmbrechts05}.}

Broadly speaking, estimation of the STDF is divided in two approaches: parametric and semi-parametric. The former is based on parametric copula functions, the logistic and $t$ copulas are commonly used (see, for example, \cite{LoginSolnik01} for an application to extreme correlation of international equity markets, and \cite{CholletePenaLu11} for its use in investment diversification across major stock market indexes). The latter is based on higher order statistics (see \cite{Embrechts2000} for a theoretical treatment, and \cite{Straetmans04} for an application to asset market linkages in crisis periods).\footnote{\cite{LoginSolnik01} exploit, in the bivariate case, the explicit and simple relation between the tail dependence coefficient of the Gumbel copula and the linear correlation coefficient for extreme events, also known as exceedance correlation. In a similar vein, \cite{CizeauPottersBouchaud01} introduce the quantile correlation, i.e. the sample correlation between observations that are contained in a ball around a given joint quantile.} As pointed out in \cite{Straetmans08}, estimation of the STDF has the weakness that it presupposes tail dependence, besides that the parametric copula rests on parametric assumptions. An alternative, proposed by \cite{Ledford96}, consists of testing for tail dependence through the tail index of an auxiliary variable (the cross-sectional minimum of the random vector). \cite{Hartmann2005} uses it for studying the banking system, \cite{PoonRockingerTawn04} for estimating the tail dependence among risky asset returns, and \cite{Straetmans08} for studying US sectoral stock indexes around 9/11, respectively.

While EVT-based literature is semi-parametric (in the sense that only the asymptotic tail is parametrized), another branch of the literature parametrises the whole distribution while assuming heavy tails -- see \cite{BLV} for a survey. The scale mixtures of multinormal distributions are particularly relevant since the elliptical and the normal mean-variance mixture distributions are used in this article. These distributions have the feature that the dependence between two random variables is given by the dispersion matrix and a single tail index (and the vector of skewness in the case of the normal mean-variance mixture). \cite{FW} proposes a new family of distributions, coined multiple scaled distributions, with the main feature that there are as many tail indexes as the dimension of the random vector. Another relevant fully parametric class of distributions are copulas, where the meta-elliptical copula of \cite{FangFangKotz2002} plays a dominant role. Beyond ellipticity, \cite{LeeLong} propose a multivariate conditional variance model that allows to model conditional correlation and dependence separately and simultaneously with nonelliptically distributed dependent errors.

\subsubsection*{This paper}

We introduce TailCoR, a measure for dependence that 
can be computed under tails that are fatter, equal, or thinner than Gaussian. In a nutshell, TailCoR between two random variables is based on the projection of the two variables onto a line, and then a tail Inter Quantile Range is computed. 
TailCoR has the following useful features: it is simple and no optimizations are needed, it performs well in small samples, and under ellipticity (i.e., the probability contours are ellipsoids) TailCoR disentangles easily between the above mentioned linear and nonlinear components.

We also show an extension that differentiates between the downside and upside TailCoRs. It is often the case, like in risk management, that the interest lies in the tail of one region of the distribution, which leads to TailCoR when the probability contours are asymmetric.



The empirical illustration is on daily returns of 21 financial market indexes across the globe. Data runs from January 2000 to July 2020, and hence it incorporates the two crises mentioned at the beginning of the Introduction. For comparison purposes we compute the upper and lower exceedance correlations of \cite{LoginSolnik01}, and the parametric and non-parametric tail dependence coefficients (the former with a $t$ copula, and the latter as in \cite{Straetmans08}). Results show that TailCoR, exceedance correlations and tail dependence coefficients behave similarly.

To study the dynamic behaviour we use windows of 3 years. 
The estimated TailCoRs are in line with the financial and economic events that happened during the sample. TailCoR increases in crises periods, when global dependence increases, as it happened in 2007-2008 (and the aftermath) as well as recently because of COVID-19. 
The evolution of the other metrics is heterogeneous. While the exceedance correlations could not be computed because of lack of extreme observations in the windows, the tail dependence coefficients move along with TailCoR though they show a greater deal of variability.

The remaining sections are laid out as follows. Section 2 introduces the notation, assumptions, definition, and representations of TailCoR. It also shows a calibration exercise and the asymptotic properties of the estimators. Section 3 covers a brief Monte Carlo study. The extension to asymmetry is touched upon in Section 4.
The illustration to the market indexes is presented in Section 5. Section 6 concludes and proofs and lengthy tables are relegated to the appendixes.

\section{TailCoR}

\subsection{Definition}


Let $\mathbf{X}_t$, $t=1,\ldots,T,$ be a random vector of size $N$ at time $t$ satisfying the following assumption

\begin{itemize}
\item[\bf G1] 
(a) The random process $\left\{\mathbf{X}_t\right\}$ is a strongly stationary sequence of random vectors, (b) the unconditional distribution of $\mathbf{X}_t$ is unimodal, and (c)
$\mathbf{X}_t$ is $S$-mixing.
\end{itemize}

Assumption {\bf G1}(a) is standard in time series analysis and {\bf G1}(b) is due to \cite{Sherman55}. Both relate to the unconditional distribution of $\mathbf{X}_t$. Assumption {\bf G1}(c) relates to the conditional distribution and it specifies the time dependence of $\mathbf{X}_t$. The purpose of unimodality is twofold. First, it rules out distributions with several modes (but not asymmetry). Second, we consider the Gaussian and uncorrelated (hence independent) process as the benchmark. Regarding  {\bf G1}(c), assuming a mixing condition instead of a particular type of dynamic model makes TailCoR applicable to a wide array of processes. The conditions for $S$-mixing, introduced by \cite{BerkesHormannSchauer09}, apply to a large number of processes often used in economics and finance, including GARCH models and its extensions, linear processes (like ARMA models), and stochastic volatility among others. 

\begin{figure}[!ht]
  \caption{Diagrammatic representation of TailCoR}
  \begin{center}
    \subfloat[Linear relation]{\includegraphics[angle=0,width=.45\linewidth]{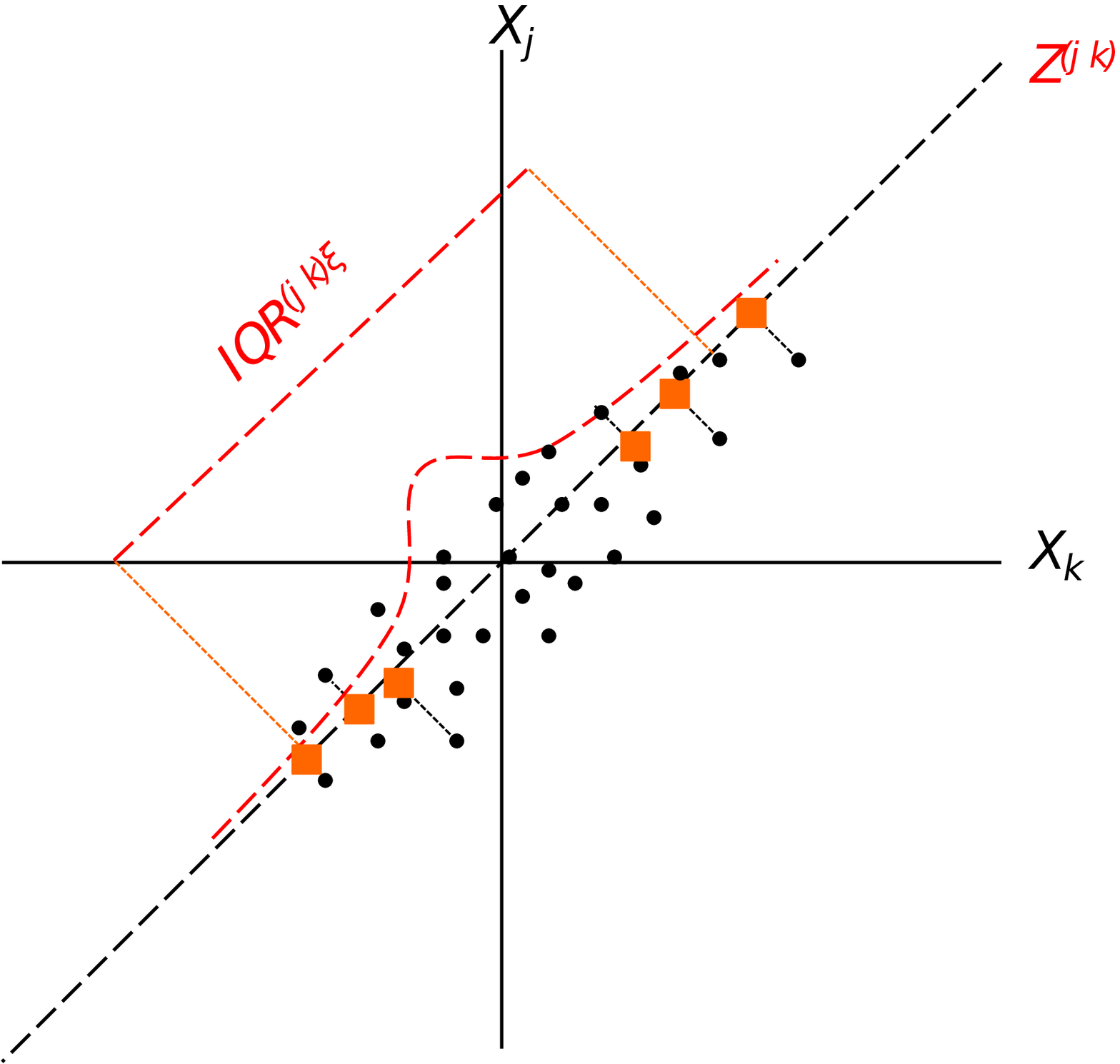}}
    \hspace{0.6cm}
    \subfloat[nonlinear relation]{\includegraphics[angle=0,width=.45\linewidth]{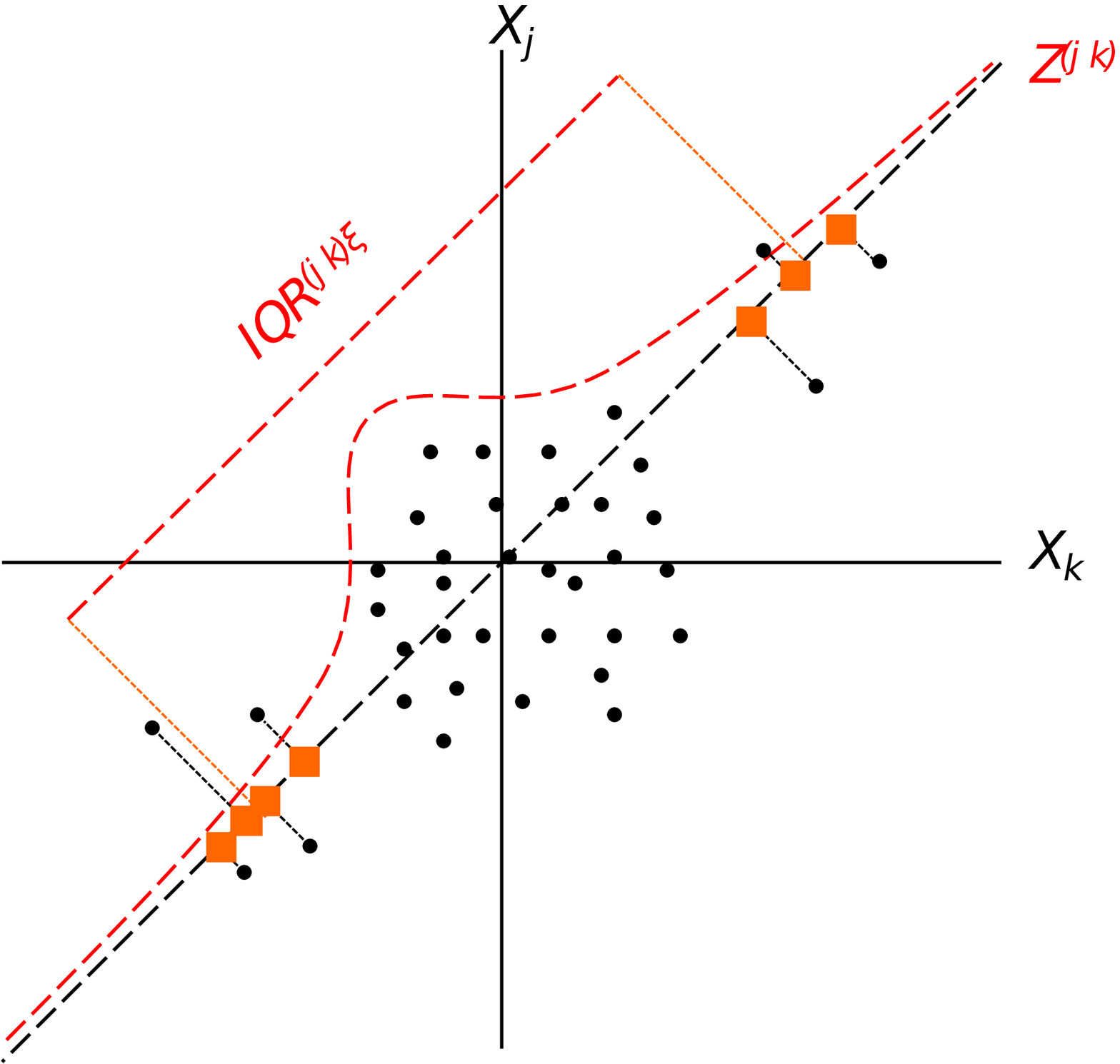}} \\
    \subfloat[Linear and nonlinear relation]{\includegraphics[angle=0,width=.45\linewidth]{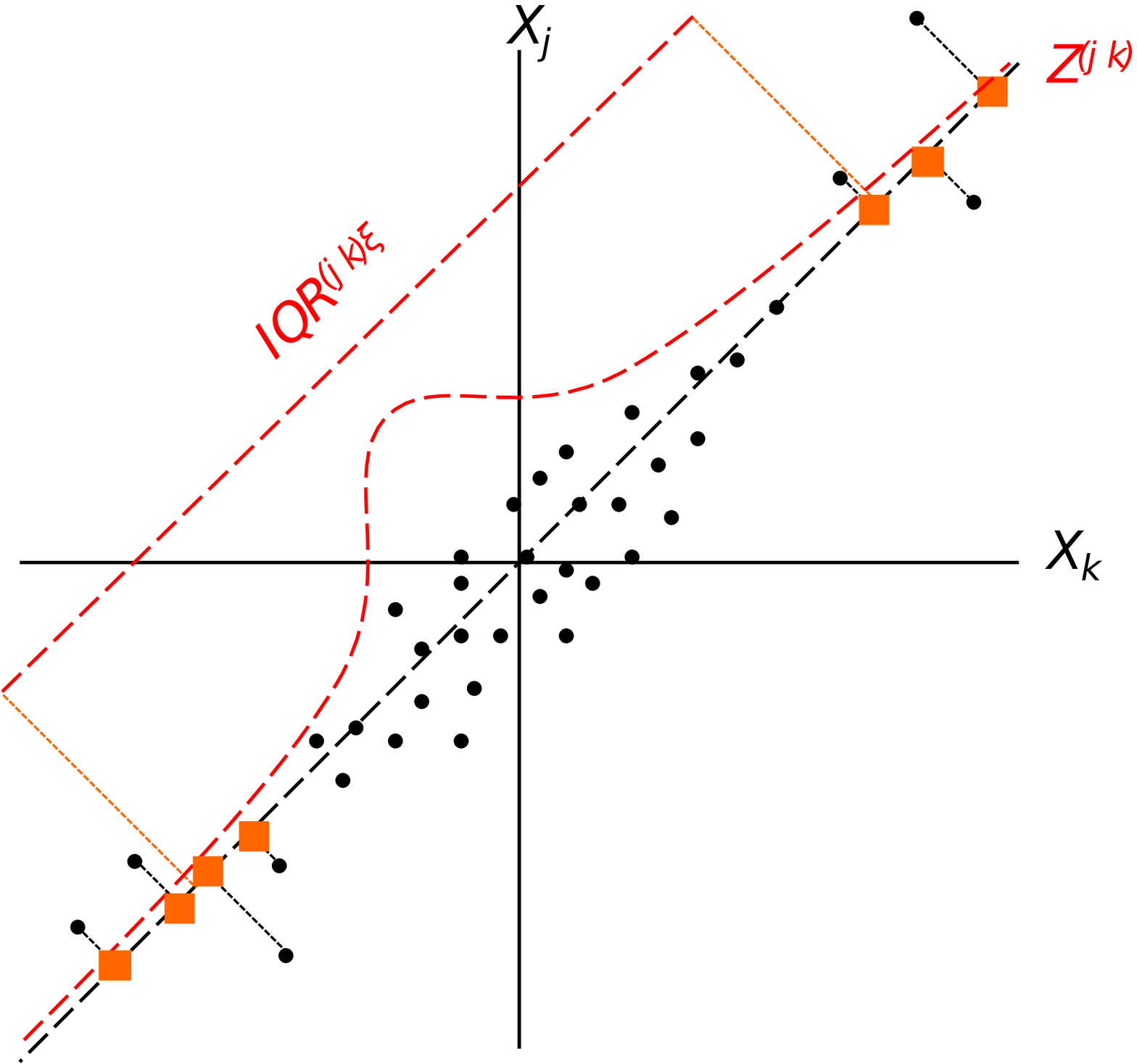}} \\
\begin{tabular}{p{14.5cm}}
\scriptsize Scatter plots where  $X_j$ and $X_k$ are positively related (the pairs are depicted with circles). Projecting the observations onto the $\phi$-degree line produces the random variable $Z^{(j\,k)}$, depicted with squares. Because of representation purposes we show the projection only for the observations on the tails but the reader should keep in mind that the projection is done for all the observations. Panel (a) shows the case of linear relation only. Panel (b) displays the case of nonlinear relation only. Panel (c) shows the scenario of both linear and nonlinear relations.
\\ 
\end{tabular}
  \end{center}
  \label{fig:scatterplots}
\end{figure}

TailCoR is based on the following simple idea, shown in Figure \ref{fig:scatterplots}. If two random variables $X_j$ and $X_k$ (properly standardized) are positively related (either linear and/or nonlinearly), most of the times the pairs of observations (depicted with circles) have the same sign, in the sense that most of them concentrate in the north-east and south-west quadrants of the scatter plot. Now, consider the $\phi$-degree line that crosses these quadrants (we illustrate the figure with $\phi=\pi/4$ or the 45-degree line) and project all the pairs on this line, producing a new random variable $Z^{(j\,k)}$, depicted with squares.\footnote{Because of representation purposes we show the projection only for the observations that are far from the origin, but the reader should keep in mind that the projection is done for all the observations.} Since the two random variables are positively related, the squares -- that are sitting on the $\phi$-degree line -- are dispersed all over the line.\footnote{In the case of negative relation, the dots mostly concentrate in the north-west and south-east quadrants, and the projection is on the corresponding $\phi$-degree line, as explained in detail later.} The extend of the dispersion depends on the strength of the relation between $X_j$ and $X_k$. If weak, the cloud of dots is concentrated around the origin without a well defined direction
The dispersion of the squares is therefore small. By contrast, if the relation is strong, the cloud of dots is stretched around the $\phi$-degree line, and hence the squares are very dispersed. 


TailCoR is equal -- up to a normalization -- to the difference between the upper and lower tail quantiles of $Z^{(j\,k)}$. This {\it tail interquantile range} can be large because of two reasons. First, if $X_j$ and $X_k$ are highly linearly correlated (panel (a) of Figure \ref{fig:scatterplots}). Second, if the  correlation between $X_j$ and $X_k$ only happens on the tails, while the observations around the origin form a cloud with undefined direction (panel b). These two situations are not mutually exclusive and both may happen, which is actually the most likely case in practice (panel c). Either way, TailCoR is large, in a sense to be precisely defined below. But first we put this intuition at work.

Let $X_{j\,t}$ be the $j$th element of the random vector $\mathbf{X}_t$. Denote by $\mbox{Q}^{\tau}_j$ its $\tau$th quantile for $0<\tau<1$, and let $\mbox{IQR}^{\tau}_{j}=\mbox{Q}^{\tau}_j-\mbox{Q}^{1-\tau}_j$ be the $\tau$th interquantile range. A typical value of $\tau$ is 0.75. Let $Y_{j\,t}$ be the standardized version of $X_{j\,t}$:
\begin{equation}\label{eq:standard}
Y_{j\,t} = \frac{X_{j\,t} - \mbox{Q}^{0.50}_j}{\mbox{IQR}^{\tau}_{j}}.
\end{equation}
Likewise for $Y_{k\,t}$. In our context of potential heavy tails, the median and the interquantile range are used for the standardization. The mean of $Y_{j\,t}$ is not necessarily zero and its variance is not one, if they exist. This is not an issue since the aim of \eqref{eq:standard} is to have the pair $(Y_{j\,t},Y_{k\,t})$ centered around the same number and with the same scale. As \eqref{eq:standard} is based on marginal quantiles, we need the following technical assumption
\begin{itemize}
\item[\bf G2]  The random variable $X_{j\,t}$ has $F(x_{j})$ as cumulative distribution function  with density $f(x_{j})$ that is continuous and non-zero in a neighbourhood of $\mbox{Q}^{\tau}_j$ for $\tau \in [0,1]$. Likewise for $X_{k\,t}$.
\end{itemize}


Alternatively to \eqref{eq:standard}, $X_{j\,t}$ and $X_{k\,t}$ could be standardized with the marginal cumulative distribution functions, i.e. $Y_{j\,t}$ and $Y_{k\,t}$ would be the probability integral transforms that are distributed uniformly on $(0,1)$.
This is advantageous if there are marginal dependencies beyond the location and the scale (like in the tails), or if we extend our method to copulas (as explained in the conclusions).

By standard trigonometric arguments, the projection of $(Y_{j\,t},Y_{k\,t})$ on the $\phi$-degree line is
\begin{equation}\label{eq:proj}
Z^{(j\,k)}_{t} = Y_{j\,t}\cos \phi + Y_{k\,t} \sin \phi,
\end{equation}
and the tail interquantile range of $Z^{(j\,k)}_{t}$ is
\begin{equation*}
\mbox{IQR}^{(j\,k)\,\xi}=\mbox{Q}^{(j\,k)\,\xi} - \mbox{Q}^{(j\,k)\,1-\xi},
\end{equation*}
where $\mbox{Q}^{(j\,k)\,\xi}$ is the $\xi$th quantile of $Z^{(j\,k)}_{t}$ and $\xi$ is typically beyond $0.90$.
The larger $\xi$ is, the further we explore the tails. We define TailCoR as follows.
\begin{description}
\item[{\bf Definition 1}] Under {\bf G1} -- {\bf G2}, TailCoR between $X_{j\,t}$ and $X_{k\,t}$ is
\begin{equation} \label{eq:definition}
\mbox{TailCoR}^{(j\,k)\,\xi}  := s_0(\xi,\tau)\mbox{IQR}^{(j\,k)\,\xi},
\end{equation}
where $s_0(\xi,\tau)$ is a normalization such that under independence $\mbox{TailCoR}^{(j\,k)\,\xi}=1$, the reference value.
\end{description}

Four remarks are in order. First, the meaning of correlation in TailCoR is not the traditional one, as TailCoR is not bounded between -1 and 1, nor centered at zero. In the next sub-section, we deal with this issue in two ways. We first examine the bounds more in detail, and then we propose an alternative representation of TailCoR that is bounded between -1 and 1, and equals 0 under independence. 

Second, $ s_0(\xi,\tau)$ equals the inverse of the $\mbox{IQR}^{(j\,k)\,\xi}$ under independence. Also in the next sub-section, we show that in the case of ellipticity, the $\mbox{IQR}^{(j\,k)\,\xi}$ can be computed from a standard Gaussian distribution.

Third, TailCoR is not an asymptotic (in the tail) dependence measure. Instead, the dependence between $X_{j\,t}$ and $X_{k\,t}$ can be computed precisely for any $\xi$, and, since TailCoR does not rely on assumptions of the asymptotic behavior of the tail, it can be used for distributions with tails that are Pareto, exponential, or even with finite end points.



Last, the angle $\phi$ has to be chosen. An optimality criterion would be to choose the angle between $0$ and $\pi$ that maximizes the $\mbox{IQR}^{(j\,k)\,\xi}$. In practice this can be done with a grid search on $\phi$, computing the projection and the tail interquantile range for each angle of the grid. Since no optimizations are involved, this grid search is computationally inexpensive.


\subsection{Disentangling the linear and nonlinear components}

\subsubsection*{Ellipticity and assumptions}

Though definition (\ref{eq:definition}) is simple and intuitive, it does not allow to understand how much of TailCoR is due to the linear and nonlinear dependencies. This is possible however if we assume ellipticity.
\begin{itemize}
\item[\bf E1] The unconditional distribution of $\mathbf{X}_t$ belongs to the elliptical family, given by the stochastic representation $\mathbf{X}_t =_d \bm{\mu} + \mathcal{R}_{\alpha\,t}\bm{\Lambda}\mathbf{U}_t$.
\end{itemize}
The $N \times 1$ random vector $\mathbf{U}_t$ under {\bf E1} is uniformly distributed on the unit sphere. The scaling matrix $\bm{\Lambda}$ produces the ellipticity and is such that $\bm{\Sigma}=\bm{\Lambda} \bm{\Lambda}^{\prime }$, a positive definite symmetric dispersion matrix -- often called the shape matrix with generic $(j\,k)$ element $\sigma_{X_j\,X_k}$, and for convenience we denote the $(j\,j)$ diagonal element by $\sigma_{X_j}^2$. The non-negative and continuous random variable $\mathcal{R}_{\alpha\,t}$ generates the tail thickness through a so-called radial density depending on the shape parameter $\alpha$, and is stochastically independent of $\mathbf{U}_t$. From now on, we will denote $\alpha$ as the tail index, in the sense that this parameter explains the decay of the tails (the smaller $\alpha$ the thicker the tails), but not necessarily according to a power law.\footnote{The random variable $\mathcal{R}_{\alpha\,t}$ may not only depend on the tail index but on a vector of shape parameters. We do not consider the latter since most of the elliptical distributions used in practice only depend on $\alpha$ (see examples after assumption {\bf E2}). The extension is however straightforward.} The vector $\bm{\mu}$ re-allocates the center of the distribution. Let $\bm{\theta}=(\bm \mu, \bm \Sigma, \alpha) \in \bm\Theta$ denote the vector of unknown parameters satisfying the following standard assumption.
\begin{itemize}
\item[\bf E2] (a) The parameter space $\bm{\Theta}$ is a non-empty and compact set on $\mathbb{R}^{N+\frac{N(N+1)}{2}+1}$. (b) The true parameter value $\bm{\theta}_0$ belongs to the interior of $\bm{\Theta}$.
\end{itemize}

Note that assumption {\bf E1} implies that the unconditional distribution of $\mathbf{X}_t$ belongs to the elliptical family, but no specific distributional assumption is made. The elliptical family nests, among others, the Gaussian, Student-$t$, elliptical stable (ES henceforth), Cauchy, Laplace, power exponential, and Kotz probability laws.\footnote{See \cite{Hashorva08} and \cite{Hashorva10} for tail theory within the elliptical family.} For a given vector of locations and a dispersion matrix, the difference between two elliptical distributions is the random quantity $\mathcal{R}_{\alpha\,t}$ with tail index $\alpha$, which plays a central role here. Another feature of the elliptical family is its closeness under location and scale shifts, which implies that $Z_t^{(j\,k)}$ is elliptical with the same tail index $\alpha$.

We further need the existence of the mean and the variance-covariance matrix:
\begin{itemize}
\item[\bf E3] The unconditional moments up to order 2 are finite, i.e. $E(\mathbf{X}_t^p) < \infty, \mbox{ for } p \leq 2$.
\end{itemize}


Next, we substitute assumptions {\bf G1} and {\bf G2} for

\begin{itemize}
\item[\bf E4] (a) The random process $\left\{\mathbf{X}_t\right\}$ is a weakly stationary sequence of random vectors, and (b) $\mathbf{X}_t$ is $S$-mixing.
\item [\bf E5] The random variable $\mathcal{R}_{\alpha\,t}$ has $P(r)$ as  cumulative distribution function   with density $p(r)$ that is  continuous and non-zero in a neighbourhood of $\mbox{Q}^{\tau}_j$ for $\tau \in [0,1]$.
\end{itemize}


\subsubsection*{The optimal projection}

Ellipticity allows to compute the optimal angle $\phi$ without the grid search mentioned above. The interquantile range used in the standardization \eqref{eq:standard} can be written as $k(\tau,\alpha)\sigma_{X_j}$, where $k(\tau,\alpha)$ is a non-random positive constant. Likewise $\mbox{IQR}^{\tau}_{k}=k(\tau,\alpha)\sigma_{X_k}$.\footnote{\cite{McC1986} introduced this relation between the IQR and the scale parameter in the context of stable distributions.} Then, by {\bf E1} the bivariate random vector $Y^{j\,k}_t=(Y_{j\,t},Y_{k\,t})$ is elliptically distributed with a $2 \times 2$ shape matrix $k(\tau,\alpha)^{-2}\bm{\mbox{R}}$. The matrix $\bm{\mbox{R}}$ has diagonal elements $\rho_{11}=\rho_{22}=1$ and off diagonal element $\rho_{12}=\sigma_{X_j\,X_k}/\sigma_{X_j}\sigma_{X_k}$ (3.3.2 of \cite{McNeilFreyEmbrechts05}).

The probability contours of $Y^{j\,k}$ are therefore ellipsoids with axes that are in the direction of the eigenvectors of $k(\tau,\alpha)^2\bm{\mbox{R}}^{-1}$, and their lengths are proportional to the reciprocals of the square roots of the eigenvalues of $k(\tau,\alpha)^2\bm{\mbox{R}}^{-1}$. Since $\rho_{11}=\rho_{22}=1$, the first eigenvalue is $k(\tau,\alpha)^{-2}(1+\rho_{12})$ with associated eigenvector $(1/\sqrt{2},1/\sqrt{2})$, while the second eigenvalue is $k(\tau,\alpha)^{-2}(1-\rho_{12})$ with associated eigenvector $(1/\sqrt{2},-1/\sqrt{2})$.

\begin{figure}
  \caption{A diagrammatic representation of TailCoR for negative relation}\label{fig:negtailcor}
  \begin{center}
    \subfloat[Projection in the 45-degree line]{\includegraphics[angle=0,width=.45\linewidth]{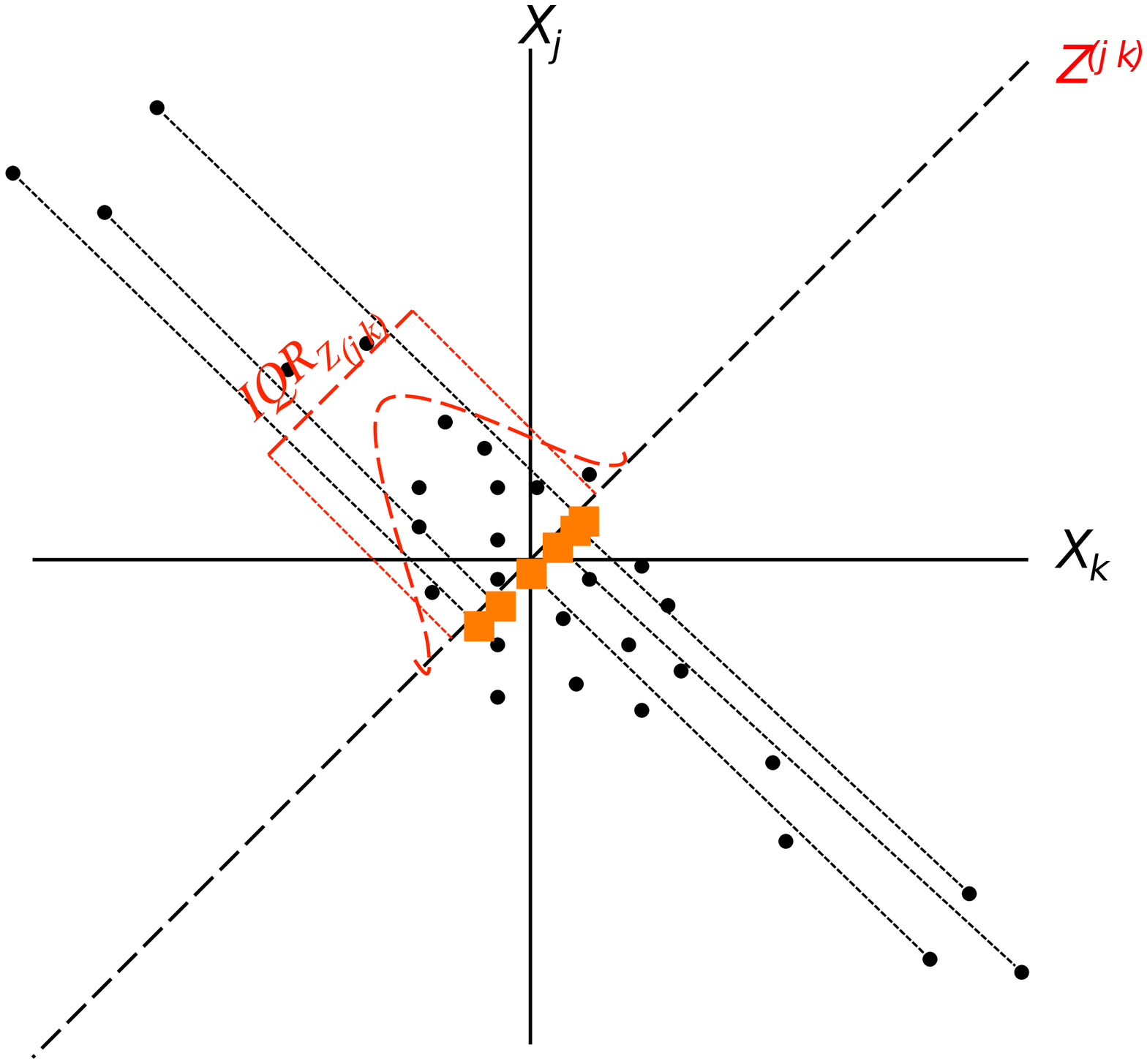}}
    \hspace{0.6cm}
    \subfloat[Projection in the 135-degree line]{\includegraphics[angle=0,width=.45\linewidth]{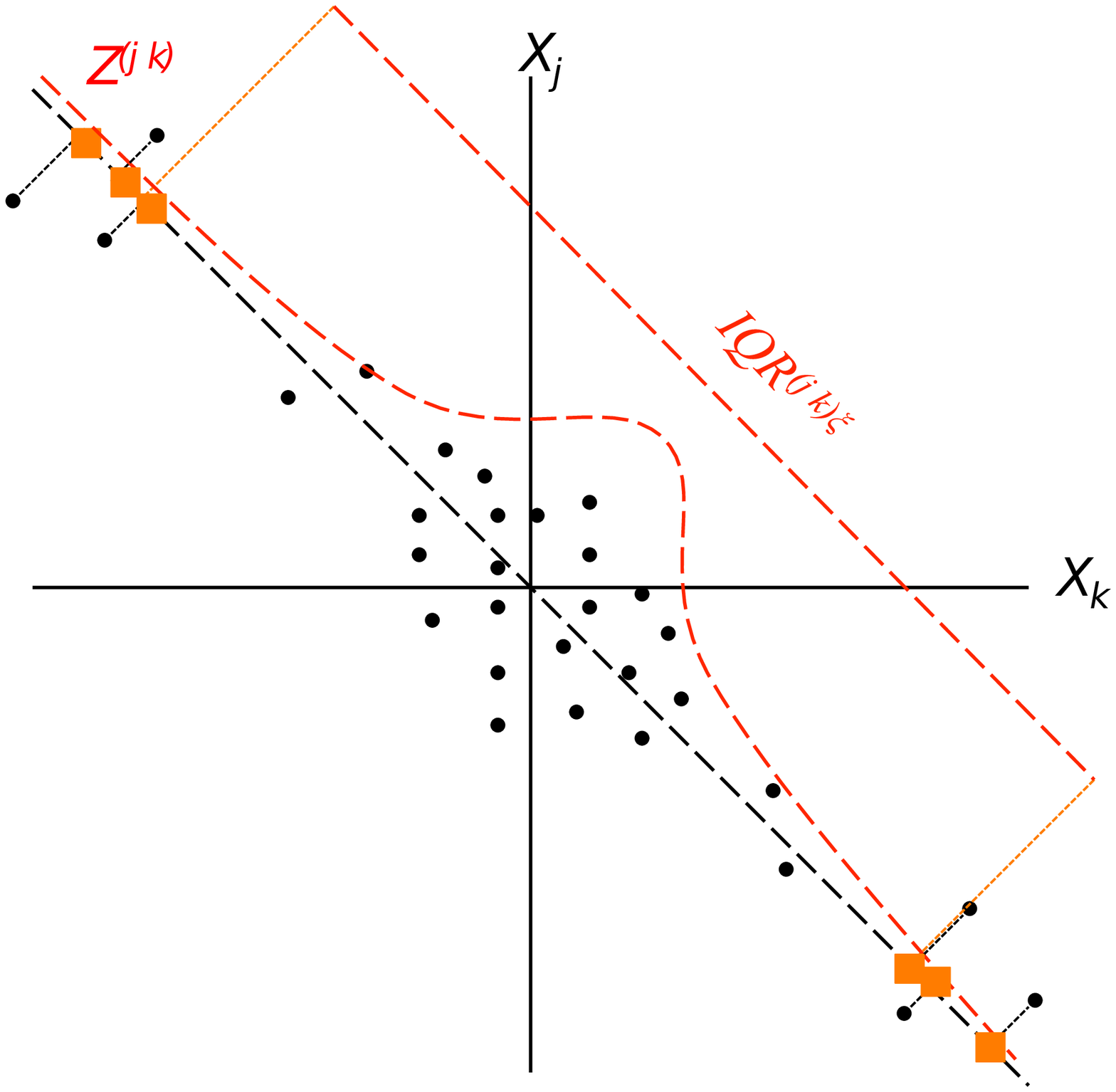}} \\
\begin{tabular}{p{14.5cm}}
\scriptsize Scatter plots where $X_j$ and $X_k$ are negatively related and the projection is done with the wrong and right angles (panels (a) and (b) respectively).\\
\end{tabular}
  \end{center}
\end{figure}

If the relation between $X_{j\,t}$ and $X_{k\,t}$ is positive, then $\rho_{12}$ is positive and $k(\tau,\alpha)^{-2}(1+\rho_{12})$ is the largest eigenvalue. Hence the projection on the 45-degree line is optimal:
\begin{equation}\label{eq:proj}
Z^{(j\,k)}_{t} = \frac{1}{\sqrt{2}}(Y_{j\,t} + Y_{k\,t}).
\end{equation}
If $\rho_{12}$ is negative, then $k(\tau,\alpha)^{-2}(1-\rho_{12})$ is the largest eigenvalue and the projection in the 135-degree line is optimal:
\begin{equation}\label{eq:proj}
Z^{(j\,k)}_{t} = \frac{1}{\sqrt{2}}(Y_{j\,t} - Y_{k\,t}).
\end{equation}
Either way, $Z^{(j\,k)}_{t}$ is the first principal component of $Y^{j\,k}$. The choice between the 45- and the 135-degree lines depends on the sign of $\rho_{12}$. If wrongly chosen, conclusions may be misleading, as exemplified in Figure \ref{fig:negtailcor}.
Panel (a) shows the projection on the 45-degree line when the relation is negative. The projection $Z^{(j\,k)}$ is concentrated around the origin and $\mbox{TailCoR}^{(j\,k)\,\xi}$ can even be smaller than one, leading to the false conclusion that tails are thinner than Gaussian.
Projecting on the 135-degree line, as shown in panel (b), captures correctly the negative relation between $X_{j\,t}$ and $X_{k\,t}$.


\subsubsection*{TailCoR under ellipticity}

We are now ready to disentangle $\mbox{TailCoR}^{(j\,k)\,\xi}$ into the linear and nonlinear components.

\begin{description}
\item[{\bf Theorem 1}] Let $X_{j\,t}$ and $X_{k\,t}$ be two elements of the random vector $\mathbf{X}_t$ that fulfills assumptions {\bf E1} -- {\bf E5}. Let $\rho_{j\,k}$ be the linear correlation, and let $s(\xi,\tau,\alpha)$ be a continuous and monotonically decreasing function of $\alpha$. Then
\begin{equation*}
\mbox{TailCoR}^{(j\,k)\,\xi} = s_g(\xi,\tau)s(\xi,\tau,\alpha)\sqrt{1+|\rho_{j\,k}|},
\end{equation*}
where $s_g(\xi,\tau)$ is a normalization such that under independence $\mbox{TailCoR}^{(j\,k)\,\xi}=1$, the reference value.

\end{description}

\begin{description}
\item[{\bf Proof}] See Appendix P.
\end{description}

The rightmost element, $\sqrt{1+|\rho_{j\,k}|}$, captures the linear dependency to $\mbox{TailCoR}^{(j\,k)\,\xi}$, while $s(\xi, \tau,\alpha)$ captures the nonlinear dependency as it depends on the tail index $\alpha$. We denote these dependencies as linear and nonlinear components.

The normalization has become now $s_g(\xi,\tau)$ where the subindex $g$ is for Gaussianity. Indeed, under ellipticity, independence translates into Gaussianity and linear uncorrelation and hence $s_g(\xi,\tau)=\frac{\Phi^{-1}(\tau)}{\Phi^{-1}(\xi)}$ where $\Phi(\cdot)$ is the cumulative distribution function of a standardized Gaussian distribution.\footnote{This can be shown using results of Theorem 1. Under Gaussianity and linear uncorrelation, $Z^{(j\,k)}_{t} \sim \frac{1}{2\Phi^{-1}(\tau)} N(0,1)$. Hence $\mbox{IQR}^{(j\,k)\,\xi}= \frac{\Phi^{-1}(\xi)}{\Phi^{-1}(\tau)}$. Since $s_g(\xi,\tau)\mbox{IQR}^{(j\,k)\,\xi}=1$, then $s_g(\xi,\tau)=\frac{\Phi^{-1}(\tau)}{\Phi^{-1}(\xi)}$.} Table \ref{tab:sg} in Appendix T shows values of $s_g(\xi,\tau)$ for a grid of reasonable values for $\tau$ and $\xi$.

\begin{figure}
  \begin{center}  \caption{Sensitivity of $s_g(\xi,\tau)s(\xi,\tau,\alpha)$ to $\alpha$ and $\rho$}\label{fig:nonlinear}
  \vspace{0.5cm}
\subfloat[Sensitivity to $\alpha$]{\includegraphics[angle=0,width=0.43\linewidth]{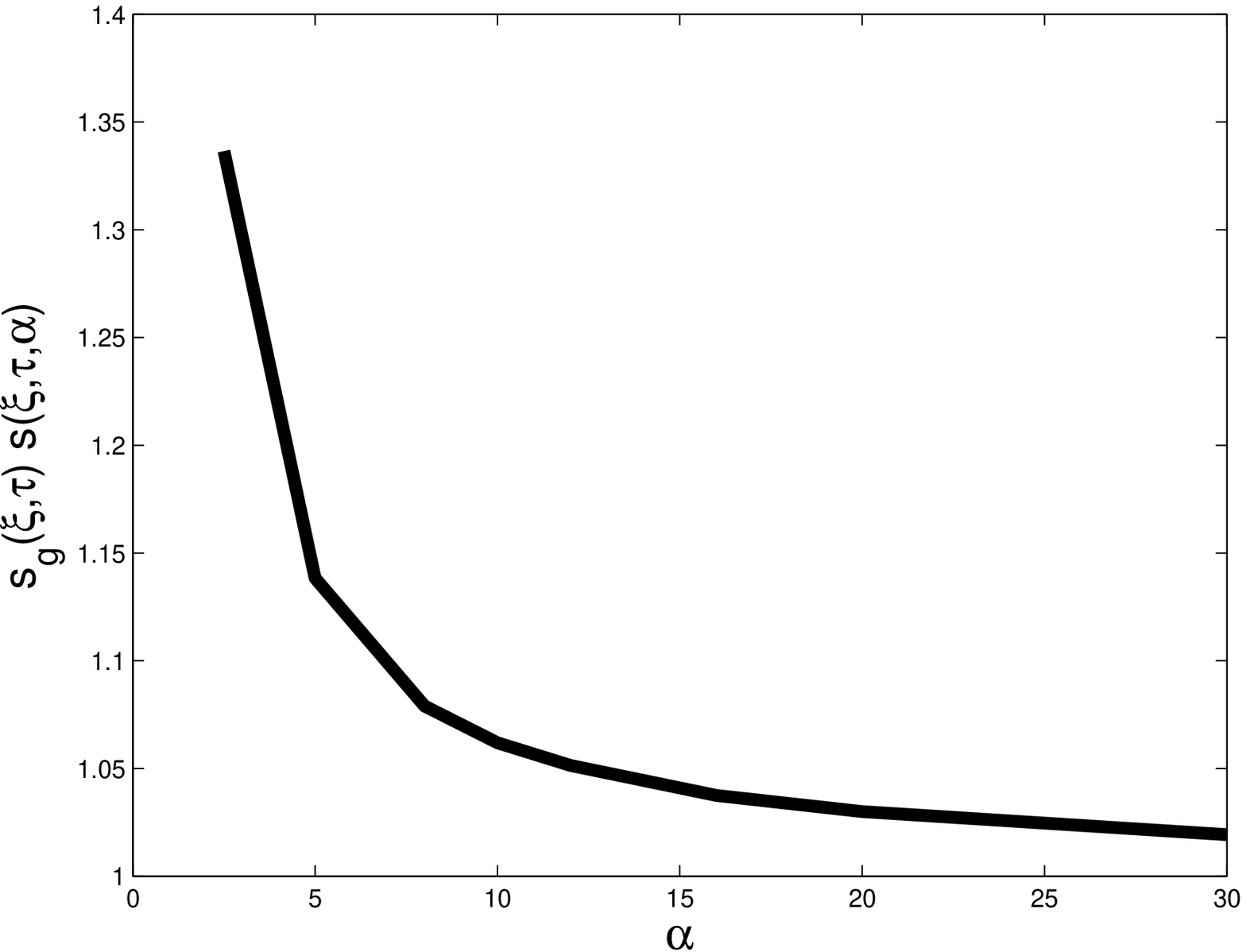}}
\subfloat[Sensitivity to $\rho$]{\includegraphics[angle=0,width=0.42\linewidth]{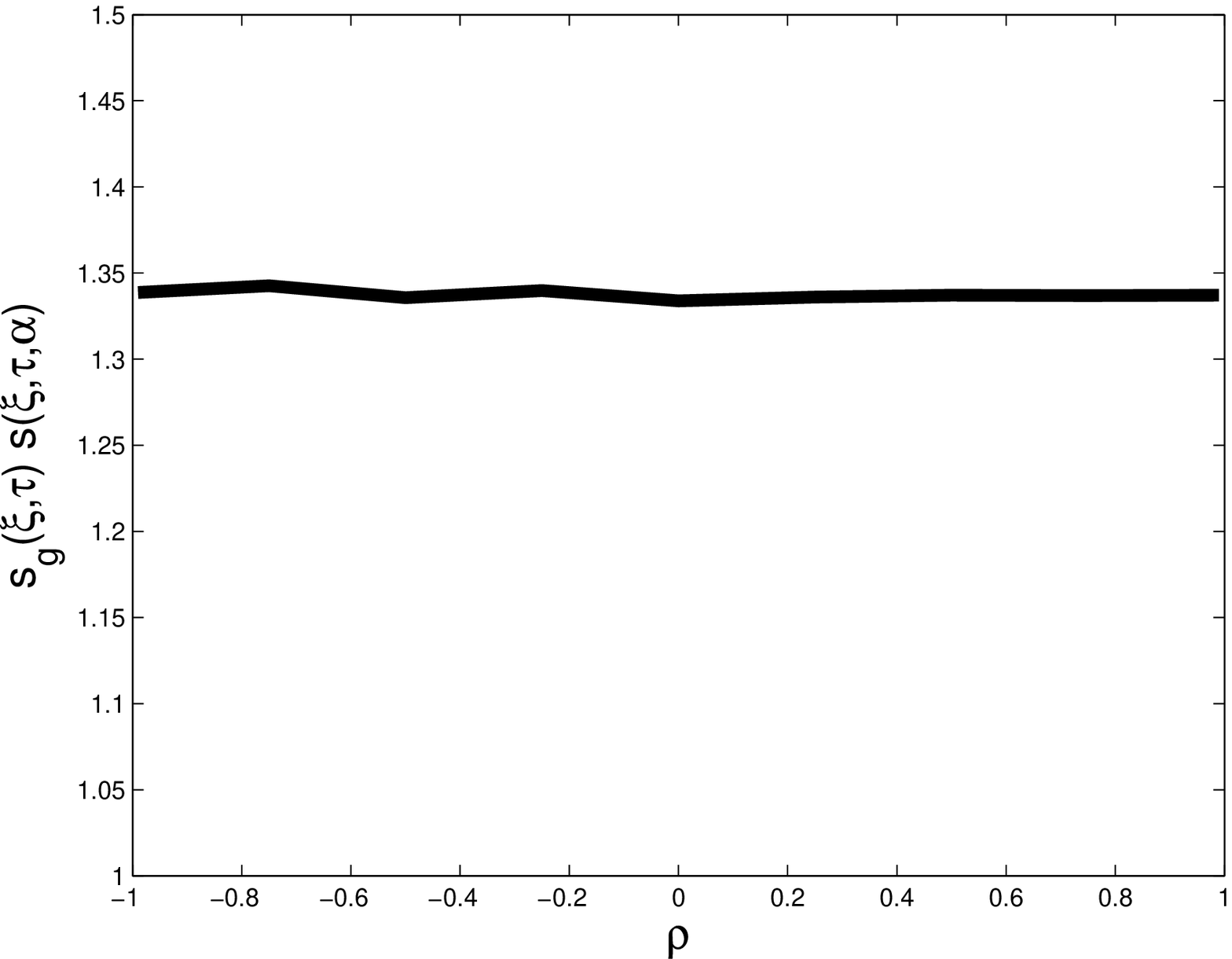}}\\
\begin{tabular}{p{12.5cm}}
\scriptsize Panel (a) shows the sensitivity of the nonlinear component to $\alpha$. The tail index varies from 2.5 to 30. Panel (b) shows the sensitivity to $\rho$ (for $\alpha=2.5$). Both plots are for $\tau=0.75$ and $\xi=0.95$.\\ 
\end{tabular}
  \end{center}
\end{figure}

Panel (a) of Figure \ref{fig:nonlinear} displays $s_g(\xi, \tau)s(\xi, \tau,\alpha)$ as a function of $\alpha$ and assuming a Student-$t$ distribution (and for $\xi=0.95$ and $\tau=0.75$). The tail index varies from 2.5 to 30. The nonlinear component decreases as $\alpha$ increases, and $s_g(\xi, \tau)s(\xi, \tau,\alpha)$ approaches  1 as $\alpha$ goes to $30$, or when the distribution is indistinguishable from the Gaussian. 
Panel (b) shows the sensitivity of $s_g(\xi, \tau)s(\xi, \tau,\alpha)$ to $\rho$ (with $\alpha=2.5$).
The nonlinear component is not affected by $\rho$, which confirms that $\sqrt{1+|\rho_{j\,k}|}$ and $s(\xi, \tau,\alpha)$ capture different aspects of $\mbox{TailCoR}^{(j\,k)\,\xi}$.



$\mbox{TailCoR}^{(j\,k)\,\xi}$ has numerous properties.
First, it captures nonlinear dependencies if tails deviate from Gaussianity,
i.e. $s(\xi,\tau,\alpha)s_g(\xi,\tau)$ can be greater, equal, or smaller than 1 (corresponding to tails fatter, equal or thinner than the Gaussian distribution). 
Second, even if $X_j$ and $X_k$ are linearly uncorrelated, $\mbox{TailCoR}^{(j\,k)\,\xi}$ is different from 1 if $X_j$ and $X_k$ are non-Gaussian. This is akin to the coefficient of tail dependence stemming from copula theory.\footnote{For instance, the tail dependence coefficient of a bivariate Student-$t$ copula is $2 t_{\alpha+1}\left(-\sqrt{\frac{(\alpha+1)(1-\rho)}{1+\rho}}\right)$, where $t_{\alpha+1}(\cdot)$ is a standardized Student-$t$ cumulative distribution function with tail index $\alpha+1$. Even if $\rho=0$ tail dependence is positive (unless $\alpha \rightarrow \infty$).} Third, if 
$\mathbf{X}_t$ is Gaussian, $s(\xi,\tau,\alpha)s_g(\xi,\tau)=1$ and $\mbox{TailCoR}^{(j\,k)\,\xi} = \sqrt{1+|\rho_{j\,k}|}$, \mbox{i.e.} the only source of dependence is linear.

\begin{figure}
  \begin{center}\caption{Sensitivity of $s_g(\xi,\tau)s(\xi,\tau,\alpha)$ to $\alpha$ and $\xi$}\label{fig:nonlinear2}
\includegraphics[angle=0,width=0.6\linewidth]{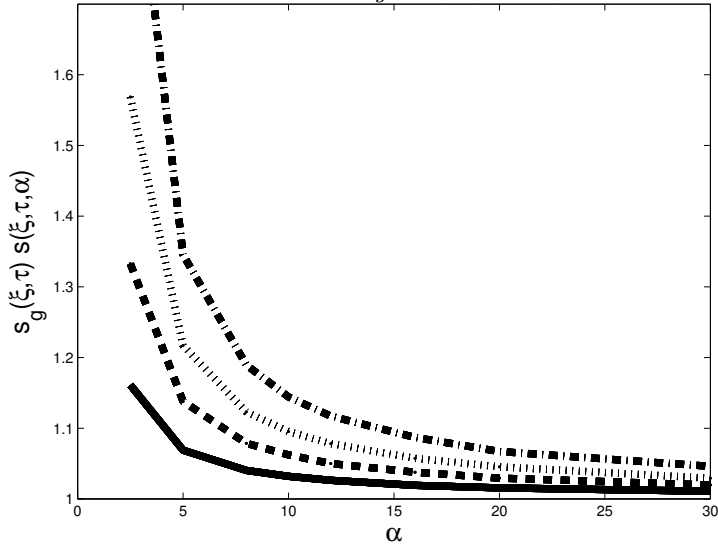}
\begin{tabular}{p{9cm}} 
\scriptsize Sensitivity of the nonlinear correlation to $\alpha$ for the Student-$t$ distribution. The tail index varies from 2.5 to 30. Each line is for a value of $\xi$: 0.90 (solid line), 0.95 (thick dashes), 0.975 (thin dashes) and 0.99 (thick and thin dashes).\\ 
\end{tabular}
  \end{center}
\end{figure}

This last property shows that under Gaussianity the upper bound is $\sqrt{2}$. Otherwise, the upper bound depends on $\alpha$ and $\xi$.
Figure \ref{fig:nonlinear2} displays similar curves to those in panel (a) of Figure \ref{fig:nonlinear} for values of $\xi$ typically used in practice: 0.90 (solid line), 0.95 (thick dashes), 0.975 (thin dashes), and 0.99 (thick and thin dashes). The further we explore the tails, the larger is $s_g(\xi, \tau)s(\xi, \tau,\alpha)$, and so is TailCoR. The range of values is $1-1.5$ (except if $\xi$ is close to 1 and tails are very heavy), which in turn translates into a range of values of TailCoR between 1 and 2.12 (e.g., if $\rho=1$, TailCoR equals $\sqrt{2}\times 1.5=2.12$).

We close this section with the issue that TailCoR is not a correlation in the traditional sense, as it is not centered at zero nor bounded above. A simple transformation consisting on subtracting 1, scaling by the value of TailCoR under perfect correlation, and multiplying by the sign of the correlation, leads to the following alternative expression
\begin{equation}\label{eq:alternative}
\mbox{TailCoR}^{(j\,k)\,\xi}_{alt} = \mbox{sign}(\rho_{j\,k})\frac{\mbox{TailCoR}^{(j\,k)\,\xi} - 1}{s_g(\xi, \tau)s(\xi, \tau,\alpha)\sqrt{2} - 1},
\end{equation}
which is bounded above and below by 1 and -1 and centered around 0, provided that TailCoR is larger than one. TailCoR smaller than one is pathological, as it occurs when tails are extremely light or when tails are light and the linear correlation is very low.

The only cases when $\mbox{TailCoR}^{(j\,k)\,\xi}_{alt}$ can be used even if TailCoR is smaller than one is when $X_{j\,t}$ and $X_{k\,t}$ are perfectly correlated. Then, $|\rho_{j\,k}|=1$ and \eqref{eq:alternative} equals 1 if $\rho_{j\,k}=1$, and -1 if $\rho_{j\,k}=-1$ regardless of the value of TailCoR.

If $\rho_{j\,k}=0$, \eqref{eq:alternative} becomes (assuming $\mbox{sign}(\rho_{j\,k})=1$ for $\rho_{j\,k}=0$)
\begin{equation*}
\mbox{TailCoR}^{(j\,k)\,\xi}_{alt} = \frac{s_g(\xi,\tau)s(\xi,\tau,\alpha) - 1}{s_g(\xi, \tau)s(\xi, \tau,\alpha)\sqrt{2} - 1},
\end{equation*}
which is zero only under Gaussianity; otherwise it is between -1 and 1.

\begin{figure}
  \begin{center}\caption{Alternative TailCoR}\label{fig:altTailCoR}
\includegraphics[angle=0,width=0.8\linewidth]{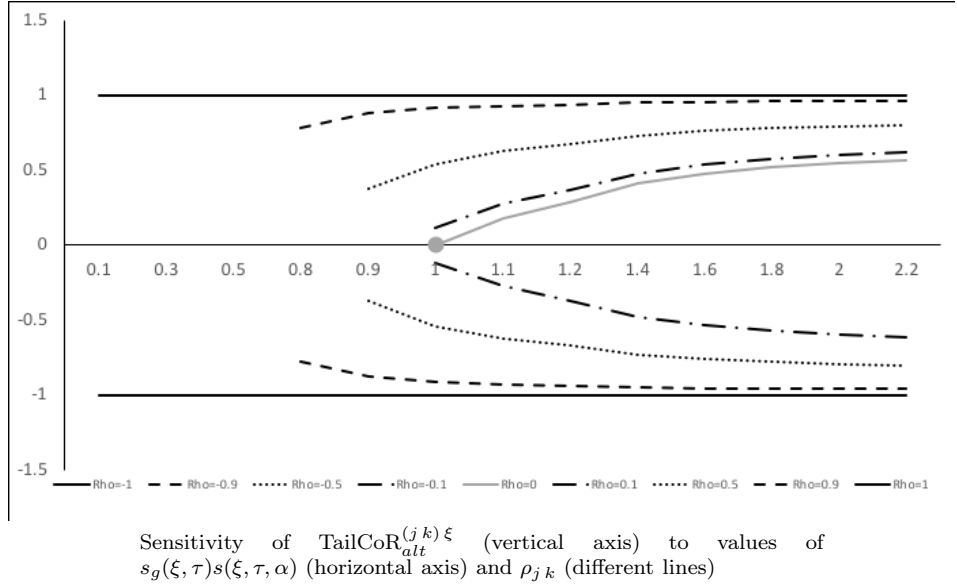}
\begin{tabular}{p{9cm}} 
\scriptsize Sensitivity of $\mbox{TailCoR}^{(j\,k)\,\xi}_{alt}$ (vertical axis) to values of $s_g(\xi,\tau)s(\xi,\tau,\alpha)$ (horizontal axis) and $\rho_{j\,k}$ (different lines)\\ 
\end{tabular}
  \end{center}
\end{figure}

Figure \ref{fig:altTailCoR} shows the value of $\mbox{TailCoR}^{(j\,k)\,\xi}_{alt}$ for different choices of $s_g(\xi,\tau)s(\xi,\tau,\alpha)$ and $\rho_{j\,k}$. The nonlinear component is in the horizontal axis and the nine lines represent values of $\rho_{j\,k}$: $\{1, 0.9, 0.5, 0.1, 0, -0.1, -0.5, -0.9, -1\}$. The horizontal solid lines at 1 and -1 are for $\rho_{j\,k}$ equal to 1 and -1 respectively, confirming what was mentioned above. As $\rho_{j\,k}$ gets closer to zero, the alternative TailCoR decreases, reaching zero when $s_g(\xi,\tau)s(\xi,\tau,\alpha)=1$ and $\rho_{j\,k}=0$ as represented by the grey dot. Note that, as explained earlier, $\mbox{TailCoR}^{(j\,k)\,\xi}_{alt}$ is not available for very light tails (e.g., $s_g(\xi,\tau)s(\xi,\tau,\alpha) \leq 0.5$ in the figure) or a combination of light tails and low correlation (e.g., $s_g(\xi,\tau)s(\xi,\tau,\alpha) = 0.9$ and $\rho_{j\,k}=0.1$ in the figure).



\subsection{Estimation}

Estimation under {\bf G1} -- {\bf G2} is straightforward and  divided in three simple steps.
\begin{description}
\item[Step 1] Standardize $X_{j\,t}$ with the sample median $\hat{\mbox{Q}}^{0.50}_j$ and the sample interquantile range $\hat{\mbox{IQR}}^{\tau}_{j}.$  Likewise for $X_{k\,t}$. Then compute the projection $Z_t^{(j\,k)}$.  
\item[Step 2] Estimate the IQR of the projection: $\hat{\mbox{IQR}}^{(j\,k)\,\xi}_{T}$.
\item[Step 3] Compute the normalization $s_g(\xi,\tau)$ and $\hat{\mbox{TailCoR}}^{(j\,k)\,\xi}_{T}(\hat{\mbox{Q}}^{0.50}_j,\hat{\mbox{Q}}^{0.50}_k,\hat{\mbox{IQR}}^{\tau}_{j},\hat{\mbox{IQR}}^{\tau}_{k}) = s_g(\xi,\tau)\hat{\mbox{IQR}}^{(j\,k)\,\xi}_{T}$.\footnote{This somehow cumbersome notation emphasises that TailCoR is a function of estimated quantiles (step 1)}

\end{description}


Under {\bf E1} -- {\bf E5} the linear correlation $\rho_{j\,k}$ is estimated with a robust method. \cite{LindskogMcNeilSchmock03} introduce a robust estimator that is invariant in the class of elliptical distributions. Let $\hat{\kappa}_{j\,k,T}$ be the estimator of  Kendall's correlation.
Then
\begin{equation*}
\hat{\rho}_{j\,k,T} = \sin \left(\frac{\pi}{2}\hat{\kappa}_{j\,k,T} \right),
\end{equation*}
and $\sqrt{1+|\hat{\rho}_{j\,k,T}|}$ follows. Given the $\hat{\mbox{IQR}}^{(j\,k)\,\xi}_{T}$ obtained in step 2 above, the estimator of the nonlinear component is
\begin{equation*}
\hat{s}(\xi,\tau,\alpha)_T = \frac{\hat{\mbox{IQR}}^{(j\,k)\,\xi}_{T}}{\sqrt{1+|\hat{\rho}_{j\,k,T}|}}.
\end{equation*}

We now see the computational advantages of $\mbox{TailCoR}^{(j\,k)\,\xi}$.
It can be estimated exactly for any probability level $\xi$
and no optimizations are needed, as it is based on simple steps each requiring a few lines of programming code. This makes $\mbox{TailCoR}^{(j\,k)\,\xi}$ fast to compute.
Moreover, estimation of the tail index is not required.

The following theorem shows the consistency of $\hat{\mbox{TailCoR}}^{(j\,k)\,\xi}_{T}$.

\begin{description}
\item[{\bf Theorem 2}]
Let $X_{j\,t}$ and $X_{k\,t}$ be two elements of the random vector $\mathbf{X}_t$ that fulfills assumptions {\bf G1} -- {\bf G2}. Then, as $T \rightarrow \infty$ 
 \begin{equation*}
     \hat{\mbox{TailCoR}}^{(j\,k)\,\xi}_{T}(\hat{\mbox{Q}}^{0.50}_j,\hat{\mbox{Q}}^{0.50}_k,\hat{\mbox{IQR}}^{\tau}_{j},\hat{\mbox{IQR}}^{\tau}_{k}) - \mbox{TailCoR}^{(j\,k)\,\xi}=o_p(1).
 \end{equation*}
\end{description}
\begin{description}
\item[{\bf Proof}] See Appendix P.
\end{description}

Next, we prove asymptotic normality. Admittedly, the theorem ignores the effect of the estimated median and the IQR in the standardization of $X_{j\,t}$ and $X_{k\,t}$ (step 1). Monte Carlo results shown below indicate that the effect of step 1 is negligible. 
 
\begin{description}
\item[{\bf Theorem 3}] Let $X_{j\,t}$ and $X_{k\,t}$ be two elements of the random vector $\mathbf{X}_t$ that fulfills assumptions {\bf E1} -- {\bf E5}. Let $f_{(j\,k)}(\cdot)$ and $F_{(j\,k)}(\cdot)$ be the probability density and cumulative distribution functions of $Z^{(j\,k)}_t$, and let $I_{\{\cdot\}}$ be an indicator function that takes value 1 if its argument is true. Then, as $T \rightarrow \infty$
\begin{equation*}
\sqrt{T}\left(\hat{\mbox{TailCoR}}^{(j\,k)\,\xi}_{T}(\mbox{Q}^{0.50}_j,\mbox{Q}^{0.50}_k,\mbox{IQR}^{\tau}_{j},\mbox{IQR}^{\tau}_{k}) - \mbox{TailCoR}^{(j\,k)\,\xi}\right) \rightarrow_d \mathcal{N}\left(0,4s_g(\xi,\tau)^2\frac{\Gamma(Q^{(j\,k)\xi})}{f^2_{(j\,k)}(F^{-1}_{(j\,k)}(\xi))}\right),
\end{equation*}
where
\begin{eqnarray*}
\Gamma(Q^{(j\,k)\xi}) &=& \sum_{t=-\infty}^{+\infty} {\rm E}\left(W_0(Q^{(j\,k)\xi})W_t(Q^{(j\,k)\xi})\right), \\\\ 
W_t(Q^{(j\,k)\xi}) &=& I_{\{Z^{(j\,k)}_t \leq Q^{(j\,k)\xi}\}}-P(Z^{(j\,k)}_t \leq Q^{(j\,k)\xi}),
\end{eqnarray*}
and $\rightarrow_d$ stands for convergence in distribution.
\end{description}

\begin{description}
\item[{\bf Proof}] See Appendix P.
\end{description}

We make the following remarks about the theorem. First, the univariate density $f_{(j\,k)}(\cdot)$ in the denominator is symmetric,
and therefore easy to compute. Second, $\Gamma(Q^{(j\,k)\xi})$ is the long-run component of the variance that accounts for the time dependence.
Third, the asymptotic variance can be computed by bootstrap, as we do in the empirical application (we use block bootstrap of length 50 and 500 replications). Last, it is possible to derive an equivalent asymptotic distribution under {\bf G1} -- {\bf G2},
as shown in the following corollary.  

\begin{description}
\item[{\bf Corollary 1}]
Let $X_{j\,t}$ and $X_{k\,t}$ be two elements of the random vector $\mathbf{X}_t$ that fulfills assumptions {\bf G1} -- {\bf G2}. Let $f_{(j\,k)}(\cdot)$ and $F_{(j\,k)}(\cdot)$ be the probability density and cumulative distribution functions of $Z^{(j\,k)}_t$, and let $I_{\{\cdot\}}$ be an indicator function that takes value 1 if its argument is true. Then, as $T \rightarrow \infty$
\begin{equation*}
\sqrt{T}\left(\hat{\mbox{TailCoR}}^{(j\,k)\,\xi}_{T}(\mbox{Q}^{0.50}_j,\mbox{Q}^{0.50}_k,\mbox{IQR}^{\tau}_{j},\mbox{IQR}^{\tau}_{k}) - \mbox{TailCoR}^{(j\,k)\,\xi}\right) \rightarrow_d \mathcal{N}\left(0,s_g(\xi,\tau)^2\Upsilon\right),
\end{equation*}
where 
\begin{eqnarray*}
&&\Upsilon = 
\frac{\Gamma(Q^{(j\,k)\xi})}{f^2_{(j\,k)}(F^{-1}_{(j\,k)}(\xi))} + \frac{\Gamma(Q^{(j\,k)1-\xi})}{f^2_{(j\,k)}(F^{-1}_{(j\,k)}(1-\xi))} - 2 \frac{\Gamma(Q^{(j\,k)\xi},Q^{(j\,k)1-\xi})}{f_{(j\,k)}(F^{-1}_{(j\,k)}(\xi))f_{(j\,k)}(F^{-1}_{(j\,k)}(1-\xi))}, \vspace{0.2cm}\\\\
&&\Gamma(Q^{(j\,k)\xi},Q^{(j\,k)1-\xi}) = \sum_{t=-\infty}^{+\infty} {\rm E}\left(W_0(Q^{(j\,k)\xi})W_t(Q^{(j\,k)1-\xi})\right), \mbox{ and } \\\\
&& W_t(Q^{(j\,k)\xi}) = I_{\{Z^{(j\,k)}_t \leq Q^{(j\,k)\xi}\}}-P(Z^{(j\,k)}_t \leq Q^{(j\,k)\xi}).
\end{eqnarray*}
\end{description}

\begin{description}
\item[{\bf Proof}] If follows from the proof of Theorem 3 but with $Q^{(j\,k)\xi} \neq - Q^{(j\,k)1-\xi}$.
\end{description}


\subsection{Multivariate TailCoR}

So far we only considered the pair $(j\,k)$ of random variables while the random vector $\mathbf{X}_t$ is of dimension $N$. Considering all the pairs leads to a $N \times N$ symmetric matrix of TailCoRs (including TailCoR of a random variable with itself). For the ease of exposition let $\tilde{N}=N(N+1)/2$. We denote by $\xi_{(j\,k)}$ the probability level at which we compute the IQR for the $(j\,k)$ projection (that is chosen optimally as explained earlier). We assume $\xi_{(j\,k)}=\xi$ $\forall j,k$. This assumption can be easily relaxed, at the expense of notation. 

\begin{description}
\item[{\bf Definition 2}] Under {\bf G1}-{\bf G2}, the matrix of TailCoRs is defined as follows
\begin{equation}\label{eq:vectortailcor}
\mbox{\bf TailCoR}^{\xi} := s_0(\xi,\tau)\mbox{\bf IQR}^{\xi},
\end{equation}
where $\mbox{\bf IQR}^{\xi}$ is a matrix of IQRs of the $N \times N$ projections.
\end{description}

Under {\bf E1} -- {\bf E5}, (\ref{eq:vectortailcor}) becomes
\begin{equation}\label{eq:vectortailcorellip}
\mbox{\bf TailCoR}^{\xi} = \sqrt{2}s_g(\xi,\tau)s(\xi,\tau,\alpha)\mathbf \Psi,
\end{equation}
where the matrix $\mathbf \Psi$ has $(j\,k)$ element $\sqrt{\frac{1+|\rho_{j\,k}|}{2}}$. This matrix is symmetric, with unitary diagonal elements, and off-diagonal elements bounded above and below by 1 and $\sqrt{1/2}$ respectively.
It is invariant to location-scale shifts of $\mathbf{X}_t$, and positive definite.

Estimation follows the same steps as in the univariate case under {\bf G1} -- {\bf G2}: $\hat{\mbox{\bf TailCoR}}^{\xi}_{T} = s_g(\xi,\tau)\hat{\mbox{\bf IQR}}^{\xi}$. Under {\bf E1} -- {\bf E5}, a new step is added, as $s(\xi,\tau,\alpha)$ is the same for all pairs. Let us rewrite $\hat{s}(\xi,\tau,\alpha)^{(j\,k) \, T}$ in terms of its $\tilde{N}$  components as $\hat{s}(\xi,\tau,\alpha)_{h , T}$, $h=1,\ldots,\tilde{N}$. The nonlinear correlation is estimated by pooling the pairwise estimators:
\begin{equation*}
\hat{s}(\xi,\tau,\alpha)_{T} = \frac{1}{\tilde{N}}\sum_{h=1}^{\tilde{N}} \hat{s}(\xi,\tau,\alpha)_{h ,T}.
\end{equation*}
Estimating tail indices by cross-sectionally averaging can be found in, for instance, \cite{Nolan10} and \cite{DominicyOgataVeredas12}.

The asymptotic distribution requires the use of results in \cite{DominicyHormannOgataVeredas12} on covariances between the sample quantiles of marginal distributions under S-mixing.
The following theorem shows the asymptotic distribution of the vectorized $\hat{\mbox{\bf TailCoR}}^{\xi}_{T}$.

\begin{description}
\item[{\bf Theorem 4}] 
Let $\mathbf{X}_t$ be a random vector that fulfills assumptions {\bf E1} -- {\bf E5}. 
Let
$$vech \hat{\mbox{\bf TailCoR}}^{\xi}_{T}(\mbox{\bf Q}^{0.50}_j,\mbox{\bf Q}^{0.50}_k,\mbox{\bf IQR}^{\tau}_{j},\mbox{\bf IQR}^{\tau}_{k})$$
be the $\tilde{N} \times 1$ half-vectorized matrix of $\hat{\mbox{\bf TailCoR}}^{\xi}_{T}(\mbox{\bf Q}^{0.50}_j,\mbox{\bf Q}^{0.50}_k,\mbox{\bf IQR}^{\tau}_{j},\mbox{\bf IQR}^{\tau}_{k})$. Likewise for $vech \mbox{\bf TailCoR}^{\xi}$. Let $f_{j}(\cdot)$ and $F_{j}(\cdot)$ be the probability density and cumulative distribution functions of the $j$th projection ($j=1,\ldots,\tilde{N}$), and let $I_{\{\cdot\}}$ be an indicator function that takes value 1 if its argument is true. Then, as $T \rightarrow \infty$
\begin{equation*}
\sqrt{T}\left(vech \hat{\mbox{\bf TailCoR}}^{\xi}_{T}(\mbox{\bf Q}^{0.50}_j,\mbox{\bf Q}^{0.50}_k,\mbox{\bf IQR}^{\tau}_{j},\mbox{\bf IQR}^{\tau}_{k}) - vech \mbox{\bf TailCoR}^{\xi}\right) \rightarrow_d \mathcal{N}\left(0,4s_g(\xi,\tau)^2\bm\Omega\right),
\end{equation*}
where $\bm\Omega$ is a $N \times N$ matrix with $(j\,j)$ element
\begin{eqnarray*}
\Omega_{j\,j} &=& \frac{\Gamma_{j\,j}(Q^{\xi}_j)}{f^2_j(F_j^{-1}(\xi))}, \mbox{ where} \\
\Gamma_{j\,j}(Q^{\xi}_j) = \sum_{t=-\infty}^{\infty} {\rm E}(W_0(Q^{\xi}_j),W_t(Q^{\xi}_j)) &\mbox{ and }& W_t(Q^{\xi}_j) = I_{\{Z_{j\,t} \leq Q^{\xi}_j \}} - P(Z_{j\,t} \leq Q^{\xi}_j).
\end{eqnarray*}
The $(j\,k)$ element of $\bm\Omega$ is
\begin{eqnarray*}
\Omega_{j\,k} &=& \frac{\Gamma_{j\,k}(Q^{\xi}_j,Q^{\xi}_k)}{f_j(F_j^{-1}(\xi))f_k(F_k^{-1}(\xi))} \forall j \neq k, \mbox{ where} \\
\Gamma_{j\,k}(Q^{\xi}_j) &=& \sum_{t=-\infty}^{\infty} {\rm E}(W_0(Q^{\xi}_j,Q^{\xi}_k),W_t(Q^{\xi}_j,Q^{\xi}_k)) \mbox{ and }\\
W_t(Q^{\xi}_j,Q^{\xi}_k) &=& I_{\{Z_{j\,t} \leq Q^{\xi}_j,Z_{k\,t} \leq Q^{\xi}_k \}} - P(Z_{j\,t} \leq Q^{\xi}, Z_{k\,t} \leq Q^{\xi}_k).
\end{eqnarray*}
\end{description}

\begin{description}
\item[{\bf Proof}] See Appendix P.
\end{description}

\section{A Monte Carlo simulation study}

We analyze the finite sample properties of TailCoR with three bivariate elliptical distributions: Gaussian, Student-$t$ with $\alpha=2.5$, and ES with $\alpha=1.5$. The most heavy tailed distribution is the Student-$t$, followed by the ES and the Gaussian. The location parameters are set to zero and the dispersion matrix has unitary diagonal elements and off-diagonal elements $0.50$. We consider three sample sizes $T=\{1000, 5000, 10000\}$ and two replication sizes $H=\{1000,10000\}$. In the sequel we show results for $T=10000$ and $H=10000$, unless otherwise stated. Results for other configurations are alike and they are available upon request.


\begin{figure}
  \begin{center}\caption{TailCoR for different distributions} 
\includegraphics[angle=0,width=0.6\linewidth]{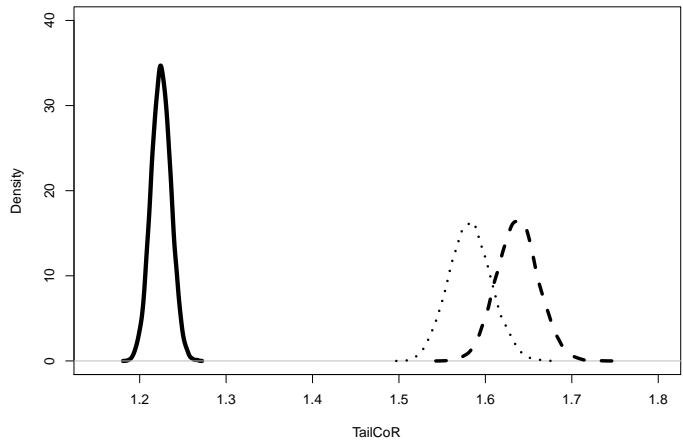}\label{fig:distributions}
\begin{tabular}{p{9cm}} 
\scriptsize Distribution of 10000 estimated TailCoR's for $\xi=0.95$ and three distributions: Gaussian (solid line), Student-$t$ with $\alpha=2.5$ (dashed line), ES with $\alpha=1.5$ (dotted line).\\ 
\end{tabular}
  \end{center}
\end{figure}

Figure \ref{fig:distributions} shows the finite sample distributions of the TailCoR estimates for $\xi=0.95$ and for the three distributions (solid line for the Gaussian, dashed for the Student-$t$, and dotted for the ES). In all cases, TailCoR is larger than one. The estimated TailCoR is more precise under Gaussianity than under heavy tails, as it only depends on the linear correlation. Moreover, the median is around 1.22, very close to the true value $\sqrt{1+0.50}=1.225$. Estimates under the Student-$t$ and the ES have higher medians (1.58 for the ES and 1.64 for the Student-$t$), reflecting the nonlinear dependencies.



\begin{figure}
  \caption{Sensitivity of TailCoR to $\xi$}\label{fig:xi}
  \begin{center}
    \subfloat[Gaussian]{\includegraphics[angle=0,width=.35\linewidth]{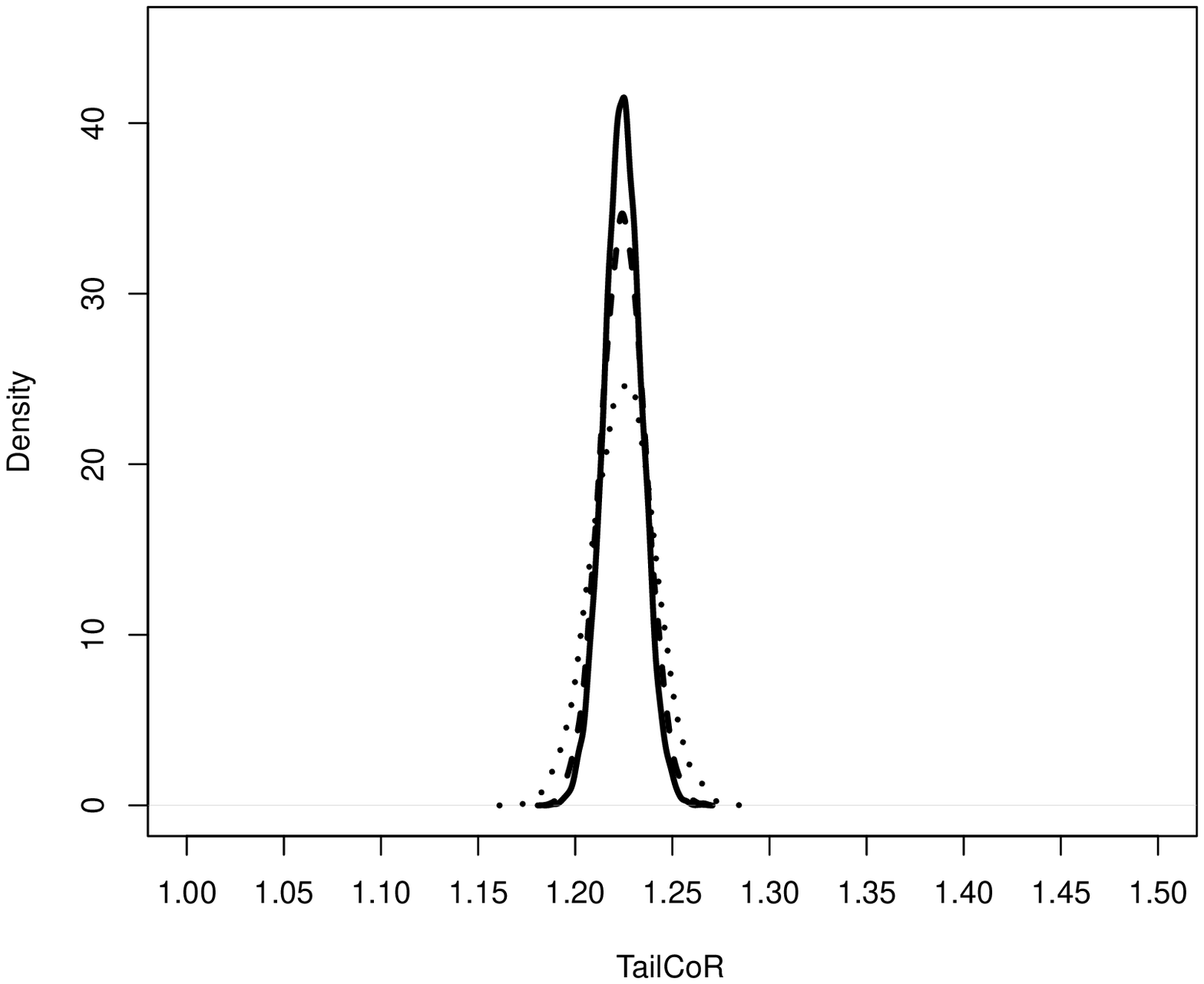}}
    \subfloat[Student-$t$]{\includegraphics[angle=0,width=.35\linewidth]{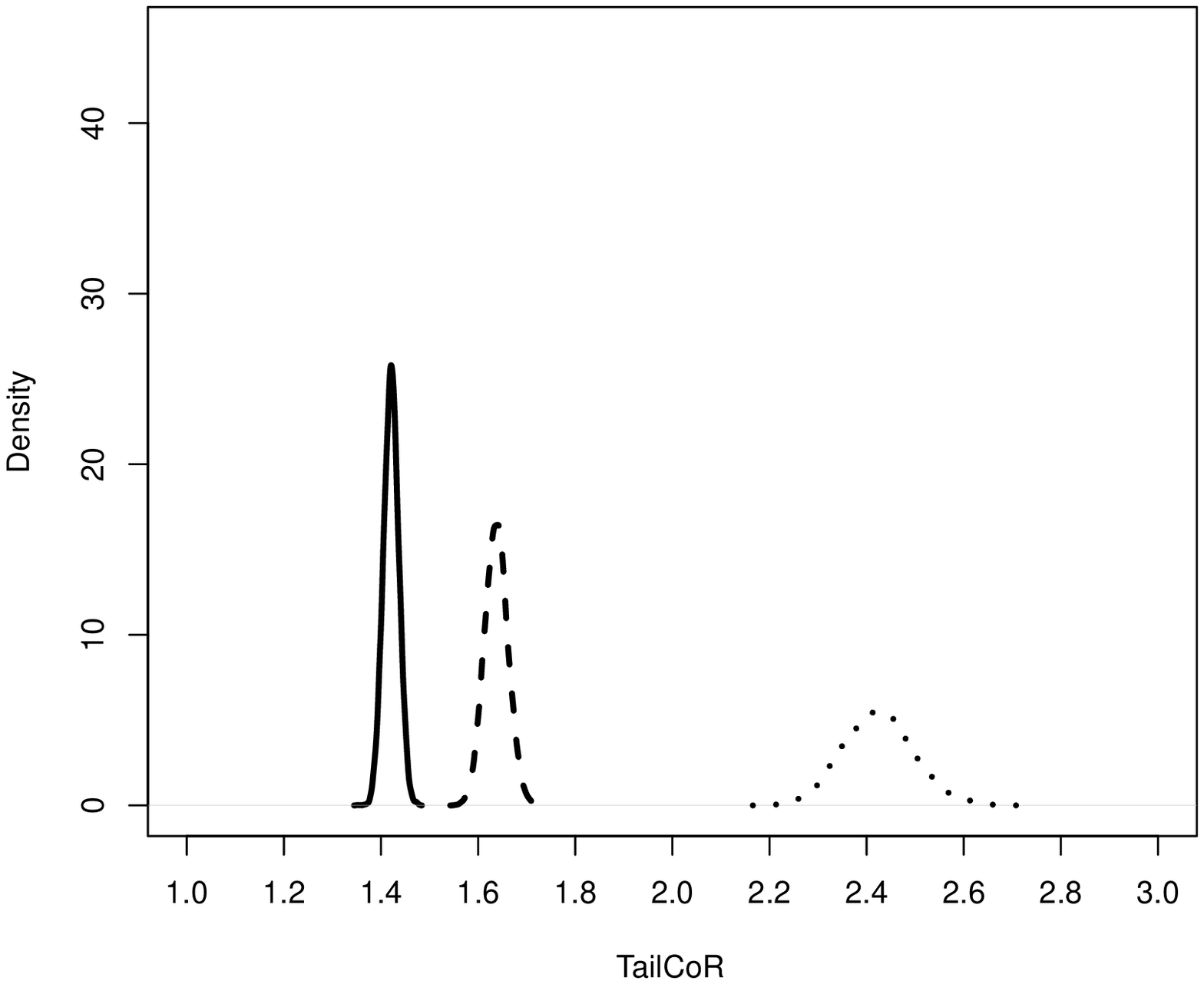}}\\
    \subfloat[ES]{\includegraphics[angle=0,width=.35\linewidth]{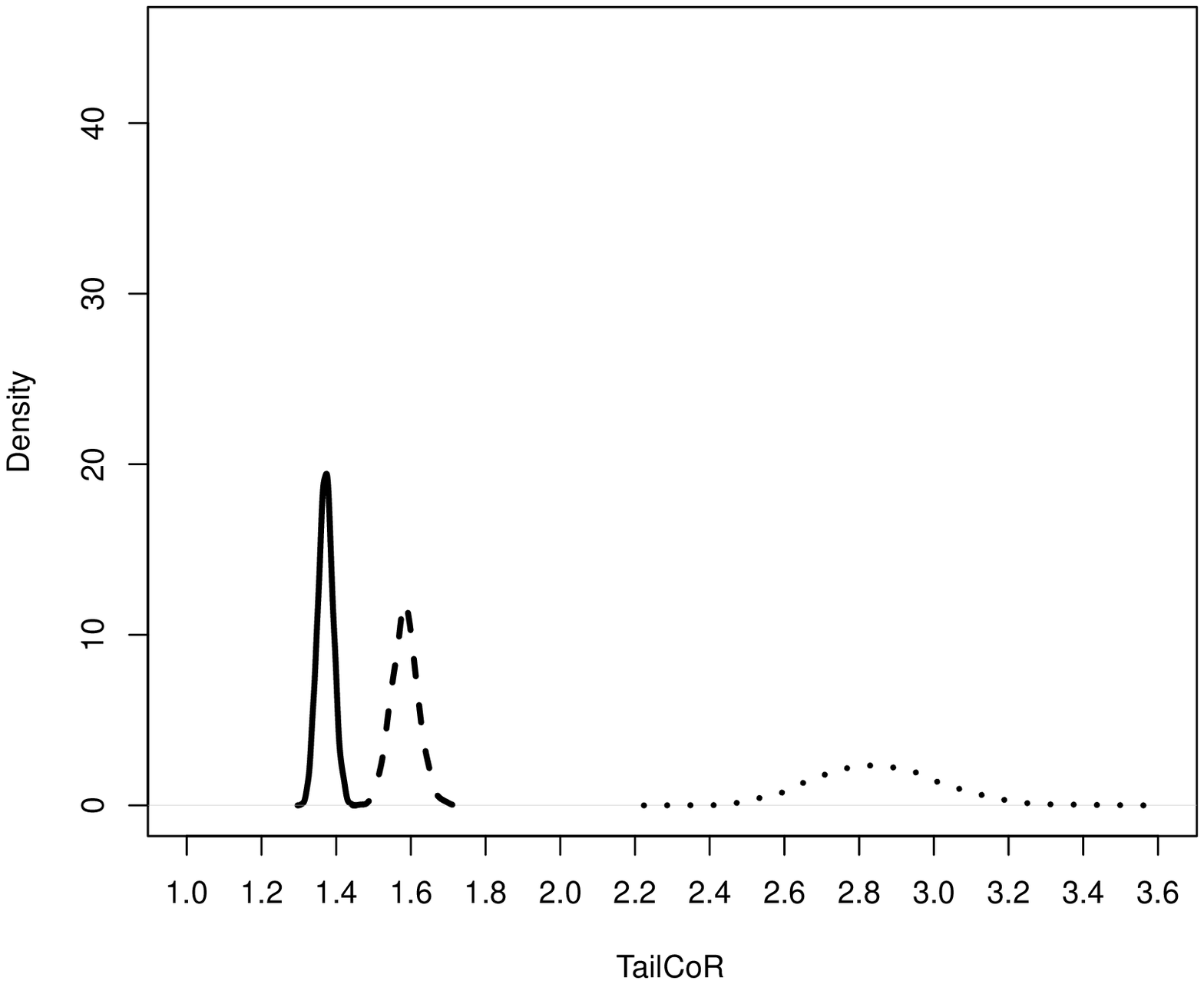}} \\
\begin{tabular}{p{11cm}}
\scriptsize Sensitivity of TailCoR to $\xi$ for the Gaussian (panel a), Student-$t$ with $\alpha=2.5$ (panel b), and ES with $\alpha=1.5$ (panel c) distributions. Each line is the density of the 10000 estimates of TailCoR for different values of $\xi$: 0.90 (solid line), 0.95 (dashed) and 0.99 (dotted).\\ 
\end{tabular}
  \end{center}
\end{figure}

Figure \ref{fig:xi} shows the sensitivity of TailCoR to $\xi$ for the Gaussian (panel (a)), Student-$t$ with $\alpha=2.5$ (panel (b)), and ES with $\alpha=1.5$ (panel (c)). Solid lines are for $\xi=0.90$, dashed for $\xi=0.95$, and dotted for $\xi=0.99$.
The densities overlap for the Gaussian distribution since $\mbox{TailCoR}^{(j\,k)\,\xi}$ does not depend on $\xi$ (it equals $\sqrt{1+|\rho_{j\,k}|}$).
The small differences are due to finite sample discrepancies. Regarding the other distributions, results show that, as expected, TailCoR increases with $\xi$.

\begin{figure}
  \caption{Convergence in distribution}\label{fig:KDE}
  \begin{center}
    \subfloat{\includegraphics[angle=0,height =5cm,width=.30\linewidth]{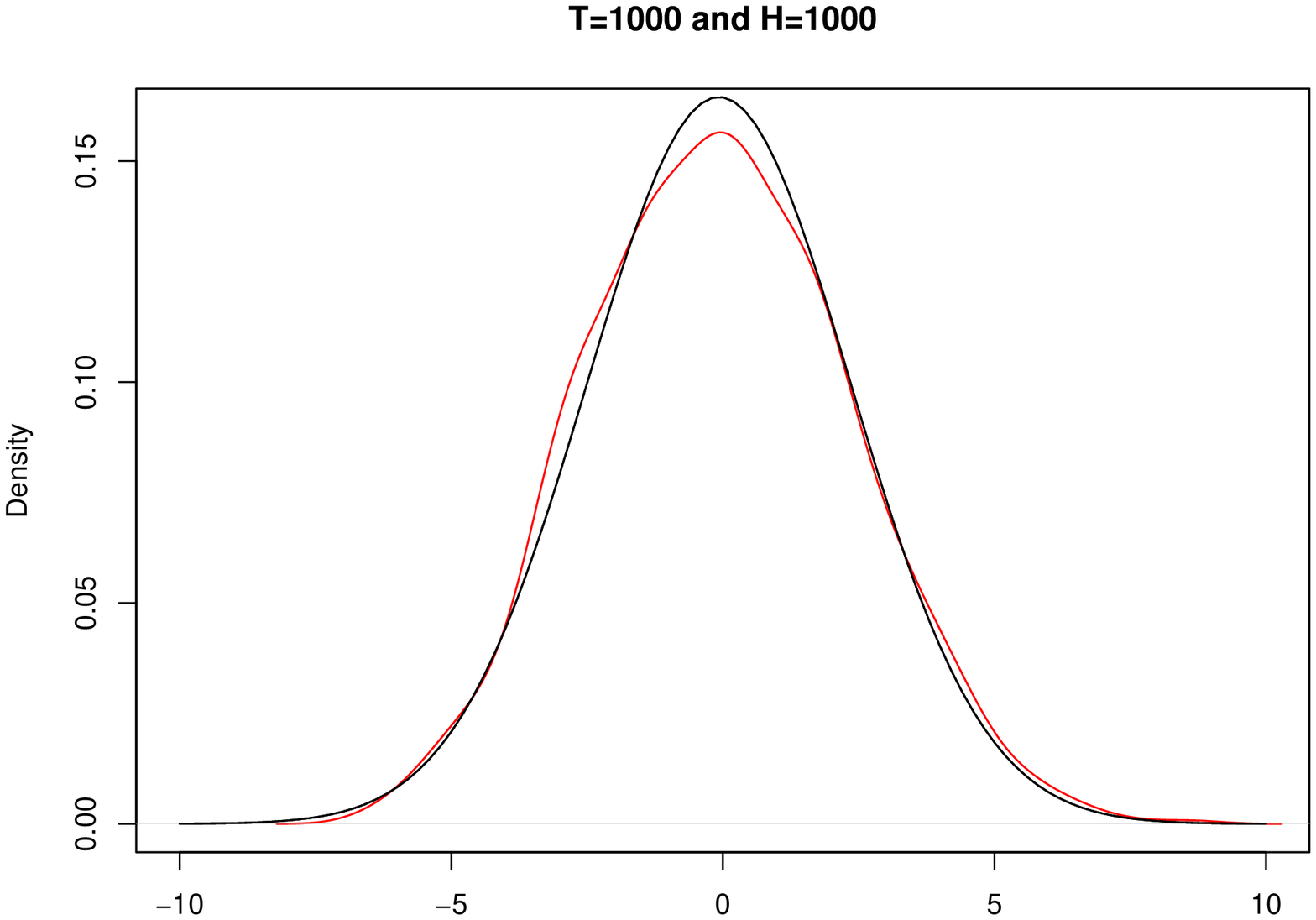}}
    \subfloat{\includegraphics[angle=0,height =5cm,width=.30\linewidth]{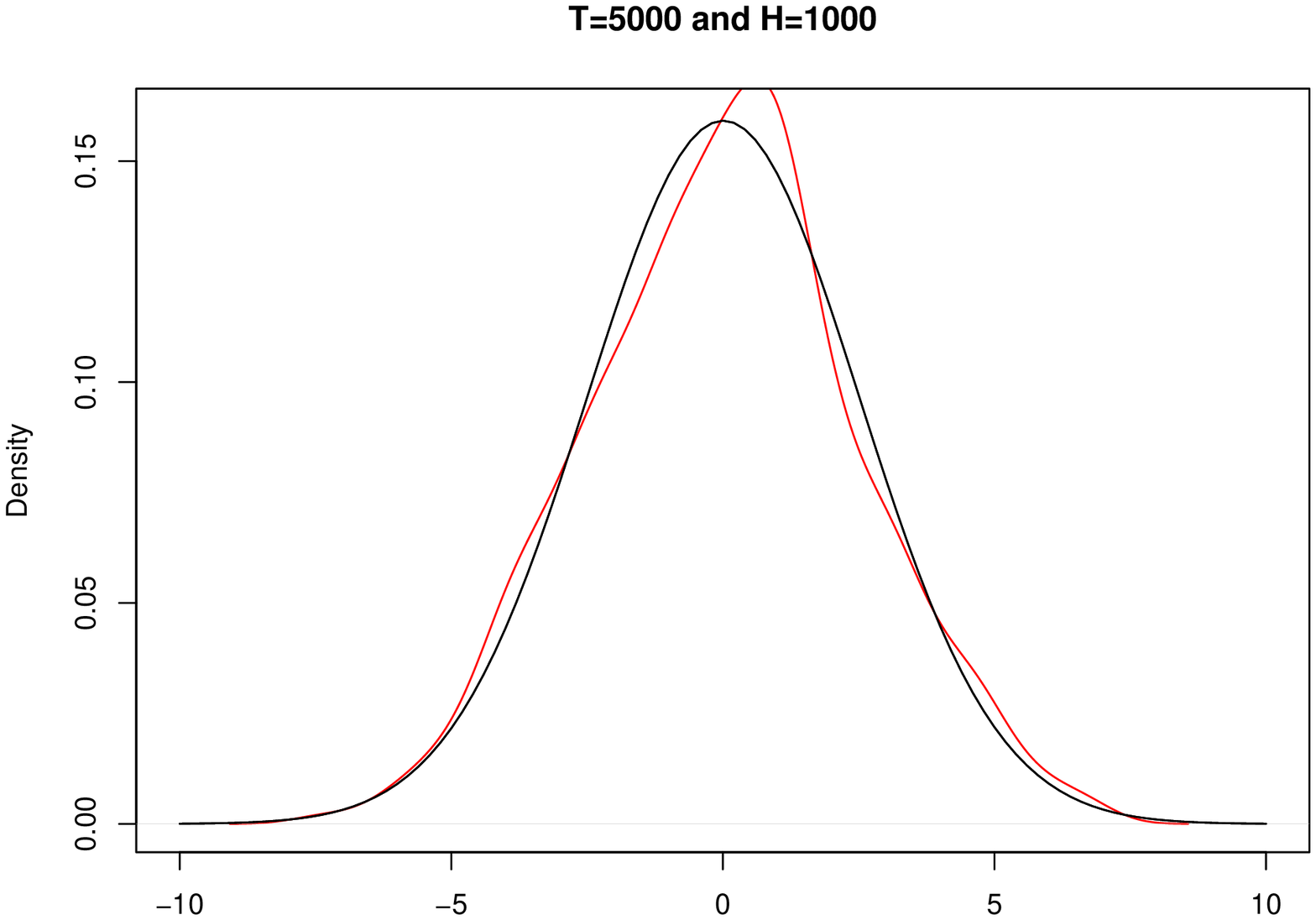}}
    \subfloat{\includegraphics[angle=0,height =5cm,width=.30\linewidth]{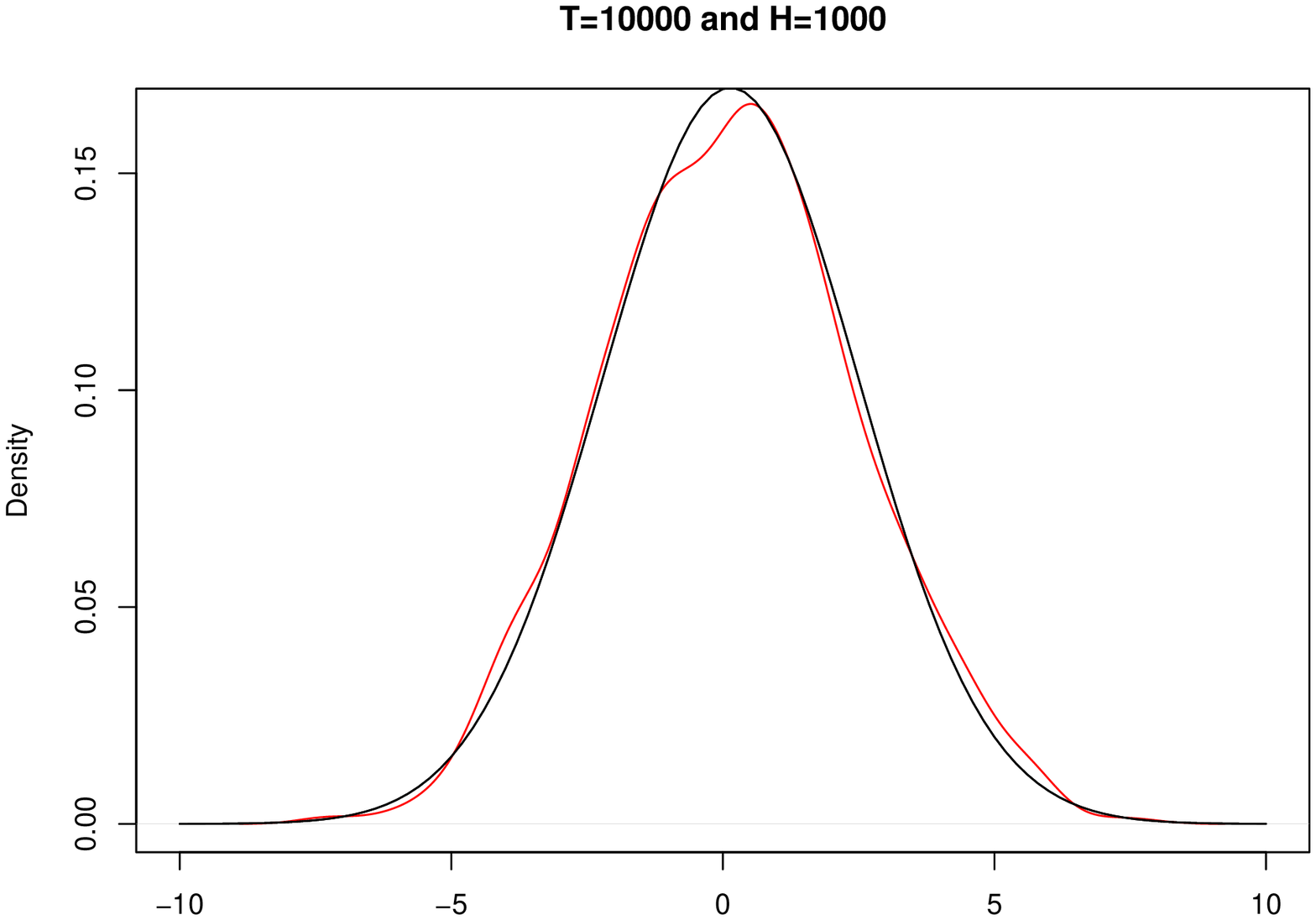}}\\
    \subfloat{\includegraphics[angle=0,height =5cm,width=.30\linewidth]{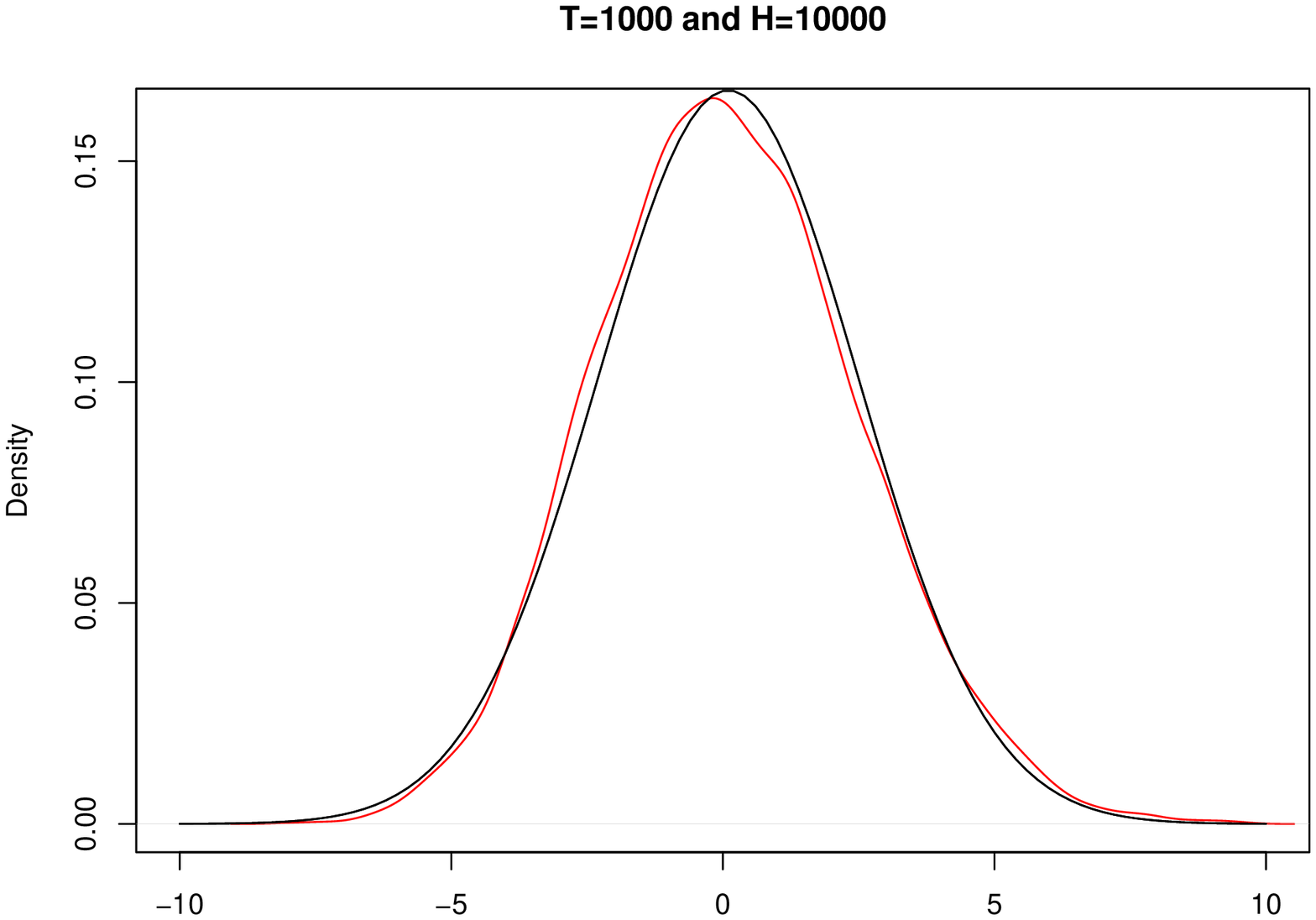}} 
    \subfloat{\includegraphics[angle=0,height =5cm,width=.30\linewidth]{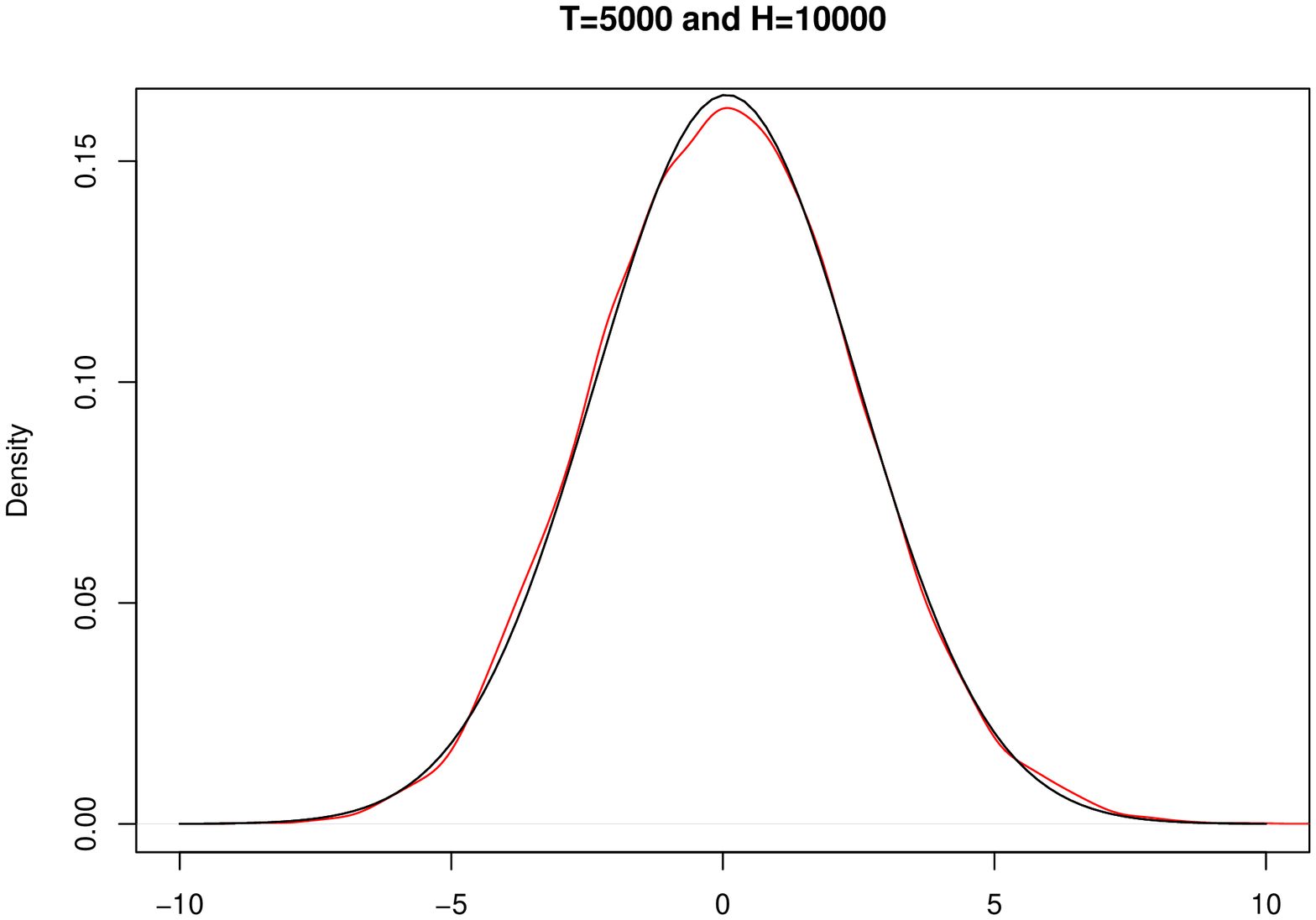}} 
    \subfloat{\includegraphics[angle=0,height =5cm,width=.30\linewidth]{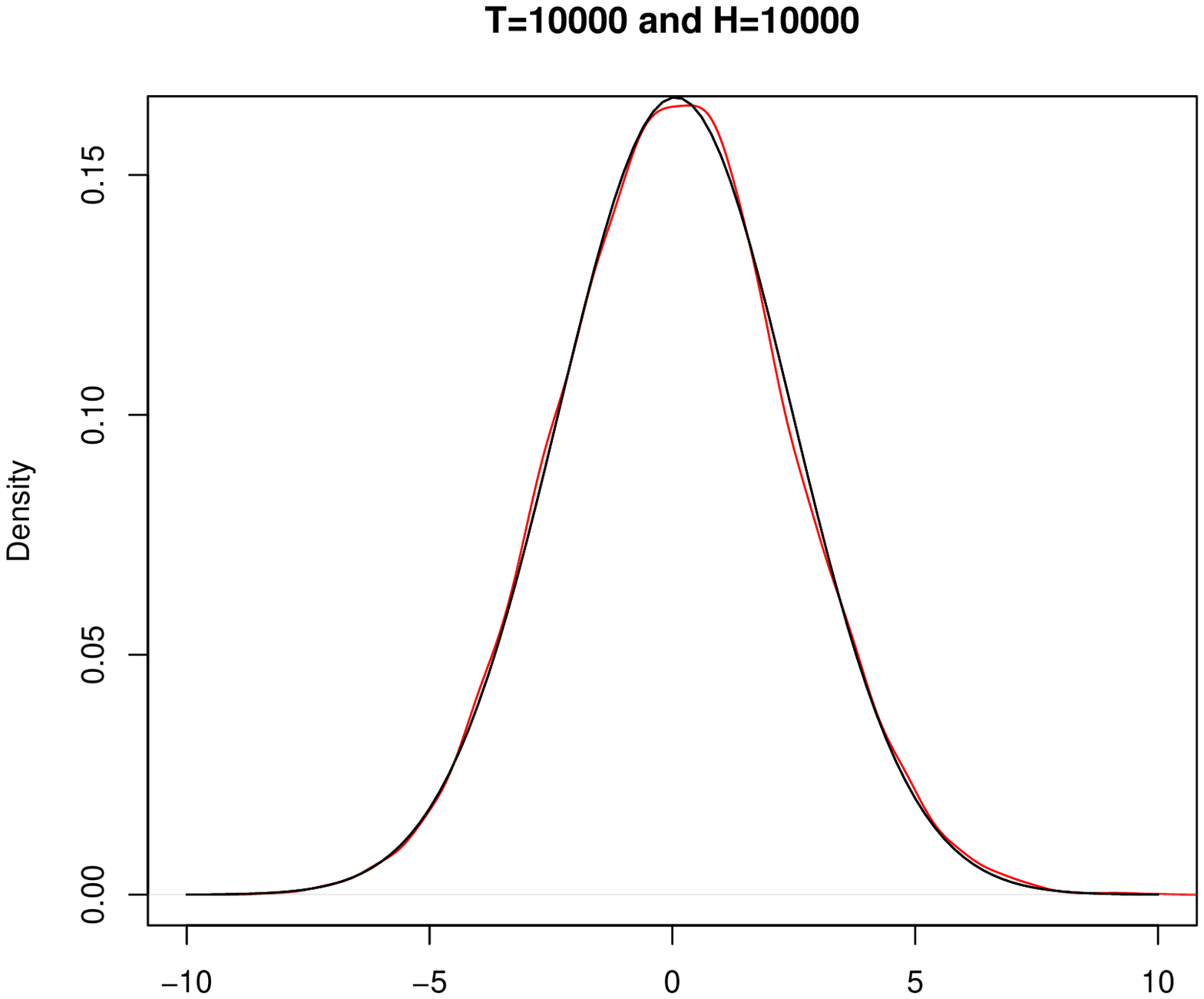}} \\
\begin{tabular}{p{13cm}}
\scriptsize Kernel densities (red line)  of the centered estimates against a corresponding Gaussian distribution (black line) for all the combinations of the sample size and replications (indicated in the top of each plot).\\ 
\end{tabular}
  \end{center}
\end{figure}

The precision and convergence in distribution of the estimator are shown in Figure \ref{fig:KDE} and in Table \ref{tab:step1}  for $\xi=0.95$. Red lines in the figure are the kernel densities of the centered estimates for different sample sizes and replications (indicated at the top of each plot), and the black lines are the corresponding Gaussians. As expected, the kernel densities approach to the limiting distribution as the sample size and the number of replications increase. The table shows the effect of the estimation error of step 1. It compares TailCoR and its components when the population and the sample quantiles are used in step 1 (as shown in the row "Step 1"). Estimations are done for the Gaussian and Student-$t$ distributions, two sample sizes (1000 and 10000), 1000 replications, and $\xi=0.95$. Mean estimates are very close to the true values, indicating unbiasedness. The standard errors are very small and, more importantly, using the sample quantiles in step 1 does not affect significantly the accuracy of the estimates in step 2.

\begin{table}
\begin{center}
\caption{Estimation uncertainty step 1} \label{tab:step1}
\begin{tabular}{lccccc}
\hline\hline
Step 1:          &                  & Popul.       & Sample             & Popul.          & Sample\\
\hline
& & \multicolumn{2}{c}{Gaussian} & \multicolumn{2}{c}{Student-$t$} \\
\cmidrule(lr){3-4} \cmidrule(lr){5-6}
& & \multicolumn{4}{c}{TailCoR} \\
                     & True          & 1.225         &  1.225        &     1.637      &    1.637   \\
$T=1000$     & Mean         & 1.223         &  1.224        &    1.639       &   1.635    \\
                     & SD                   &  0.034        &  0.037       &      0.082     & 0.077   \\
$T=10000$   & Mean         & 1.225         &  1.225       &     1.637      &     1.637  \\
                     & SD                   &  0.011        &  0.011       &     0.024       &   0.024  \\
\cmidrule(lr){3-4} \cmidrule(lr){5-6}
& & \multicolumn{4}{c}{$s(\xi,\tau,\alpha)$} \\
                     & True          & 2.438         &  2.438        &     3.259      &    3.259    \\
$T=1000$     & Mean         & 2.435         &  2.438        &   3.266        &    3.257   \\
                     & SD                   & 0.059         &  0.072       &      0.152     & 0.147  \\
$T=10000$   & Mean         & 2.439         &  2.438        &    3.260       &     3.259  \\
                     & SD                   & 0.020         &  0.023       &      0.045      &   0.046   \\
\cmidrule(lr){3-4} \cmidrule(lr){5-6}
& & \multicolumn{4}{c}{$\sqrt{1+|\rho_{j\,k}|}$} \\
                     & True          & 1.225         &  1.225        &     1.225      &  1.225     \\
$T=1000$     & Mean         & 1.225         &  1.224        &     1.224      &     1.224  \\
                     & SD                   &  0.011        &  0.011       &      0.012      &0.012    \\
$T=10000$   & Mean         & 1.225         &  1.225        &     1.224      &    1.225   \\
                     & SD                 &  0.003        &  0.003        &        0.004      &   0.004  \\
\hline \hline
\multicolumn{6}{p{10cm}}{\footnotesize{Means and standard deviations of estimates of TailCoR and its components when using population (Popul. in top row) or sample marginal quantiles in step 1. All results are for H=1000 and $\xi=0.95$}}
\end{tabular}
\end{center}
\end{table}


\section{Downside and upside TailCoR under asymmetry}

It is often the case that the interest lies in a particular region of the distribution. For instance, in risk management the region of interest is where $Y_{j\,t}$ and $Y_{k\,t}$ are negative. This translates into the negative side of the distribution of $Z^{(j\,k)}_{t}$, which leads to the following definition.

\begin{description}
\item[{\bf Definition 3}] Under {\bf G1} -- {\bf G2} we define downside- and upside-TailCoR as follows:
\begin{eqnarray*}
\mbox{TailCoR}^{(j\,k)\,\xi\,-} &:=& s_0(\xi,\tau)\mbox{IQR}^{(j\,k)\,\xi\,-} \mbox{ and} \\
\mbox{TailCoR}^{(j\,k)\,\xi\,+} &:=& s_0(\xi,\tau)\mbox{IQR}^{(j\,k)\,\xi\,+},
\end{eqnarray*}
where $\mbox{IQR}^{(j\,k)\,\xi\,-} =  \mbox{Q}^{(j\,k)\,0.50} - \mbox{Q}^{(j\,k)\,1-\xi}$ is the difference between the median and the lower tail quantile, and $\mbox{IQR}^{(j\,k)\,\xi\,+} = \mbox{Q}^{(j\,k)\,\xi} - \mbox{Q}^{(j\,k)\,0.50}$ is the difference between the upper tail quantile and the median.
\end{description}
The value added of downside and upside TailCoRs is therefore put forward under asymmetry. We model asymmetry 
with a normal mean-variance mixture (or NMVM, see e.g. \cite{MenciaSentana09} and \cite{BLV}), i.e. $\mathbf{X}_t =_d \bm{\mu} +  \mathcal{R}_{\alpha\,t}^{2}\bm{\gamma}+ \mathcal{R}_{\alpha\,t}\bm{\Lambda}\bm{\Upsilon}_t$,
 where $\bm{\Upsilon}_t$ is a standardized multivariate Gaussian distribution, and $\bm{\gamma}$ is a vector of parameters that capture the asymmetry.
The most prominent distribution that belongs to this family is the generalized hyperbolic, as it nests the hyperbolic, normal gamma, and normal inverse Gaussian, among others. As the elliptical family, the NMVM family enjoys the property of affine invariance, i.e. it is closed under linear operations, which means that the projection $Z^{(j\,k)}_t$ is also a NMVM random variable. 

The vector of unknown parameters is now $\bm{\theta}=(\bm \mu, \bm \Sigma, \bm \gamma,\alpha)$. While assumptions {\bf E1} and {\bf E2} have to be trivially modified, {\bf E3} -- {\bf E5} remain unchanged. We re-name them as {\bf A1} -- {\bf A5}. In order to disentangle the linear and nonlinear correlations, uniform asymmetry is needed, as stated in the following assumption.

\begin{itemize}
\item[\bf A6] $\gamma=\gamma_j=\gamma_k$, $\forall j,k=1,\ldots,N$. 
\end{itemize}

This assumption does not imply symmetry, which is a particular case when $\gamma=0$. Rather, it implies that all the random variables are equally asymmetric. Though uniform asymmetry may seem restrictive, empirical results -- see for instance \cite{Jondeau10} -- show that the estimated asymmetries do not show large discrepancies, which is also corroborated in the empirical illustration.

Under asymmetry, the relation between $X_{j\,t}$ and $X_{k\,t}$ is not the same over the domain of the observations. For instance, \cite{LoginSolnik01} and \cite{Jondeau10} find that the dependence across asset returns may be stronger in bearish than in bullish conditions. These asymmetric dependencies are captured with 
semi--variances, semi--interquantile ranges, and semi--correlations. The prefix "semi" means that we use observations that are in a certain range. For instance, the positive semi--interquantile range is the difference between the upper quantile and the median, as used in the upside-TailCoR.

Let $\sigma_j^{2\,+}$ be the positive semi--variance and $\mbox{IQR}^{\tau\,+}_j = \mbox{Q}^{\tau}_j - \mbox{Q}^{0.50}_j$ be the positive semi--interquantile range of $X_{j\,t}$ of order $\tau$. The relation between both is given by $\mbox{IQR}^{\tau\,+}_j=k(\tau,\alpha,\gamma)^{+}\sigma_j^{+}$. Similarly, $\sigma_j^{2\,-}$ and $\mbox{IQR}^{\tau\,-}_j = \mbox{Q}^{0.50}_j - \mbox{Q}^{1-\tau}_j$ are the negative semi--variance and the negative semi--interquantile range, which are related by $\mbox{IQR}^{\tau\,-}_j=k(\tau,\alpha,\gamma)^{-}\sigma_j^{-}$. Similar definitions apply for $X_{k\,t}$. We also need the positive and negative semi--correlations between $X_{j\,t}$ and $X_{k\,t}$: the former is defined as $\rho^+_{j\,k}=\sigma_{j\,k}^{+}/\sigma_j^{+}\sigma_k^{+}$, where $\sigma_{j\,k}^{+}$ is the positive semi--covariance. Likewise for the negative semi--correlation.

By the affine invariance of the NMVM family, it follows that semi--interquantile ranges in Definition 3 equal $\mbox{IQR}^{(j\,k)\, \xi\,+} =k(\xi,\alpha,\gamma)^{+}\sigma_{(j\,k)}^{+}$ and $\mbox{IQR}^{(j\,k)\, \xi\,-} =k(\xi,\alpha,\gamma)^{-}\sigma_{(j\,k)}^{-}$, where $\sigma_{(j\,k)}^{2\,+}$ and $\sigma_{(j\,k)}^{2\,-}$ are the positive and negative semi--variances of the projection $Z^{(j\,k)}_t$. Further properties of the semi--moments are found in Appendix P. 

Under the NMVM family of distributions, the optimal projection may not be on the 45- or the 135-degree lines, but on a line with angle $\phi$  that maximizes the variability of the projection. The procedures explained under {\bf G1} -- {\bf G2} apply here. The following theorem shows downside- and upside-TailCoR under asymmetry.

\begin{description}
\item[{\bf Theorem 5}]
Let $X_{j\,t}$ and $X_{k\,t}$ be two elements of the random vector $\mathbf{X}_t$ that fulfill assumptions {\bf A1} -- {\bf A6}. Let $\rho^+_{j\,k}$ and $\rho^-_{j\,k}$ be the positive and negative semi--correlations, and let $s(\xi,\tau,\alpha,\gamma)^{+}$ and $s(\xi,\tau,\alpha,\gamma)^{-}$ be two continuous and monotonically decreasing functions of $\alpha$. Then
\begin{eqnarray*}
\mbox{TailCoR}^{(j\,k)\,\xi \,-} &=&2s_g(\xi,\tau)s(\xi,\tau,\alpha,\gamma)^{-}\sqrt{1+  2 |\rho^-_{j\,k}| |\sin \phi \cos \phi|} \quad \mbox{and} \\
\mbox{TailCoR}^{(j\,k)\,\xi \,+} &=&2s_g(\xi,\tau)s(\xi,\tau,\alpha,\gamma)^{+}\sqrt{1+ 2 |\rho^+_{j\,k}| |\sin \phi \cos \phi|},
\end{eqnarray*}
where $s_g(\xi,\tau)$ is a normalization such that, under independence, we have $\mbox{TailCoR}^{(j\,k)\,\xi \,-}=\mbox{TailCoR}^{(j\,k)\,\xi \,+}=1$, the reference value.
\end{description}

\begin{description}
\item[{\bf Proof}] See Appendix P.
\end{description}

The normalization is now $2s_g(\xi,\tau)$. It makes $\mbox{TailCoR}^{(j\,k)\,\xi\,-}$ and $\mbox{TailCoR}^{(j\,k)\,\xi\,+}$ comparable with the full--fledged TailCoR. When $\phi$ equals $\frac{\pi}{4}$ or $\frac{3\pi}{4}$ (i.e., projections in the 45- and 135-degree lines), $\sqrt{1+ 2 |\rho^+_{j\,k}| |\sin \phi \cos \phi|}$ equals $\sqrt{1+ |\rho^+_{j\,k}|}$. The downside- and upside-TailCoR can differ because either the linear and/or the nonlinear correlations are different. Estimation follows the same steps as in the previous estimators except for the use of semi metrics.

\section{TailCoR of 21 financial
market indexes around the globe}

We illustrate TailCoR with an application to 20.5 years of daily stock log returns of 21 major equity market indexes that represent three geographical regions: America (S\&P500,NASDAQ, TSX, Merval, Bovespa and IPC), Europe (AEX, ATX, FTSE, DAX, CAC40, SMI and MIB), and East Asia and Oceania (HgSg, Nikkei, StrTim, SSEC, BSE, KLSE, KOSPI and AllOrd). The sample spans from 5 January 2000 to 9 July 2020 (each series contains 5369 observations). Table \ref{tab:returns} shows the countries and descriptive statistics. All medians are approximately zero, as expected.

The IQRs are a measure of volatility. If annualised, the maximum is Merval with 35\% and the minimum is KSLE with 12\%.
European markets have annualised IQRs between 24\% for MIB and 17.8\% for SMI, while the IQR for S\&P500 is 16.5\%, very close to 15\%, the average that market participants consider as the average annualized volatility. All skewness are negative (but one albeit very close to zero) which indicates that the probability distribution is symmetric or very close to it. Finally, excess kurtosis are all positive, denoting tails heavier than Gaussian.  The open and international markets have the highest excess kurtosis -- the top 5 indices are in the US, South Korea, Netherlands and Canada.

\begin{table}
\begin{center}
\caption{Descriptive statistics} \label{tab:returns}
\begin{tabular}{cccccc}
 \hline\hline
Index & Country & Median & $\text{IQR}_{0.75}$ & Kurtosis & Skewness \\   \hline
S\&P & US & 0.0006 &  0.0104 &  1.8598 & -0.0034 \\ 
  NASDAQ & US &  0.0009 &  0.0137 &  1.9299 & -0.0036 \\ 
  TSX &  Canada & 0.0007 &  0.0098 &  1.6225 & -0.0042 \\ 
  Merval & Argentina & 0.0013 &  0.0221 &  1.4193 &  0.0008 \\ 
  Bovespa & Brazil &  0.0009 &  0.0200 &  0.5942 & -0.0029 \\ 
  IPC &  Mexico & 0.0006 &  0.0126 &  1.1945 & -0.0020 \\ 
  AEX & Netherlands & 0.0006 &  0.0126 &  1.6462 & -0.0036 \\ 
  ATX & Austria & 0.0007 &  0.0141 &  1.3296 & -0.0053 \\ 
  FTSE & UK & 0.0005 &  0.0113 &  1.4377 & -0.0028 \\ 
  DAX & Germany & 0.0008 &  0.0143 &  1.3958 & -0.0055 \\ 
  CAC & France & 0.0004 &  0.0138 &  1.3717 & -0.0038 \\ 
  SMI & Spain & 0.0006 &  0.0112 &  1.2319 & -0.0035 \\ 
  MIB & Italy & 0.0007 &  0.0151 &  1.2464 & -0.0038 \\ 
  HgSg & Hong Kong & 0.0006 &  0.0142 &  1.2157 & -0.0025 \\ 
  Nikkei & Japan &  0.0006 &  0.0155 &  0.9529 & -0.0030 \\ 
  StrTim & Singapore &  0.0002 &  0.0106 &  1.5006 & -0.0036 \\ 
  SSEC & China &  0.0008 &  0.0144 &  1.5891 & -0.0050 \\ 
  BSE &  India & 0.0009 &  0.0140 &  1.4791 & -0.0045 \\ 
  KLSE & Malaysia &  0.0004 &  0.0077 &  1.4068 & -0.0012 \\ 
  KOSPI & South Korea & 0.0006 &  0.0134 &  1.8703 & -0.0036 \\ 
  AllOrd & Australia & 0.0007 &  0.0096 &  1.1289 & -0.0034 \\ 
   \hline \hline 
\multicolumn{6}{p{11.5cm}}{\footnotesize{All the metrics are quantile-based. $\mbox{IQR}_{75}$ is the interquartile range, kurtosis is in excess and computed as $\mbox{IQR}^{0.975}/\mbox{IQR}^{0.75} - 2.91$, and skewness is computed as $(Q^{0.975}-Q^{0.50})-(Q^{0.50}-Q^{0.025})$.}}
\end{tabular}
\end{center}
\end{table}




We first show full-sample results, followed by rolling window estimations to study the dynamic behaviour. The window has a size of 3 years and it is rolled every year, i.e. we start with January 2000 - December 2002, followed by January 2001 - December 2003, and so on and so forth. There are between 780 and 800 observations per window (except the last one that ends in July 2020). The choice of 3 years is motivated by having a relatively small sample size (so we can compare the performance with the much larger full sample) but not too small (that would not allow us to draw meaningful comparisons with alternative measures).

Indeed, we compare TailCoR with the upper and lower exceedance correlations of \cite{LonginSolnik2001} (denoted $\theta^+(>\xi)$ and $\theta^-(<\xi)$, respectively), and the parametric and non-parametric tail dependence coefficients (the former with a $t$-copula and the latter as in \cite{Straetmans08}).\footnote{The non-parametric tail dependence coefficient is also called the co-exceedance probability; see equation (10) in \cite{Straetmans08}.} We denote them $\tau_p$ and $\tau_{np}$, respectively. Results are for $\xi=0.975$ ($\xi=0.025$ for downside correlation). Estimations for other values of $\xi$ are available upon request.\footnote{Since $\tau_{np}$ depends on the Hill statistic, alternative methods for choosing $\xi$ are available. See \cite{DIV16} for a survey.}

\begin{figure}
  \caption{Full sample results -- TailCoR and components}\label{fig:tailcor}
  \begin{center}
    \subfloat[TailCoR]{\includegraphics[angle=0,height =6cm,width=.40\linewidth]{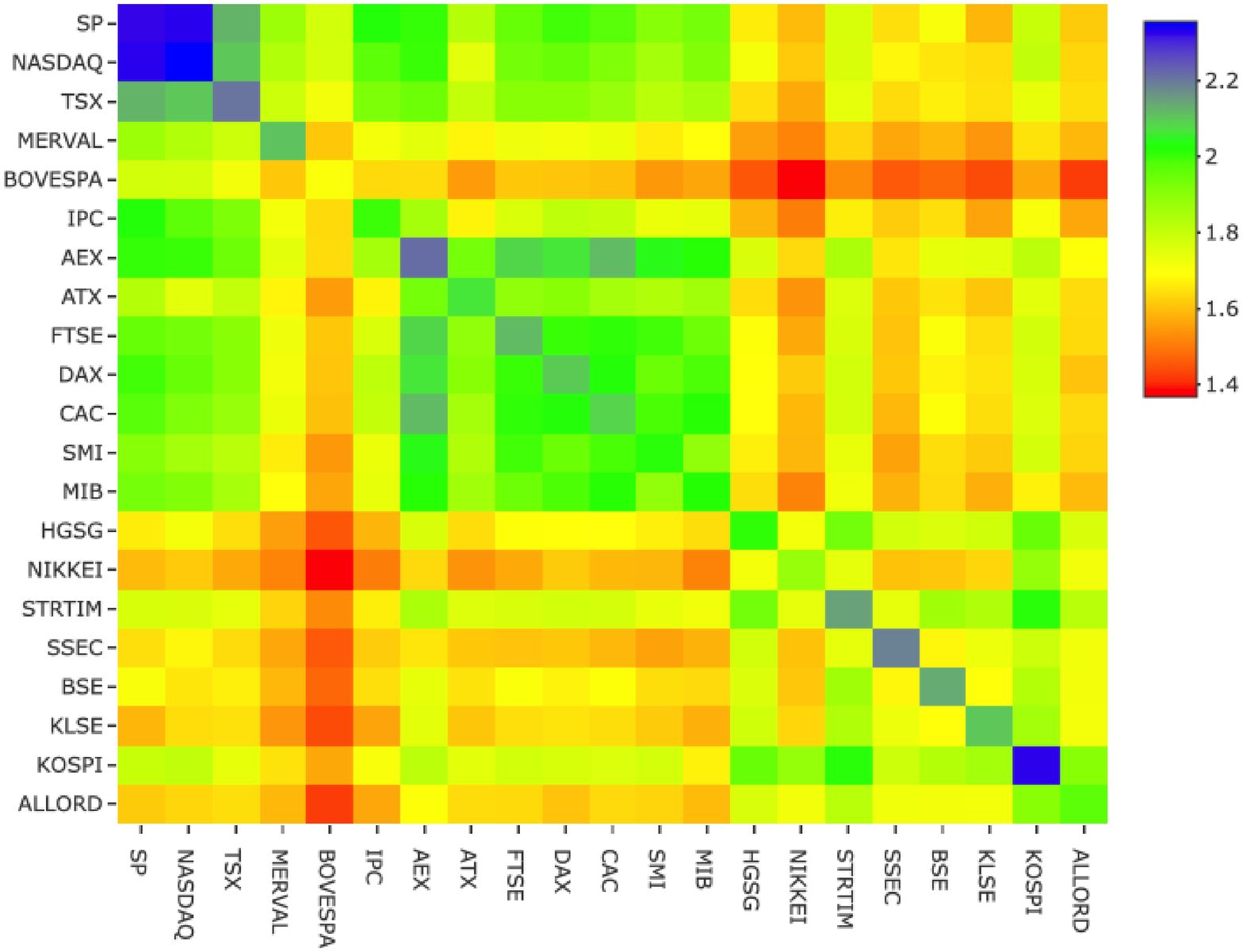}}
    \subfloat[Linear component]{\includegraphics[angle=0,height =6cm,width=.40\linewidth]{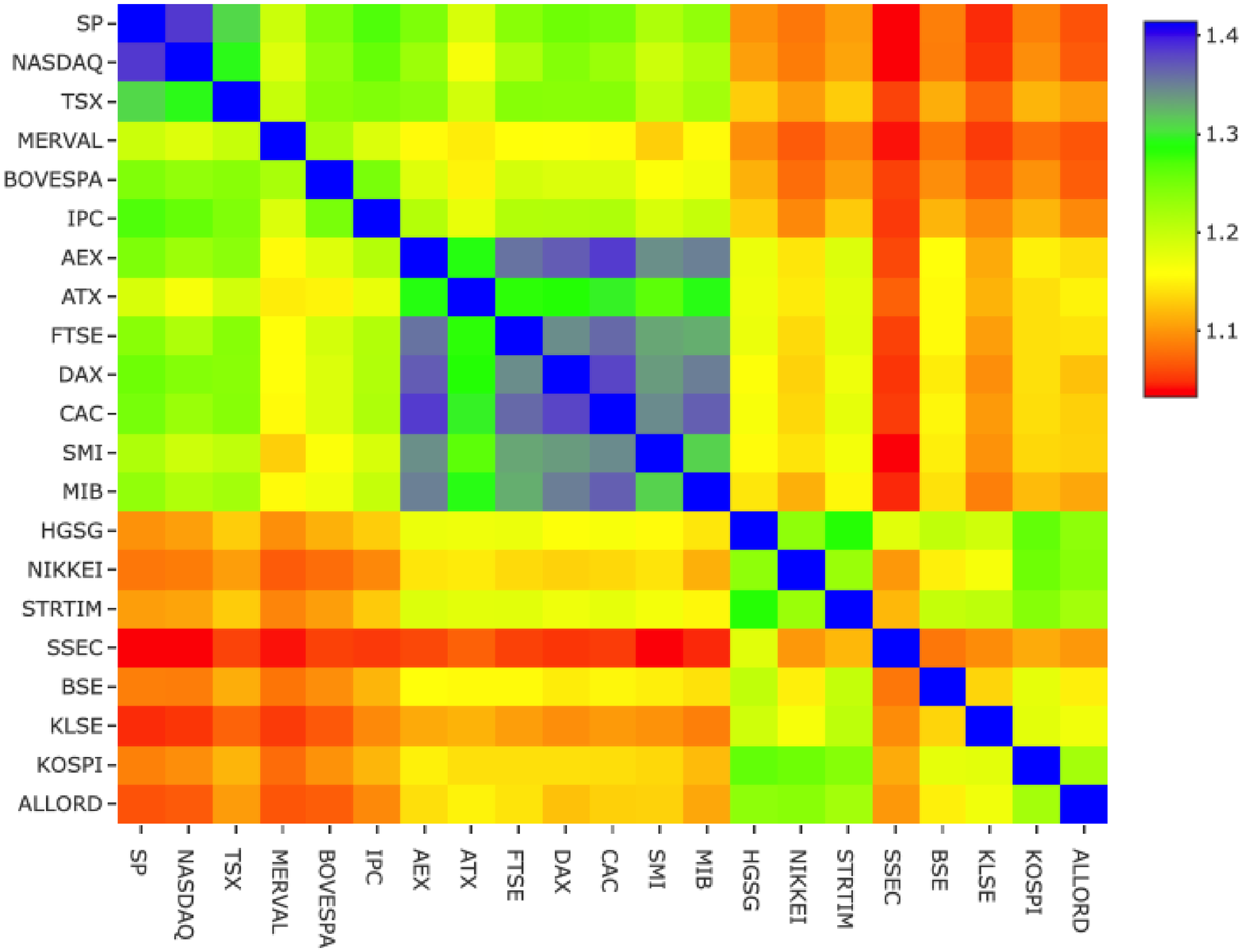}} \\
    \subfloat[Nonlinear component]{\includegraphics[angle=0,height =6cm,width=.40\linewidth]{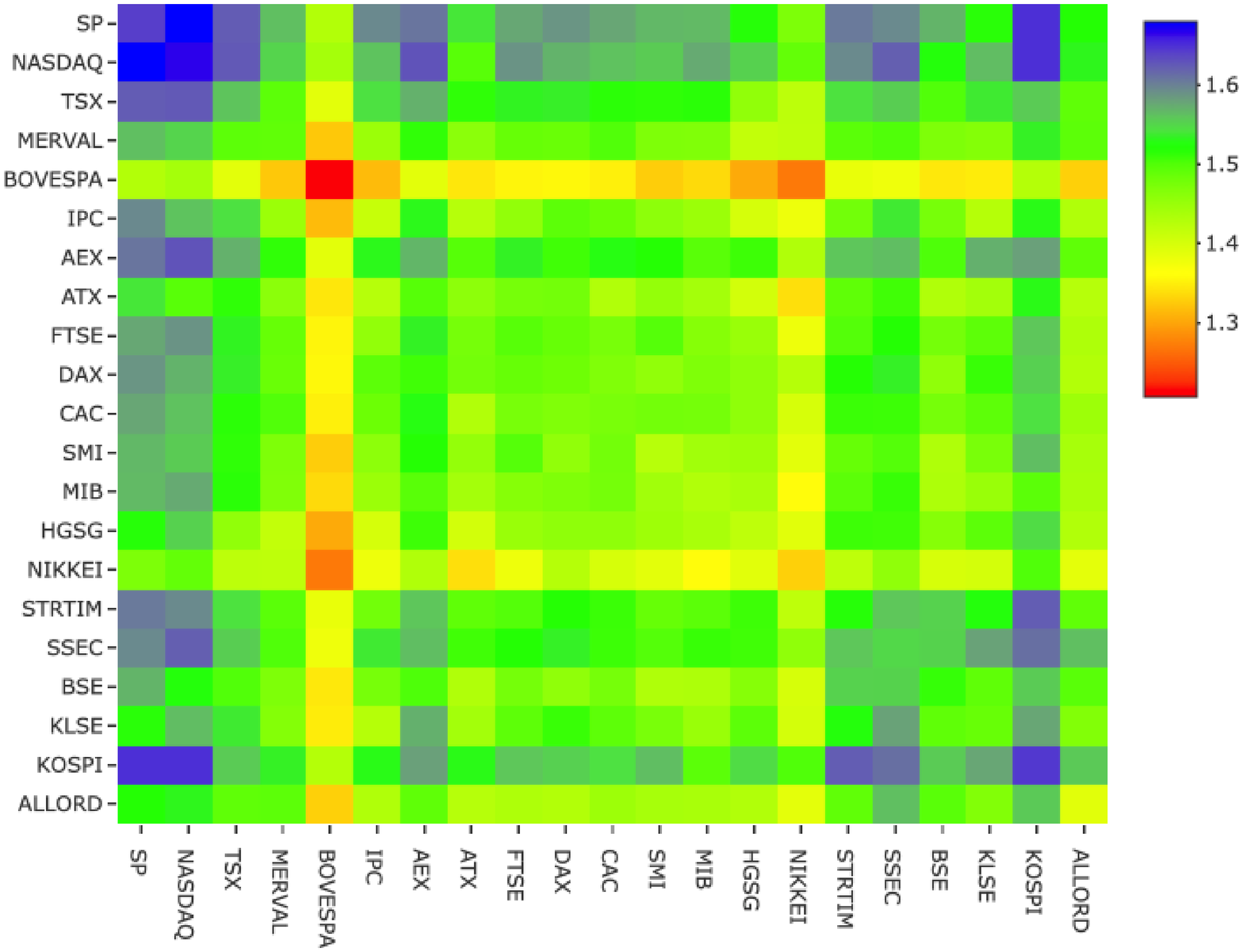}} \\
\begin{tabular}{p{13cm}}
\end{tabular}
  \end{center}
\end{figure}

\begin{figure}
  \caption{Full sample results -- comparing measures}\label{fig:competing}
  \begin{center}
    \subfloat[$\theta^-(<0.025)$]{\includegraphics[angle=0,height =6cm,width=.40\linewidth]{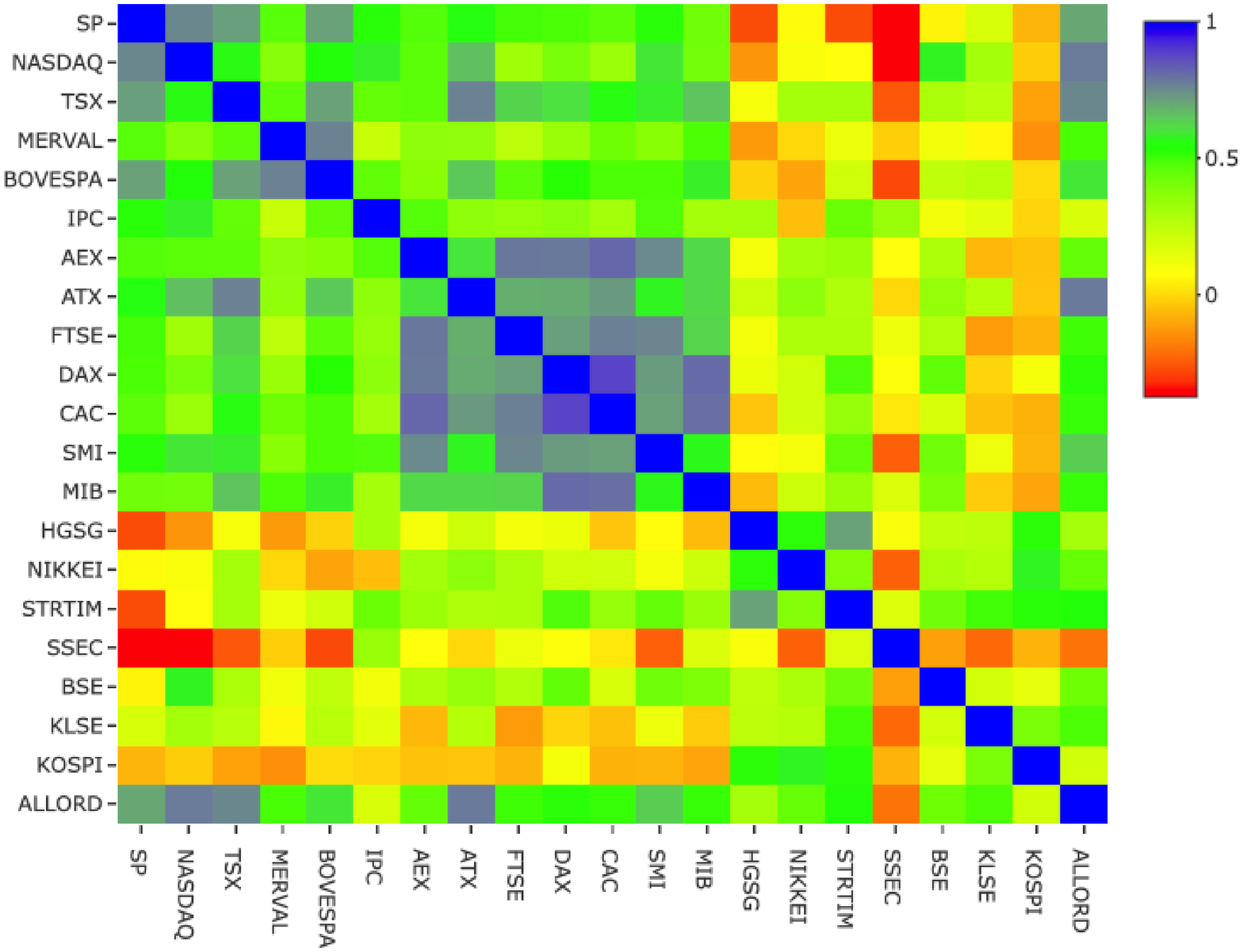}} 
    \subfloat[$\theta^+(>0.975)$]{\includegraphics[angle=0,height =6cm,width=.40\linewidth]{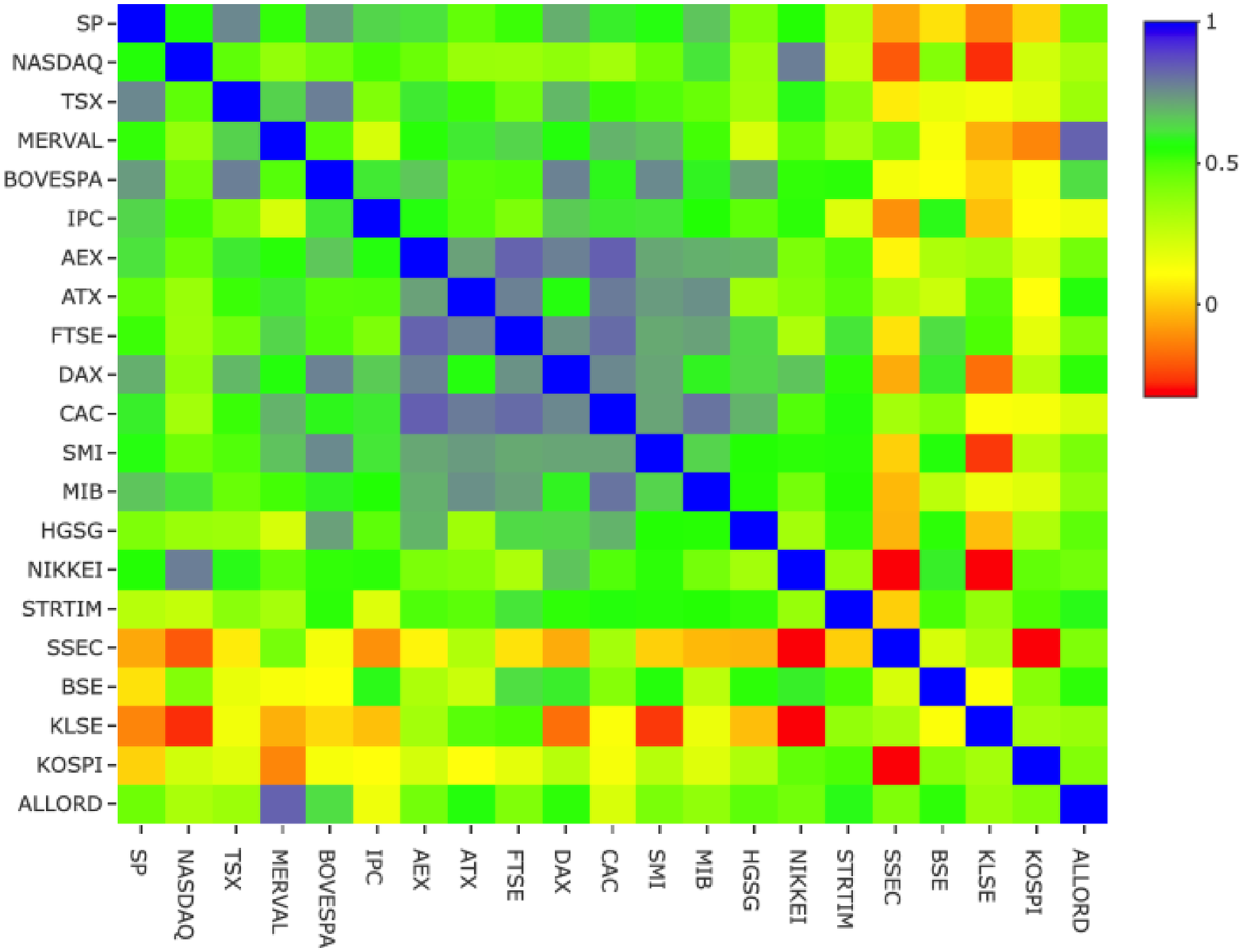}} \\
    \subfloat[$\tau_p$]{\includegraphics[angle=0,height =6cm,width=.40\linewidth]{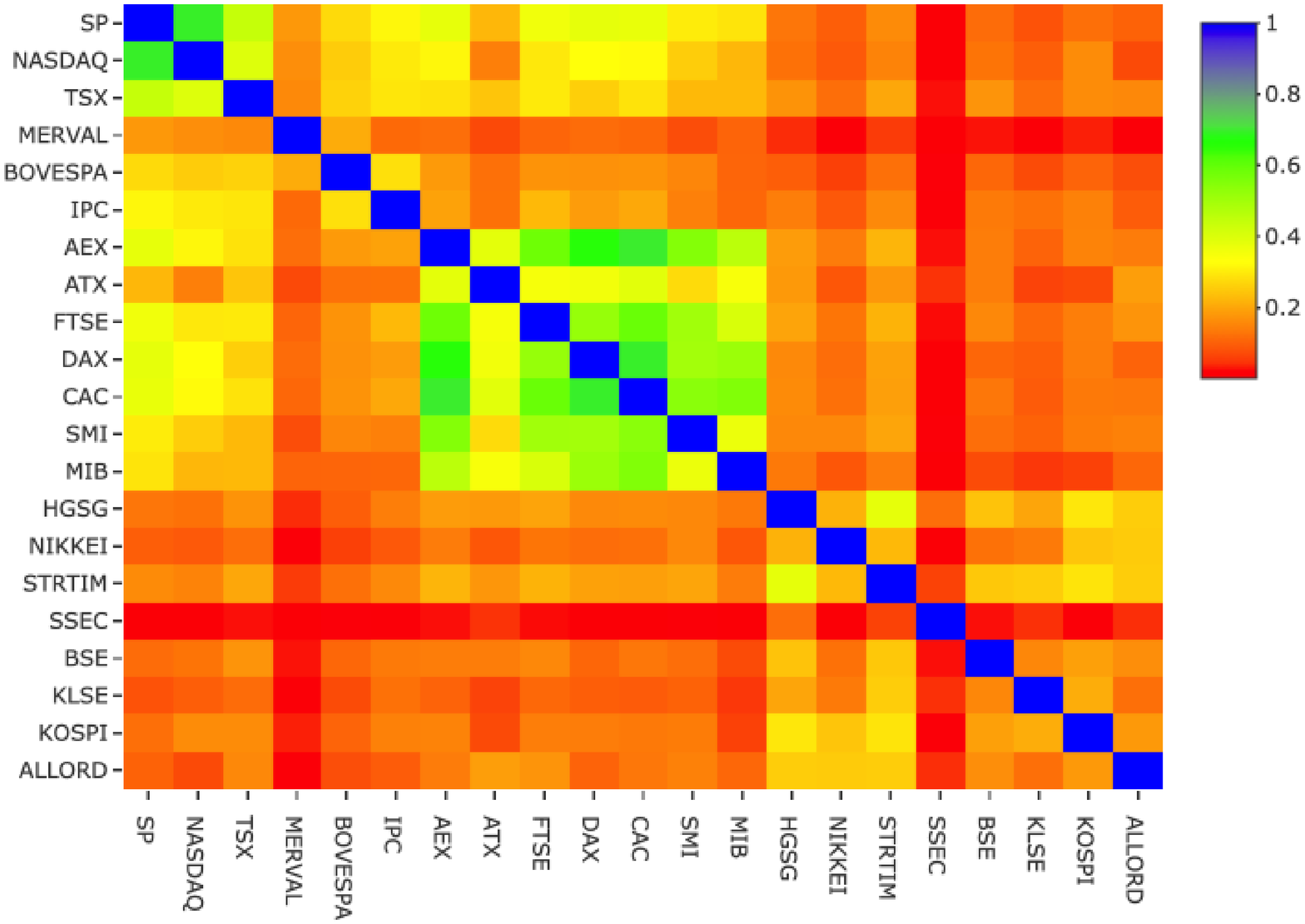}}
    \subfloat[$\tau_{np}$]{\includegraphics[angle=0,height =6cm,width=.40\linewidth]{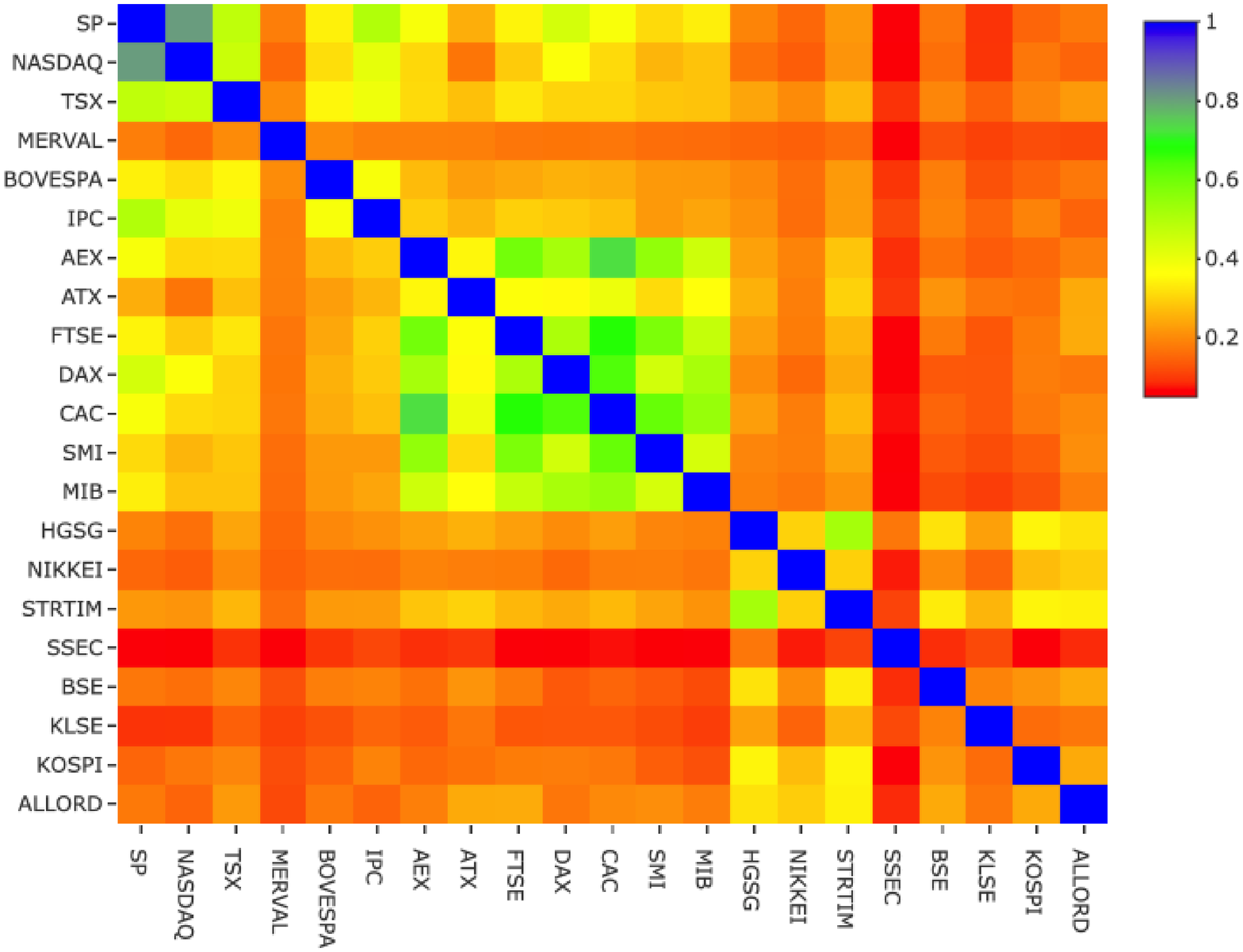}} \\
\begin{tabular}{p{13cm}}
\end{tabular}
  \end{center}
\end{figure}

Full sample results are found in Tables \ref{tab:tailcor} to \ref{tab:nonlinearComponent} for TailCoR and its components (numbers in parenthesis are standard deviations multiplied by 100 for facilitating interpretation), and Tables \ref{tab:dexceedance} to \ref{tab:non-param} show results for $\theta^+(>\xi)$, $\theta^-(<\xi)$, $\tau_p$, and $\tau_{np}$ (the rightmost column in every matrix shows the row averages). These tables are lengthy and relegated to Appendix T. Figures \ref{fig:tailcor} and \ref{fig:competing} show a summary in the form of heat maps. The market indexes are ordered as in Table \ref{tab:returns}: America on the top left, Europe in the middle, and East Asia and Oceania in the bottom right. Note that the colour grading depends on the scale, but in all cases the closer to dark blue the stronger the dependence, and the closer to dark red the weaker the dependence.

Regarding TailCoR, all estimates are well above one -- the minimum is 1.37 -- and with small standard deviations (between 0.04 and 0.13) reflecting the fact that markets are dependent and with heavy tails. Expectedly, for every market the diagonal elements are the highest. These elements measure TailCoR of one market with itself, which can be interpreted as a measure of tail risk. The average TailCoR of an index with the rest of world varies from 1.57 for Bovespa to 1.87 for AEX. 

More interestingly, the heat map shows clearly a geographical clustering. The most dependent markets are in North America (S\&P500 and NASDAQ and Toronto Stock Exchange), all in dark blue and with TailCoRs above 2.1. By contrast, the remaining American markets have TailCoR of about 1.7 (interestingly enough, South American markets are less related between them than with North American markets). The next block is Europe that is mostly in the green \& dark green range (TailCoR is in the range 1.8 -- 2.2). Among these markets, the most related is AEX -- Netherlands is a very open economy and a European cross road -- and the less ATX -- Austria is the furthest geographically. The third block is Asia and Oceania with relatively low TailCoRs. The most dependent market is KOSPI -- South Korea is another small open economy and very export oriented. 
 
TailCoR within blocks is weaker than between blocks, especially between Asia and Oceania with the rest of the world. However, the European and the North American blocks are fairly related with TailCoR around 1.9 or higher.

The heat map for the linear component reveals that the source of the geographical cluster is linear. All linear components are above one, with a minimum of 1.03 and a maximum of 1.39 (diagonal elements excluded that by definition equal 1.41), and all standard deviations are very small. Within the blocks, and on average, the European block shows the higher linear dependence (1.22), followed by North America (1.18), South America (1.15), and Asia and Oceania (1.13).

The heat map for the nonlinear component shows a different pattern. All nonlinear components are above one and standard deviations are small (though not as small as for the linear component, which is expected since this component captures the dependence on the tails of the distribution). North American markets have the highest nonlinear relations, not only between them but also with the rest of the world. Some Asian indexes also show high nonlinear components, even higher than for European indexes, like KOSPI, SSEC, and STRTIM. The main diagonal measures the thickness of the tails for each market and, in fact, the correlation of the main diagonal with the kurtosis in Table \ref{tab:returns} is 99.8\%.

Turning to the comparing measures, $\tau_p$ and $\tau_{np}$ show the same geographical clustering as TailCoR, though the contrast between and within regional blocks is sharper, in particular for $\tau_p$. This clustering is also visible in the heat map for $\theta^-(<0.025)$, though more blurred, and less visible for $\theta^+(>0.975)$. Table \ref{tab:correlations} displays the matrix of sample correlations between $\mbox{TailCoR}^{0.975} $, $\theta^+(>0.975)$, $\theta^-(<0.025)$, $\tau_p$, and $\tau_{np}$. The sample correlations are computed by vectorizing the upper triangle of the matrices of estimates (hence without the diagonal elements that are 1 by definition, and would artificially inflate the correlations), and calculating the Pearson correlations between them. The correlations of TailCoR with the other measures vary between 0.39 and 0.89. TailCoR highly correlates with $\tau_p$ and $\tau_{np}$, mildly correlates with $\theta^-(<0.025)$, and weakly correlates with $\theta^+(>0.975)$ (as a matter of consistency, one can see that $\tau_{np}$ and $\tau_p$ strongly or mildly correlate with all other measures). Overall, we conclude that for large samples TailCoR provides empirical results that are in line with equivalent measures.

\begin{table}
\begin{center}
\caption{Empirical correlations between all measures} \label{tab:correlations}
\begin{tabular}{cccccccc}
 \hline\hline
					& $\text{TailCoR}^{0.975}$& $\theta^{-}(<0.025)$	& $\theta^{+}(>0.975)$	& $\tau_p$& $\tau_{np}$  \\ 
$\text{TailCoR}^{0.975}$	& 1.00 				& 0.51				& 0.39				& 0.84	& 0.80 		\\ 
$\theta^{-}(<0.025)$		& 0.51				& 1.00				& 0.59				& 0.69	& 0.70 		\\ 
$\theta^{+}(>0.975)$		& 0.39				& 0.59				& 1.00				& 0.57	& 0.61 		\\ 
$\tau_p$				& 0.84				& 0.69				& 0.57				& 1.00	& 0.96 		\\ 
$\tau_{np}$			& 0.80				& 0.70				& 0.61				& 0.96	& 1.00 		\\ 
\hline \hline
\multicolumn{6}{p{12cm}}{\footnotesize{Empirical correlations between $\mbox{TailCoR}^{0.975} $, $\theta^+(>0.975)$, $\theta^-(<0.025)$, $\tau_p$, and $\tau_{np}$. The correlations are computed by vectorizing the upper triangle of the matrices of estimates, and calculating the Pearson correlations between them.}}
\end{tabular}
\end{center}
\end{table}

\begin{figure}
\caption{TailCoR and tail dependence coefficients}
\begin{center}
\subfloat[TailCoR]{\includegraphics[angle=0,width=0.45\linewidth]{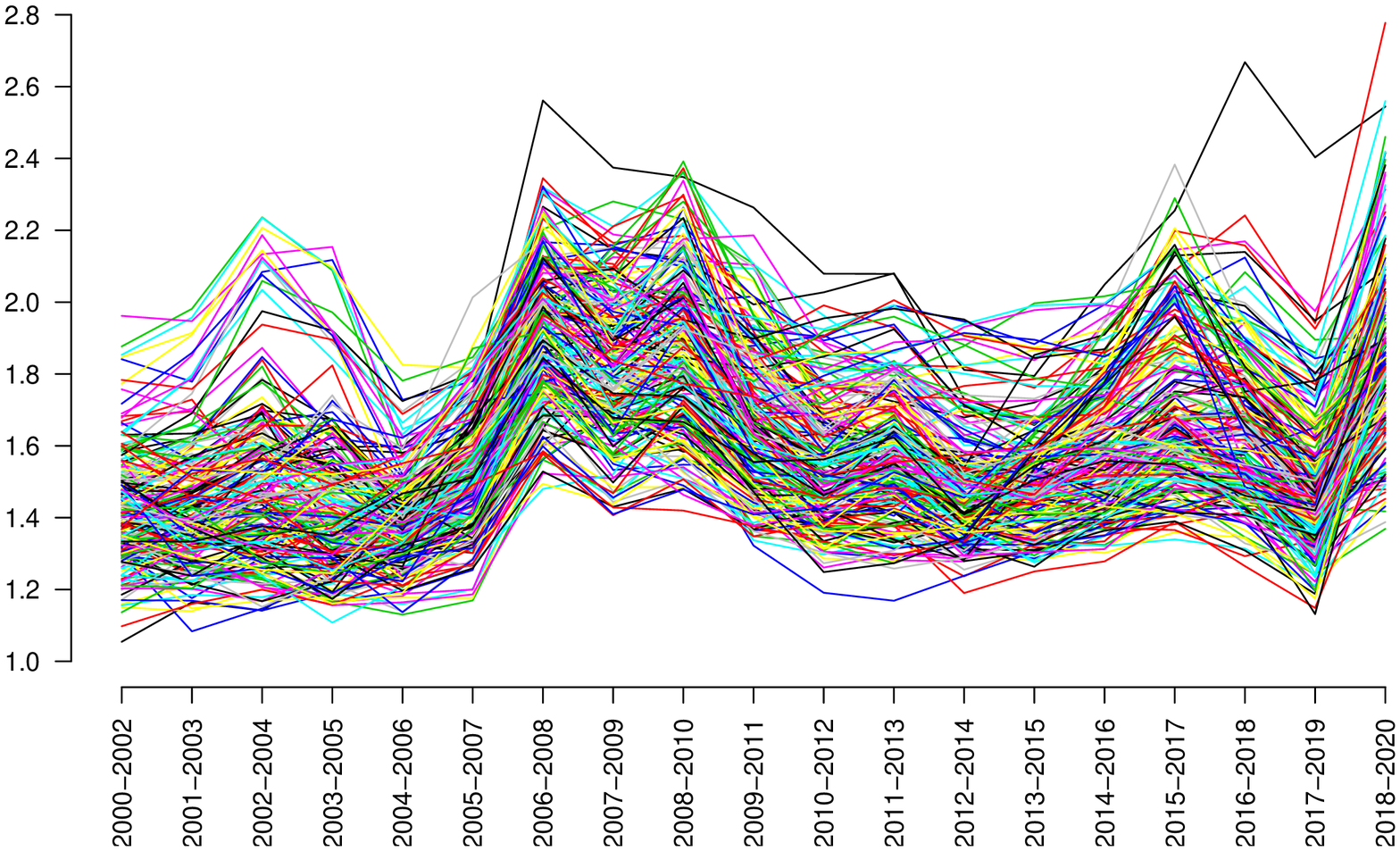}} \hspace{0.15cm}
\subfloat[$\tau_{np}$]{\includegraphics[angle=0,width=0.45\linewidth]{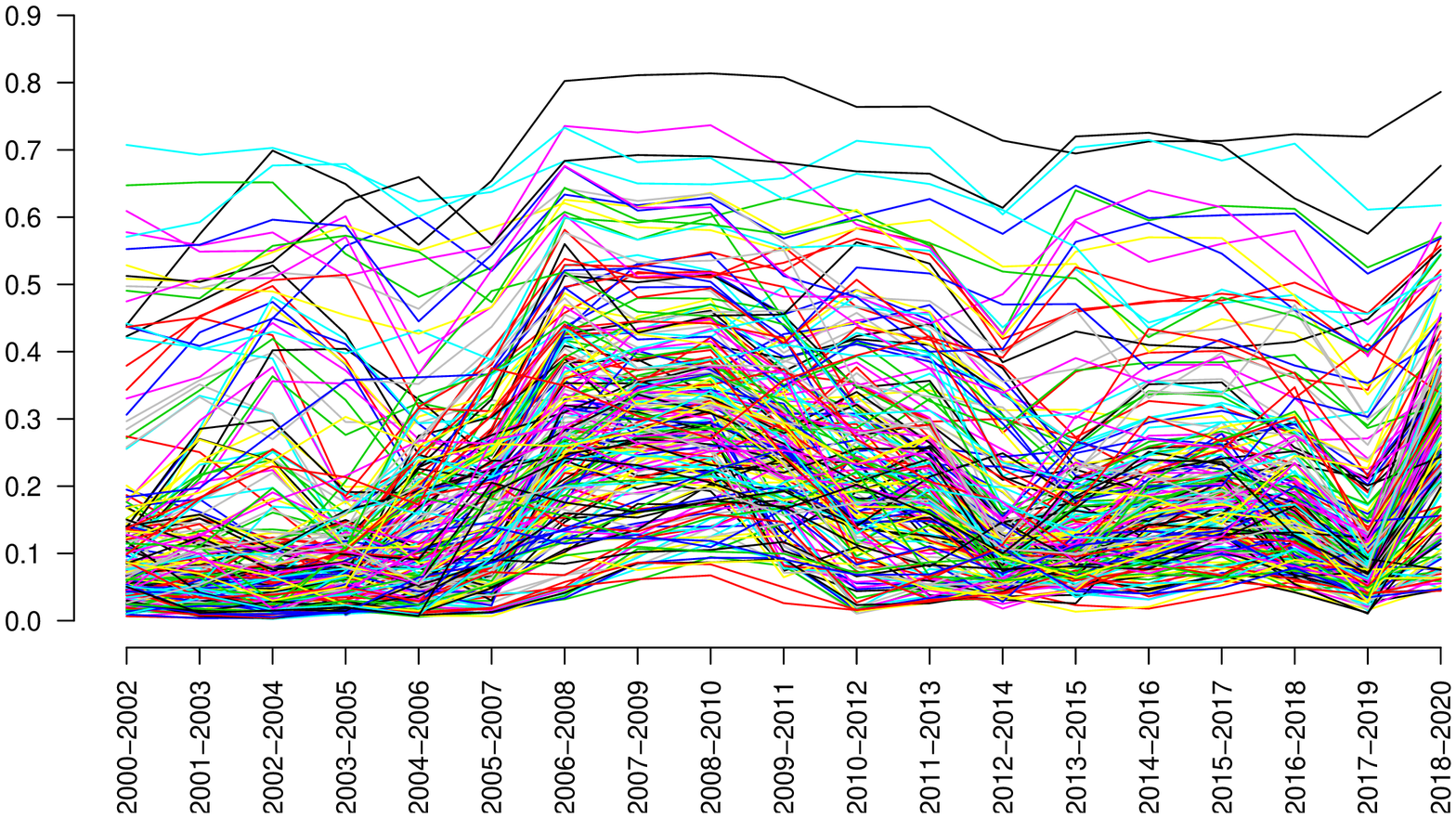}}\\
\subfloat[$\tau_p$]{\includegraphics[angle=0,width=0.45\linewidth]{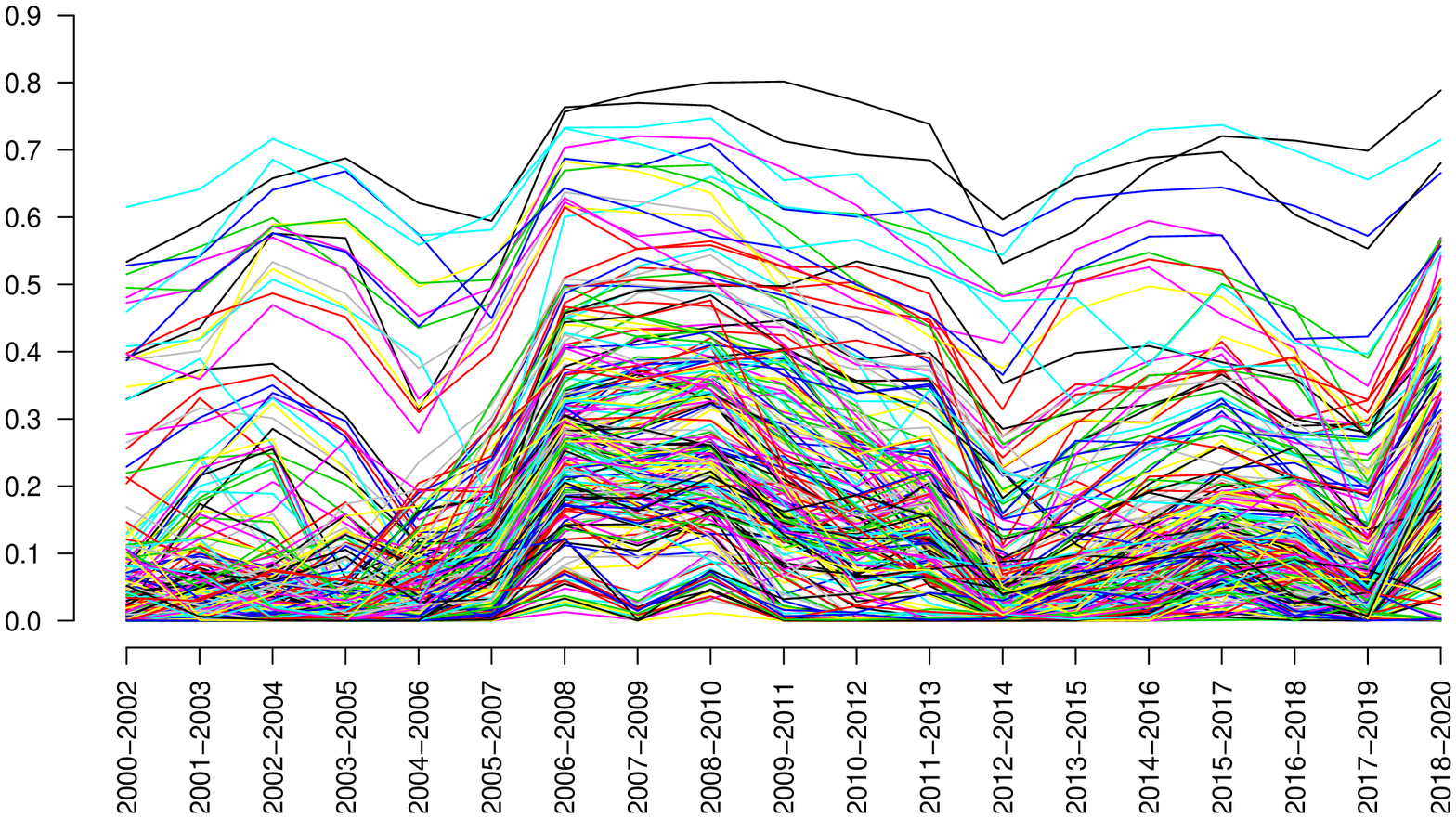}}
\end{center}
{\footnotesize \vspace{-0.35cm} Panel (a) shows the evolution of TailCoR for all pairs of market indexes, while panels (b) and (c) show the non-parametric and parametric tail dependence coefficients.}
\label{fig:mess}
\end{figure}

Next, we move to the rolling window exercise. Panels (a), (b) and (c) of Figure \ref{fig:mess} display the dynamic evolution of TailCoR, $\tau_p$, and $\tau_{np}$ respectively and for all pairs of market indexes.\footnote{Alternatively to rolling a window, one could think of a dynamic conditional TailCoR. The dynamic conditional extension would be in spirit similar to \cite{LeeLong}. This is left for future research, as explained in the conclusions.} The exceedance correlations could not be computed due to lack of extreme observations. All TailCoRs are always larger than one, and they show a pattern in line with the financial and economic events that happened during the sample period -- as explained in detail below. The tail dependence coefficients also show a recognisable pattern, though there is a larger variability across coefficients, both cross-sectionally and across time. That is, we do not observe increased clustering in times of crisis.

\begin{figure}
\caption{TailCoR and its components}
\begin{center}
\subfloat[TailCoR]{\includegraphics[angle=0,width=0.45\linewidth]{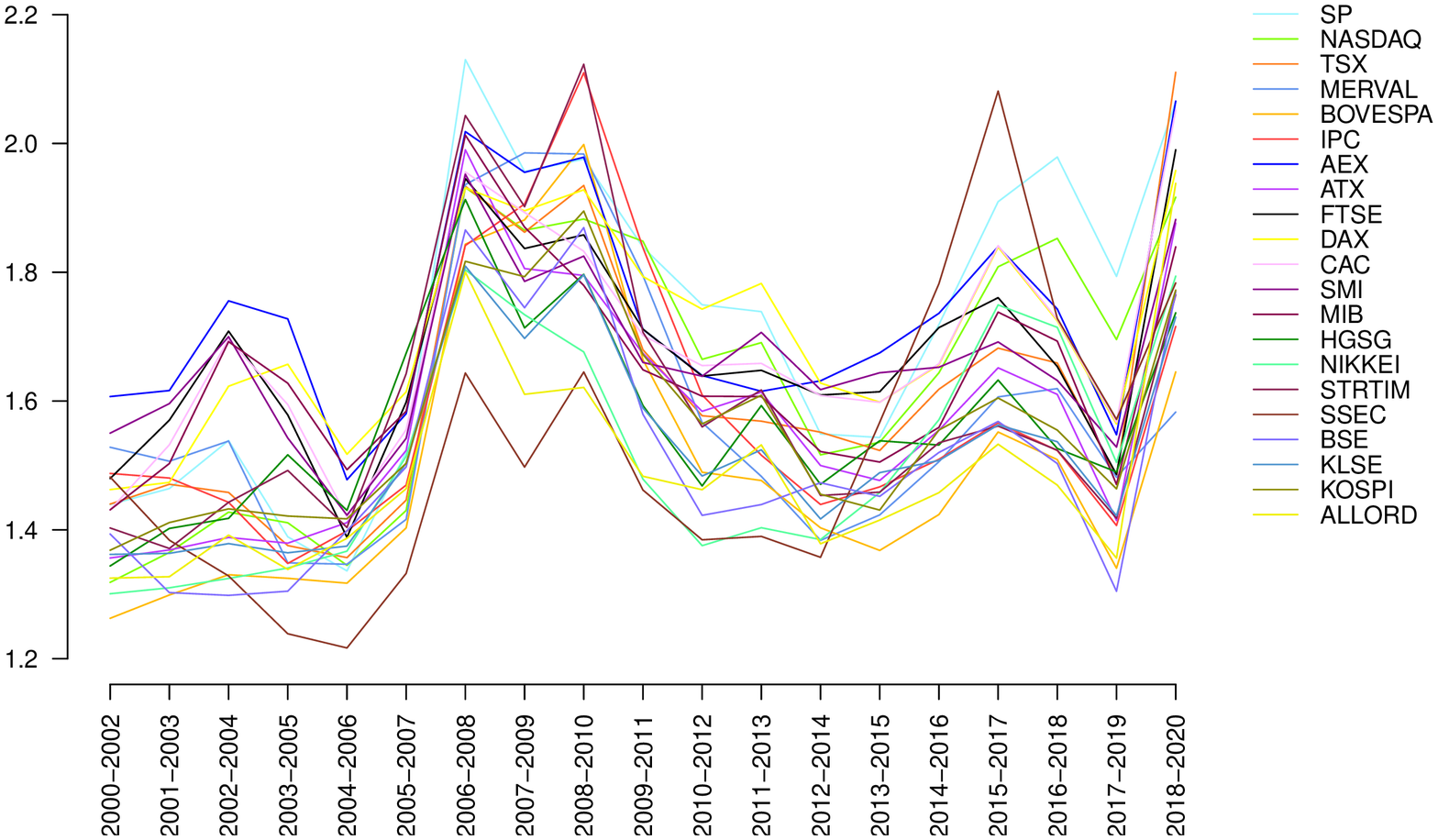}} \hspace{0.15cm}
\subfloat[Nonlinear component]{\includegraphics[angle=0,width=0.45\linewidth]{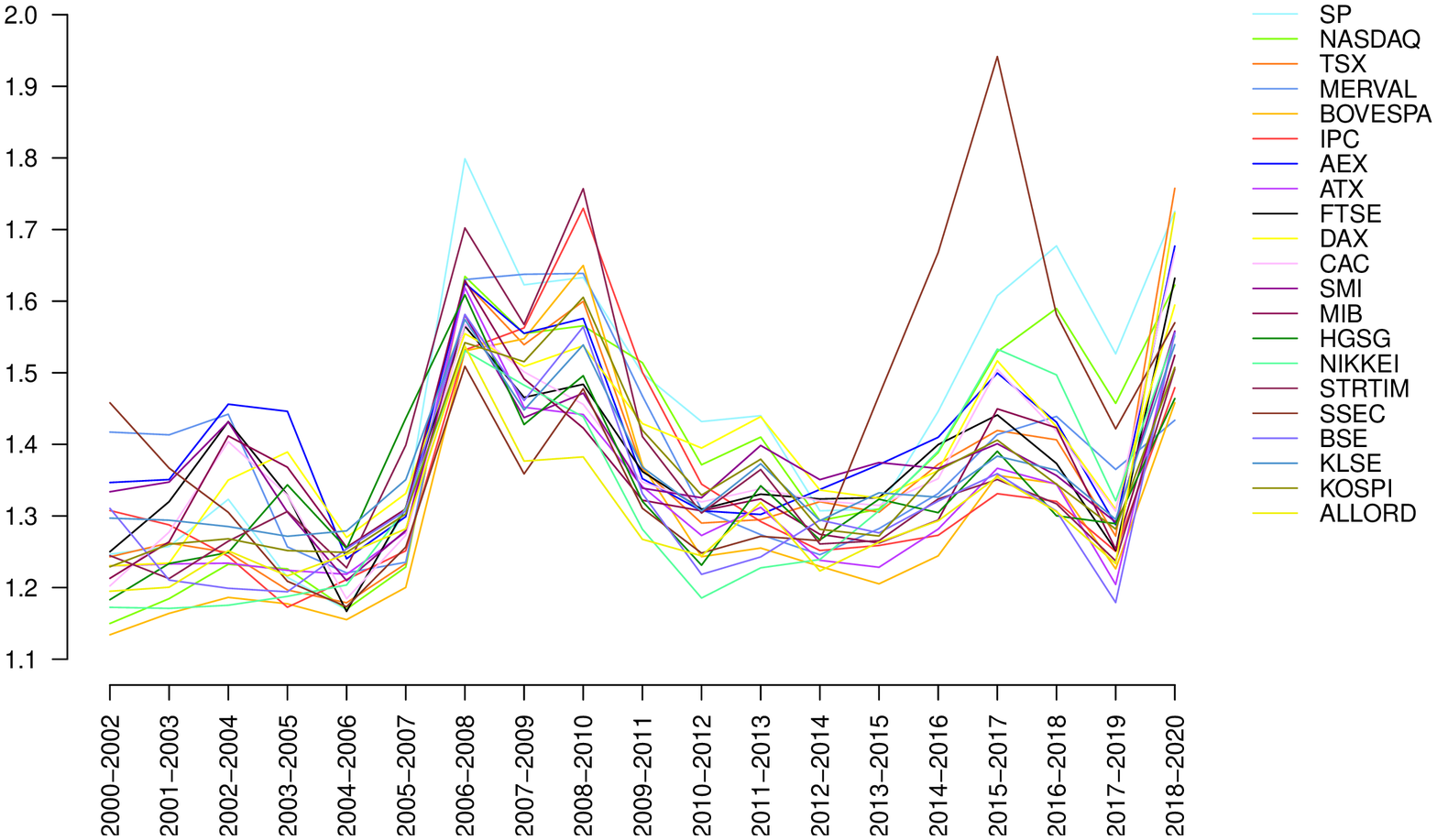}} \\
\subfloat[Linear component]{\includegraphics[angle=0,width=0.45\linewidth]{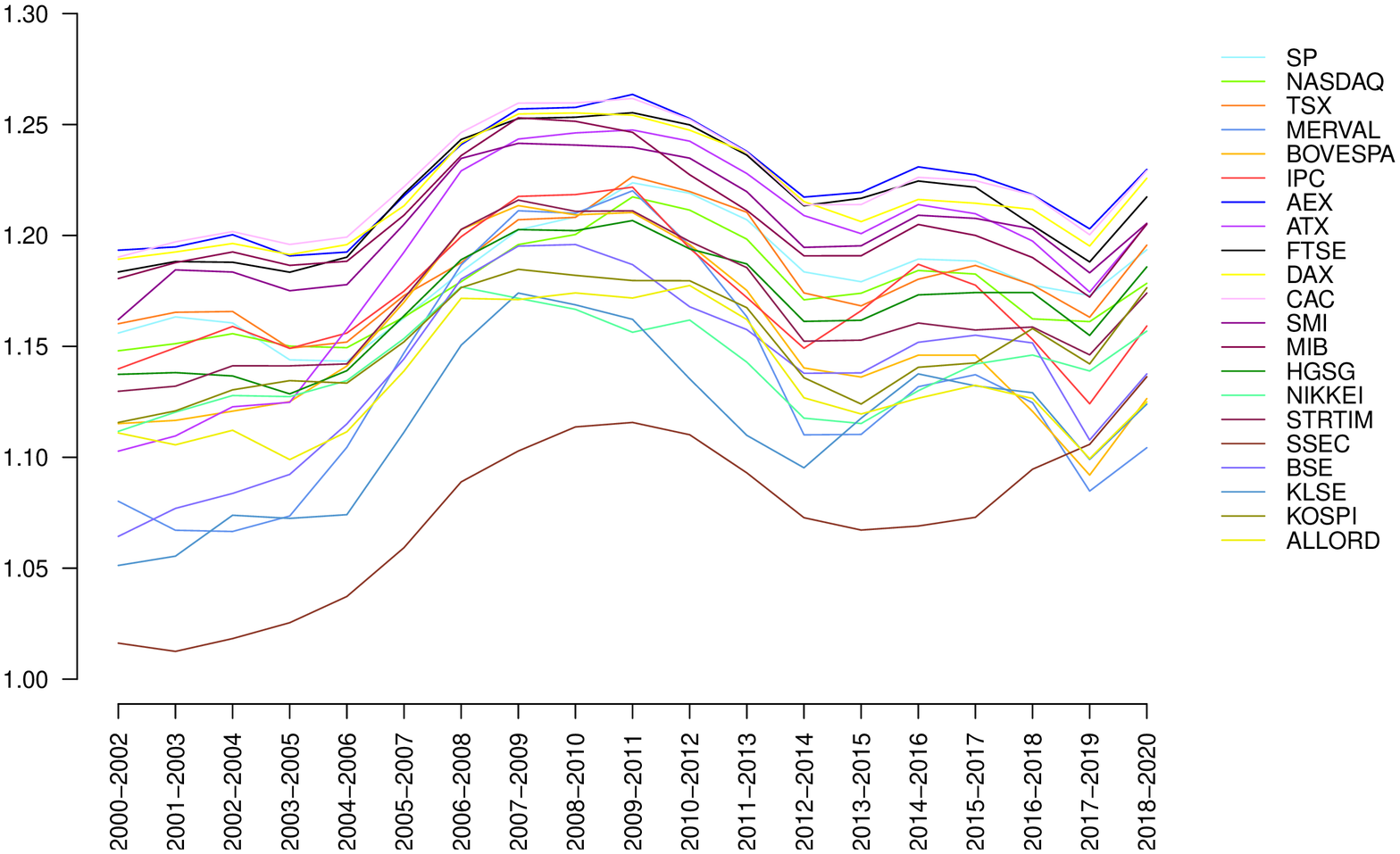}}
\end{center}
{\footnotesize \vspace{-0.35cm} Panel (a) shows the evolution of TailCoR for all market indexes. Each line is the cross-sectional average of one market index with respect to the others. Panels (b) and (c) show the evolution of the nonlinear and linear components respectively.}
\label{fig:dynamic}
\end{figure}

Figure \ref{fig:dynamic} focuses on TailCoR (panel a) and its nonlinear (panel b) and linear components (panel c). For the sake of visibility and interpretation, each line is the cross-sectional average of one index with respect to the others. For instance, the value of TailCoR for S\&P500 in 2000--2002 is the average TailCoR of S\&P500 with respect to all the other indexes on that period.

We also here observe that all TailCoRs are larger than 1 for all windows. The ups and downs show a pattern that is clearly identified with financial, economic and political events. In general TailCoR increases in crises periods, when markets become more dependent. Related with this, the variability of TailCoRs across the globe decreases in crises. During 2000-2004, the range of TailCoRs is 1.2 -- 1.7. In 2005, all TailCoRs decrease, concentrating around 1.4. This a period characterised by tails nearly Gaussian. Since then, the heaviness and the tail co-movements increase steadily. March 2007 has been identified as the beginning of the financial crisis (\cite{Acharya:2009}). TailCoR decreases between 2007 and 2008, but picks up again in 2008, when further troubles in the financial industry were made public. TailCoR then lowers for the next three years, though there is a small peak around 2011-2012 due to the European sovereign debt crisis. These are years of bonanza in financial markets and the economy overall. In 2016, there were a number of corrections worldwide that are reflected in the increase of TailCoR. The last increase, to levels comparable to the 2008 Great Financial Crisis, is due to COVID-19.

Panels (b) and (c) reveal that the pattern of TailCoR is driven by the nonlinear component. The linear component evolves  smoothly while the nonlinear component is more volatile and reacts faster to the events that drive financial markets. COVID-19 is a case in point. Though the linear component has increased by 0.05 for every index, the nonlinear component increases significantly more.

\section{Conclusions}
We have introduced TailCoR, a metric for tail correlations that is a function of linear and nonlinear components, the latter characterised by the heaviness of the tails. TailCoR is exact for any probability level, it does not depend on any specific distributional assumption, and no optimisations are needed. Monte Carlo simulations reveal its goodness in finite samples. An empirical illustration to a panel of 21 market indexes around the world shows an increase of TailCoR during the financial crisis that was due to a surge of linear and nonlinear components, while the surge during the COVID-19 crisis was mostly due to the nonlinear component.

Several extensions are possible. One is conditional TailCoR, which would be based on the conditional IQR of the projection ($\mbox{IQR}^{(j\,k)\,\xi}_t = \mbox{Q}^{(j\,k)\,1-\xi}_t-\mbox{Q}^{(j\,k)\,\xi}_t$), where the time varying quantiles are regressed on the financial and economic determinants of tail correlations.
Under ellipticity
the dynamic linear correlation $\rho_{i\,j,t}$ can be estimated with a robustified version of the DCC model (\cite{BoudtDanielssonLaurent11}).

A second extension is assuming a meta-elliptical copula (\cite{KKP08} and \cite{Krajina12}), which allows for different tail indexes for the marginal distributions and the copula. In this case, the first step should be done with a monotonically increasing transformation that standardises not only for the location and scale but also for the marginal heaviness of the tails.


\small
\section*{Acknowledgements}
We are grateful to the seminar participants at the Federal Reserve Bank of New York, the University of Valencia, the Universit\' e de Franche-Comt\' e, Riksbank, Complutense University, HEC Lausanne, University of Helsinki, University of Vienna, University of \mbox{St.} Gallen, and the Swedish House of Finance, as well as the conference participants at the CFE11 (London, December 2011), the Humboldt-Copenhagen Conference on Recent Developments in Financial Econometrics (Berlin, March 2013), the ENTER jamboree (Brussels, March 2013), the Econometric Society Australasian Meeting (Sydney, July 2013), the First Intl. Workshop on Financial Econometrics (Natal, October 2013), and the Workshop on Skewness, Heavy Tails, Market Crashes, and Dynamics (Cambridge, April 2014).

\small
\bibliographystyle{chicago}
\bibliography{RV_biblio}


\newpage
\section*{Appendix P: proofs} \label{appendixP}

\subsection*{Proof of Theorem 1}
We first consider the case when $\rho_{j\,k}>0$ and hence $Z^{(j\,k)}_t=\frac{1}{\sqrt{2}}(Y_{j\,t}+Y_{k\,t})$. The variance of $Y_{j\,t}$ is
\begin{equation*} 
\sigma^2_{Y_{j}} = \frac{\sigma^2_{X_j}}{({\rm IQR}^{\tau}_j)^2},
\end{equation*} 
and likewise for $Y_{k\,t}$. The variance of $Z^{(j\,k)}_t$ is
\begin{equation*}
\sigma^2_{(j\,k)} = \frac{1}{2}\left(  \frac{\sigma^2_{X_j} }{({\rm IQR}^{\tau}_j)^2} + \frac{ \sigma^2_{X_k}}{({\rm IQR}^{\tau}_k)^2}  + 2\sigma_{Y_j Y_k} \right), 
\end{equation*}
where $\sigma_{Y_j Y_k}$ is the covariance between $Y_{j\,t}$ and $Y_{k\,t}$. Since ${\rm IQR}^{\tau}_j=  k(\tau, \alpha)\sigma_{X_j}$ and ${\rm IQR}^{\tau}_k=  k(\tau, \alpha)\sigma_{X_k}$, we find
\begin{equation*}
\sigma^2_{(j\,k)}  = \frac{1}{2}\left(\frac{\sigma_{X_j}^2}{k(\tau, \alpha)^2  \sigma_{X_j}^2}+ \frac{\sigma_{X_k}^2}{k(\tau, \alpha)^2  \sigma_{X_k}^2} + 2 \frac{\sigma_{X_j\,X_k}}{k(\tau, \alpha)^2  \sigma_{X_j} \sigma_{X_k}} \right),
\end{equation*}
which, in a more compact form, equals
\begin{equation}
\sigma^2_{(j\,k)}  = \frac{1}{k(\tau, \alpha)^2}\left(1 + \rho_{j\,k} \right).  \nonumber 
 \label{VARZRHO}
\end{equation}
By the affine invariance of the elliptical family, ${\rm IQR}^{(j\,k)\,\xi}= k(\xi, \alpha)\sigma_{(j\,k)}$. Substituting in $\sigma^2_{(j\,k)}$
\begin{equation*}
{\rm IQR}^{(j\,k)\,\xi} =\frac{ k(\xi,\alpha)}{ k(\tau,\alpha)}  \sqrt{1 +\rho_{j\,k}}=s(\xi,\tau,\alpha)\sqrt{1+\rho_{j\,k}}.
\end{equation*}
In the Gaussian case $k(\tau,\alpha)= k(\tau)$ and $k(\xi,\alpha)= k(\xi)$. We normalize ${\rm IQR}^{(j\,k)\xi}$ by $\frac{ k(\tau)}{k(\xi)}=s_g(\xi,\tau)$ yielding
\begin{equation*}
\mbox{TailCoR}^{(j\,k)\,\xi} = s_g(\xi,\tau)s(\xi,\tau,\alpha)\sqrt{1+\rho_{j\,k}}.
\end{equation*}
The same proof follows for $\rho_{j\,k}<0$ and $Z^{(j\,k)}_t=\frac{1}{\sqrt{2}}(Y_{j\,t}-Y_{k\,t})$, except that $\sqrt{1+\rho_{j\,k}}$ is replaced by $\sqrt{1-\rho_{j\,k}}$. This change is unsubstantial since both expressions are equal ($\rho_{j\,k}$ is positive in $\sqrt{1+\rho_{j\,k}}$ and negative in $\sqrt{1-\rho_{j\,k}}$). Hence
\begin{equation*}
\mbox{TailCoR}^{(j\,k)\,\xi} = s_g(\xi,\tau)s(\xi,\tau,\alpha)\sqrt{1+|\rho_{j\,k}|}.
\end{equation*}

Q.E.D.

\subsection*{Proof of Theorem 2}
Let $\mbox{Q}_{j_T}^{0.50} := \mbox{Q}_{j}^{0.50} + T^{-\frac{1}{2}}\eta_{T}^{1}$ for some bounded sequence  $\eta_{T}^{1}$ and $\mbox{Q}_{k_T}^{0.50} := \mbox{Q}_{k}^{0.50} + T^{-\frac{1}{2}}\eta_{T}^{2}$ for some bounded sequence  $\eta_{T}^{2}$. Similarly, let $\mbox{IQR}_{j_T}^{\tau} := \mbox{IQR}_{j}^{\tau}+ T^{-\frac{1}{2}}s_{T}^{1}$ and $\mbox{IQR}_{k_T}^{\tau} = \mbox{IQR}_{k}^{\tau}+ T^{-\frac{1}{2}}s_{T}^{2}$ for some bounded sequences $s_{T}^{1}$ and $s_{T}^{2}$.

First, we show boundedness of the estimated IQR.

\begin{description}
\item[{\bf Lemma 1}]
For a given data set $X_1, \dots, X_n$ and some values $a_1, \dots, a_n$ that belong to the interval $[-\epsilon, \epsilon]$, we can define $Y_i := X_i + a_i$. Denote the interquantile range of the data set $X$ and $Y$ by $\hat{\mbox{IQR}}(X_1, \dots, X_n) $ and $\hat{\mbox{IQR}}(Y_1, \dots, Y_n)$, respectively. Then the following holds:
$$
|\hat{\mbox{IQR}}(X_1, \dots, X_n) - \hat{\mbox{IQR}}(Y_1, \dots, Y_n)| \leq 10\epsilon.
$$
\end{description}

\begin{description}
\item[{\bf Proof}] 
Let $\sigma$ be a permutation that orders our data $X$ in ascending order. That is, $X_{\sigma(1)} \leq X_{\sigma(2)} \leq \dots \leq X_{\sigma(n)}$. Similarly, $\pi$ is a permutation such that $Y_{\pi(1)} \leq Y_{\pi(2)} \leq \dots \leq Y_{\pi(n)}$. Then, for fixed values $i$ and $j$ we have that  $\hat{\mbox{IQR}}(X_1, \dots, X_n)= X_{\sigma(j)} - X_{\sigma(i)}$ and $\hat{\mbox{IQR}}(Y_1, \dots, Y_n) = Y_{\pi(j)} - Y_{\pi(i)}$. First we will show that $|X_{\sigma(i)} - Y_{\pi(i)}| \leq 5\epsilon$ for all $i$. Let $k$ be such that $\pi(k) = \sigma(i)$ and let $m$ be such that $\pi(i) = \sigma(m)$. We have two cases:
\begin{itemize}
    \item $k,m < i$ or $k, m > i$. That means that data points $X_{\sigma(i)}$ and $X_{\sigma(m)}$ changed their ordering after we added the noise. This implies that $|X_{\sigma(i)} - X_{\sigma(m)}| \leq 2\epsilon$. Therefore, $|X_{\sigma(i)} - Y_{\pi(i)}| = |X_{\sigma(i)} - X_{\sigma(m)} - a_{\sigma(m)}|\leq |X_{\sigma(i)} - X_{\sigma(m)}| + |a_{\sigma(m)}| \leq 3\epsilon$.
    \begin{center}
    \begin{tikzpicture}
  [scale=.6,auto=left,every node/.style={circle,fill=blue!20}]
\node (n11) at (1,19) {};
  \node (n12) at (3,19)  {$\sigma(m)$};
  \node (n13) at (5,19)  {};
  \node (n14) at (7,19)  {$\sigma(i)$};
  \node (n15) at (9,19)  {};
  \node (n16) at (11,19) {};
\node (n21) at (1,15)  {$\pi(k)$};
\node (n22) at (3,15)  {};
\node (n23) at (5,15)  {};
\node (n24) at (7,15)  {$\pi(i)$};
\node (n25) at (9,15)  {};
\node (n26) at (11,15)  {};
  \foreach \from/\to in {n12/n24,n14/n21}
   \draw (\from) -- (\to);
\end{tikzpicture}
\end{center}
    \item $k < i < m$ or $m < i < k$ . In this case we have that the ordering is preserved which means $X_{\sigma(m)} \geq X_{\sigma(i)} \land Y_{\pi(i)} \geq Y_{\pi(k)}$ or $X_{\sigma(m)} \leq X_{\sigma(i)} \land Y_{\pi(i)} \leq Y_{\pi(k)}$. Assume that we have the first case.
    \begin{center}
    \begin{tikzpicture}
  [scale=.6,auto=left,every node/.style={circle,fill=blue!20}]
\node (n11) at (1,19) {};
  \node (n12) at (3,19)  {$\sigma(a)$};
  \node (n13) at (5,19)  {};
  \node (n14) at (7,19)  {};
  \node (n15) at (9,19)  {};
  \node (n16) at (11,19) {$\sigma(i)$};
  \node (n17) at (13,19)  {};
  \node (n18) at (15,19)  {};
  \node (n19) at (17,19)  {$\sigma(m)$};
\node (n21) at (1,15)  {};
\node (n22) at (3,15)  {};
\node (n23) at (5,15)  {};
\node (n24) at (7,15)  {$\pi(k)$};
\node (n25) at (9,15)  {};
\node (n26) at (11,15)  {$\pi(i)$};
\node (n27) at (13,15)  {};
\node (n28) at (15,15)  {};
\node (n29) at (17,15)  {$\pi(b)$};
  \foreach \from/\to in {n12/n29,n16/n24,n19/n26}
   \draw (\from) -- (\to);
\end{tikzpicture}
\end{center}
    The ordering of the elements $X_{\sigma(i)}$ and $X_{\sigma(m)}$ is preserved but these elements shifted after adding the noise. In our example, $X_{\sigma(m)}$ was not in the first $i$ elements originally, but after adding the noise, it is there. This implies that there is an element, let us say $X_{\sigma(a)}$, that originally was in the first $i$ elements but after adding the noise it is not there anymore. That is, $a<i<m$. Therefore, there is $b$ such that $k<i<b$. We can conclude that $X_{\sigma(a)}$ has changed its ordering with respect to both $X_{\sigma(i)}$ and  $X_{\sigma(m)}$. This implies that 
     $|X_{\sigma(i)} - X_{\sigma(a)}| \leq 2\epsilon$ and  $|X_{\sigma(m)} - X_{\sigma(a)}| \leq 2\epsilon$. Hence,
  \begin{align*}
    &|X_{\sigma(i)} - Y_{\pi(i)}| = |X_{\sigma(i)} - X_{\sigma(m)} - a_{\sigma(m)}| \leq |X_{\sigma(i)} - X_{\sigma(m)}| + |a_{\sigma(m)}| \leq\\
    &|X_{\sigma(i)} - X_{\sigma(a)} + X_{\sigma(a)} - X_{\sigma(m)}| + \epsilon \leq |X_{\sigma(i)} - X_{\sigma(a)}| + |X_{\sigma(a)} - X_{\sigma(m)}| + \epsilon 
    \leq 2\epsilon + 2\epsilon + \epsilon = 5 \epsilon
    \end{align*}
\end{itemize}
Finally,
\begin{align*}
&|\hat{\mbox{IQR}}(X_1, \dots, X_n) - \hat{\mbox{IQR}}(Y_1, \dots, Y_n)| = | X_{\sigma(j)}  - X_{\sigma(i)}  - (Y_{\pi(j)} - Y_{\pi(i)}) | \leq \\
&| X_{\sigma(j)}  - Y_{\pi(j)}| + | X_{\sigma(i)}  - Y_{\pi(i)}| \leq 10\epsilon.
\end{align*}
\end{description}

Q.E.D.

\

We now turn to the proof of Theorem $2$. We need to prove that
 $$\hat{\mbox{TailCoR}}_{T}^{(jk)\xi}(\hat{\mbox{Q}}_{j}^{0.50},\hat{\mbox{Q}}_{k}^{0.50},\hat{\mbox{IQR}}_{j}^{\tau},\hat{\mbox{IQR}}_{k}^{\tau}) -\mbox{TailCoR}^{(jk)\xi}$$
is $o_p(1)$. First, we add and subtract $\hat{\mbox{TailCoR}}_{T}^{(jk)\xi}$ as a function of the population quantiles:
\begin{align}\label{eq:proof1}
    \hat{\mbox{TailCoR}}_{T}^{(jk)\xi}(\hat{\mbox{Q}}_{j}^{0.50},\hat{\mbox{Q}}_{k}^{0.50},\hat{\mbox{IQR}}_{j}^{\tau},\hat{\mbox{IQR}}_{k}^{\tau}) - \hat{\mbox{TailCoR}}_{T}^{(jk)\xi}(\mbox{Q}_{j}^{0.50},\mbox{Q}_{k}^{0.50},\mbox{IQR}_{j}^{\tau},\mbox{IQR}_{k}^{\tau})\\
    +\hat{\mbox{TailCoR}}_{T}^{(jk)\xi}(\mbox{Q}_{j}^{0.50},\mbox{Q}_{k}^{0.50},\mbox{IQR}_{j}^{\tau},\mbox{IQR}_{k}^{\tau})
    -\mbox{TailCoR}^{(jk)\xi}. \nonumber
\end{align}

We will first prove that the first part of (\ref{eq:proof1}) is $o_p(1)$, that is, as $T\rightarrow\infty$
$$
\hat{\mbox{TailCoR}}_{T}^{(jk)\xi}(\hat{\mbox{Q}}_{j}^{0.50},\hat{\mbox{Q}}_{k}^{0.50},\hat{\mbox{IQR}}_{j}^{\tau},\hat{\mbox{IQR}}_{k}^{\tau}) - \hat{\mbox{TailCoR}}_{T}^{(jk)\xi}(\mbox{Q}_{j}^{0.50},\mbox{Q}_{k}^{0.50},\mbox{IQR}_{j}^{\tau},\mbox{IQR}_{k}^{\tau}) = o_p(1).
$$
The definition of TailCoR implies that we have to prove that
$$
\hat{\mbox{IQR}}_{T}^{(jk)\xi}(\hat{\mbox{Q}}_{j}^{0.50},\hat{\mbox{Q}}_{k}^{0.50},\hat{\mbox{IQR}}_{j}^{\tau},\hat{\mbox{IQR}}_{k}^{\tau}) - \hat{\mbox{IQR}}_{T}^{(jk)\xi}(\mbox{Q}_{j}^{0.50},\mbox{Q}_{k}^{0.50},\mbox{IQR}_{j}^{\tau},\mbox{IQR}_{k}^{\tau}) = o_p(1)
$$
as $T\rightarrow\infty$. In view of Lemma 4.4 of \cite{kreiss1987adaptive}, this reduces to proving  that
$$
\hat{\mbox{IQR}}_{T}^{(jk)\xi}(\mbox{Q}_{j_T}^{0.50},\mbox{Q}_{k_T}^{0.50},\mbox{IQR}_{j_T}^{\tau},\mbox{IQR}_{k_T}^{\tau}) - \hat{\mbox{IQR}}_{T}^{(jk)\xi}(\mbox{Q}_{j}^{0.50},\mbox{Q}_{k}^{0.50},\mbox{IQR}_{j}^{\tau},\mbox{IQR}_{k}^{\tau}) = o_p(1)
$$
as $T\rightarrow\infty$, for the sequences defined at the very beginning of the proof. We add and subtract  $\hat{\mbox{IQR}}_{T}^{(jk)\xi}$ as a function of the population quantiles:
\begin{align}\label{eq:proof2}
  \hat{\mbox{IQR}}_{T}^{(jk)\xi}(\mbox{Q}_{j_T}^{0.50},\mbox{Q}_{k_T}^{0.50},\mbox{IQR}_{j_T}^{\tau},\mbox{IQR}_{k_T}^{\tau})
  -\hat{\mbox{IQR}}_{T}^{(jk)\xi}(\mbox{Q}_{j}^{0.50},\mbox{Q}_{k}^{0.50},\mbox{IQR}_{j_T}^{\tau},\mbox{IQR}_{k_T}^{\tau})\\
  +\hat{\mbox{IQR}}_{T}^{(jk)\xi}(\mbox{Q}_{j}^{0.50},\mbox{Q}_{k}^{0.50},\mbox{IQR}_{j_T}^{\tau},\mbox{IQR}_{k_T}^{\tau})
  - \hat{\mbox{IQR}}_{T}^{(jk)\xi}(\mbox{Q}_{j}^{0.50},\mbox{Q}_{k}^{0.50},\mbox{IQR}_{j}^{\tau},\mbox{IQR}_{k}^{\tau}). \nonumber
\end{align}
We now show that $  \hat{\mbox{IQR}}_{T}^{(jk)\xi}(\mbox{Q}_{j_T}^{0.50},\mbox{Q}_{k_T}^{0.50},\mbox{IQR}_{j_T}^{\tau},\mbox{IQR}_{k_T}^{\tau})
  -\hat{\mbox{IQR}}_{T}^{(jk)\xi}(\mbox{Q}_{j}^{0.50},\mbox{Q}_{k}^{0.50},\mbox{IQR}_{j_T}^{\tau},\mbox{IQR}_{k_T}^{\tau}) = 0.$ Indeed,
$
Y_{jt}^{(1)} = \frac{X_{jt} - \mbox{Q}_{j_T}^{0.50}}{\mbox{IQR}_{j_T}^{\tau}} = Y_{jt}^{(2)}  - \frac{T^{-\frac{1}{2}}\eta_{T}^{1}}{\mbox{IQR}_{j_T}^{\tau}}
$
where
$
Y_{jt}^{(2)} = \frac{X_{jt}-\mbox{Q}_{j}^{0.50}}{\mbox{IQR}_{j_T}^{\tau}}
$
and
$
Y_{kt}^{(1)} = \frac{X_{kt} - \mbox{Q}_{k_T}^{0.50}}{\mbox{IQR}_{k_T}^{\tau}} = Y_{kt}^{(2)}  - \frac{T^{-\frac{1}{2}}\eta_{T}^{2}}{\mbox{IQR}_{k_T}^{\tau}}
$
where
$
Y_{kt}^{(2)} = \frac{X_{kt}-\mbox{Q}_{k}^{0.50}}{\mbox{IQR}_{k_T}^{\tau}}.
$
Therefore,
\begin{align*}
Z_{t}^{(1)(jk)}& = Y_{jt}^{(1)}\mbox{ cos }\phi + Y_{kt}^{(1)}\mbox{ sin }\phi\\
&=\left(Y_{jt}^{(2)}  - \frac{T^{-\frac{1}{2}}\eta_{T}^{1}}{\mbox{IQR}_{j_T}^{\tau}}\right)\mbox{ cos }\phi + \left(Y_{kt}^{(2)}  - \frac{T^{-\frac{1}{2}}\eta_{T}^{2}}{\mbox{IQR}_{k_T}^{\tau}}\right)\mbox{ sin }\phi\\
&=Y_{jt}^{(2)}\mbox{ cos }\phi + Y_{kt}^{(2)}\mbox{ sin }\phi -
\frac{T^{-\frac{1}{2}}\eta_{T}^{1}}{\mbox{IQR}_{j_T}^{\tau}}\mbox{ cos }\phi -\frac{T^{-\frac{1}{2}}\eta_{T}^{2}}{\mbox{IQR}_{k_T}^{\tau}}\mbox{ sin }\phi\\
&=Z_{t}^{(2)(jk)}-\frac{T^{-\frac{1}{2}}\eta_{T}^{1}}{\mbox{IQR}_{j_T}^{\tau}}\mbox{ cos }\phi -\frac{T^{-\frac{1}{2}}\eta_{T}^{2}}{\mbox{IQR}_{k_T}^{\tau}}\mbox{ sin }\phi.
\end{align*}
This implies that the IQR of $Z_{t}^{(1)(jk)}$ equals the IQR of $Z_{t}^{(2)(jk)}$.
The next step is to prove that 
\begin{align*}
    \hat{\mbox{IQR}}_{T}^{(jk)\xi}(\mbox{Q}_{j}^{0.50},\mbox{Q}_{k}^{0.50},\mbox{IQR}_{j_T}^{\tau},\mbox{IQR}_{k_T}^{\tau})
  - \hat{\mbox{IQR}}_{T}^{(jk)\xi}(\mbox{Q}_{j}^{0.50},\mbox{Q}_{k}^{0.50},\mbox{IQR}_{j}^{\tau},\mbox{IQR}_{k}^{\tau})= o_p(1)
\end{align*}
as $T\rightarrow\infty$. The notation
$
    \hat{\mbox{IQR}}_{T}^{(jk)\xi}(\mbox{Q}_{j}^{0.50},\mbox{Q}_{k}^{0.50},\mbox{IQR}_{j_T}^{\tau},\mbox{IQR}_{k_T}^{\tau})
$ means that we have to estimate the IQR of the following sample
\begin{align*}
 \frac{X_{j1} - \mbox{Q}_{j}^{0.50}}{\mbox{IQR}_{j_T}^{\tau}}\mbox{ cos }\phi +\frac{X_{k1} - \mbox{Q}_{k}^{0.50}}{\mbox{IQR}_{k_T}^{\tau}} \mbox{ sin }\phi, \dots, \frac{X_{jT} - \mbox{Q}_{j}^{0.50}}{\mbox{IQR}_{j_T}^{\tau}}\mbox{ cos }\phi +\frac{X_{kT} - \mbox{Q}_{k}^{0.50}}{\mbox{IQR}_{k_T}^{\tau}} \mbox{ sin }\phi.
\end{align*}
Let  $T$ be large enough such that $\mbox{IQR}_{j_T}^{\tau} = \mbox{IQR}_{j}^{\tau}+ \frac{s_{T}^{1}}{\sqrt{T}}   > \frac{\mbox{IQR}_{j}^{\tau}}{2}$ and $\mbox{IQR}_{k_T}^{\tau} = \mbox{IQR}_{k}^{\tau}+ \frac{s_{T}^{2}}{\sqrt{T}}     > \frac{\mbox{IQR}_{k}^{\tau}}{2}$. For such $T$ we have that:
$$
\displaystyle\abs*{\frac{1}{\mbox{IQR}_{j}^{\tau}+ \frac{s_{T}^{1}}{\sqrt{T}}} - \frac{1}{\mbox{IQR}_{j}^{\tau}}} = \frac{1}{\sqrt{T}}\abs*{\frac{s_{T}^{1}}{\mbox{IQR}_{j}^{\tau}(\mbox{IQR}_{j}^{\tau} + \frac{s_{T}^{1}}{\sqrt{T}})}} < \frac{1}{\sqrt{T}}\frac{2|s_{T}^{1}|}{(\mbox{IQR}_{j}^{\tau}) ^ 2} \leq \frac{1}{\sqrt{T}}\frac{2C_1}{(\mbox{IQR}_{j}^{\tau}) ^ 2}.
$$
Similarly, $\displaystyle\abs[\Big]{\frac{1}{\mbox{IQR}_{k}^{\tau}+ \frac{s_{T}^{2}}{\sqrt{T}}} - \frac{1}{\mbox{IQR}_{k}^{\tau}}} \leq \frac{1}{\sqrt{T}}\frac{2C_2}{(\mbox{IQR}_{k}^{\tau}) ^ 2}$.
So, for any $t < T$ and $T$ large enough we have that 
\begin{align*}
\displaystyle&\abs*{\frac{X_{jt} - \mbox{Q}_{j}^{0.50}}{\mbox{IQR}_{j_T}^{\tau}}\mbox{ cos }\phi +\frac{X_{kt} - \mbox{Q}_{k}^{0.50}}{\mbox{IQR}_{k_T}^{\tau}} \mbox{ sin }\phi - \frac{X_{jt} - \mbox{Q}_{j}^{0.50}}{\mbox{IQR}_{j}^{\tau}}\mbox{ cos }\phi -\frac{X_{kt} - \mbox{Q}_{k}^{0.50}}{\mbox{IQR}_{k}^{\tau}} \mbox{ sin }\phi}  
\\ \leq &\frac{1}{\sqrt{T}}\frac{2C_1}{(\mbox{IQR}_{j}^{\tau}) ^ 2}\abs[\Big]{X_{jt} - \mbox{Q}_{j}^{0.50}}|\mbox{cos } \phi| +  \frac{1}{\sqrt{T}}\frac{2C_2}{(\mbox{IQR}_{k}^{\tau}) ^ 2}\abs[\Big]{X_{kt} - \mbox{Q}_{k}^{0.50}}|\mbox{sin } \phi|
 \\
\leq&\frac{1}{\sqrt{T}}\frac{2C_1}{(\mbox{IQR}_{j}^{\tau}) ^ 2}\max_{t}\abs[\Big]{X_{jt}}|\mbox{cos } \phi|  + \frac{1}{\sqrt{T}}\frac{2C_2}{(\mbox{IQR}_{k}^{\tau}) ^ 2}\max_{t} \abs[\Big]{X_{kt}}|\mbox{sin } \phi| 
\\ +&\frac{1}{\sqrt{T}}\frac{2C_1}{(\mbox{IQR}_{j}^{\tau}) ^ 2}\abs[\Big]{\mbox{Q}_{j}^{0.50}}|\mbox{cos } \phi|  + \frac{1}{\sqrt{T}}\frac{2C_2}{(\mbox{IQR}_{k}^{\tau}) ^ 2}\abs[\Big]{\mbox{Q}_{k}^{0.50}}|\mbox{sin } \phi|.
\end{align*}
Using Lemma 1 and the upper bound obtained above, we can conclude that 
\begin{align*}
  &  \abs*{\hat{\mbox{IQR}}_{T}^{(jk)\xi}(\mbox{Q}_{j}^{0.50},\mbox{Q}_{k}^{0.50},\mbox{IQR}_{j_T}^{\tau},\mbox{IQR}_{k_T}^{\tau})
  - \hat{\mbox{IQR}}_{T}^{(jk)\xi}(\mbox{Q}_{j}^{0.50},\mbox{Q}_{k}^{0.50},\mbox{IQR}_{j}^{\tau},\mbox{IQR}_{k}^{\tau})}  \\
\leq & 10\left(\frac{1}{\sqrt{T}}\frac{2C_1}{(\mbox{IQR}_{j}^{\tau})^2}\max_{t}\abs[\Big]{X_{jt}}|\mbox{cos } \phi|  + \frac{1}{\sqrt{T}}\frac{2C_2}{(\mbox{IQR}_{k}^{\tau})^2} \max_{t}\abs[\Big]{ X_{kt}  }|\mbox{sin } \phi|\right.\\
& \left.\quad\quad+\frac{1}{\sqrt{T}}\frac{2C_1}{(\mbox{IQR}_{j}^{\tau})^2}\abs[\Big]{\mbox{Q}_{j}^{0.50}}|\mbox{cos } \phi|  + \frac{1}{\sqrt{T}}\frac{2C_2}{(\mbox{IQR}_{k}^{\tau})^ 2}\abs[\Big]{\mbox{Q}_{k}^{0.50}}|\mbox{sin } \phi|\right)
\end{align*}
Since
\begin{align*}
   & \frac{1}{\sqrt{T}}\frac{2C_1}{(\mbox{IQR}_{j}^{\tau})^2}\max_{t}\abs[\Big]{X_{jt}}|\mbox{cos } \phi|  + \frac{1}{\sqrt{T}}\frac{2C_2}{(\mbox{IQR}_{k}^{\tau})^2} \max_{t}\abs[\Big]{ X_{kt}  }|\mbox{sin } \phi| +\\
 &+\frac{1}{\sqrt{T}}\frac{2C_1}{(\mbox{IQR}_{j}^{\tau})^2}\abs[\Big]{\mbox{Q}_{j}^{0.50}}|\mbox{cos } \phi|  + \frac{1}{\sqrt{T}}\frac{2C_2}{(\mbox{IQR}_{k}^{\tau})^ 2}\abs[\Big]{\mbox{Q}_{k}^{0.50}}|\mbox{sin } \phi| = o_p(1)
\end{align*}
as $T\rightarrow\infty$, it holds that 
$$
\hat{\mbox{IQR}}_{T}^{(jk)\xi}(\mbox{Q}_{j}^{0.50},\mbox{Q}_{k}^{0.50},\mbox{IQR}_{j_T}^{\tau},\mbox{IQR}_{k_T}^{\tau})
  - \hat{\mbox{IQR}}_{T}^{(jk)\xi}(\mbox{Q}_{j}^{0.50},\mbox{Q}_{k}^{0.50},\mbox{IQR}_{j}^{\tau},\mbox{IQR}_{k}^{\tau}) = o_{p}(1)
$$
as $T\rightarrow\infty$. Last, we need to prove that the second part of (\ref{eq:proof1}) is $o_p(1)$, in other words, that
$$\hat{\mbox{TailCoR}}_{T}^{(jk)\xi}(\mbox{Q}_{j}^{0.50},\mbox{Q}_{k}^{0.50},\mbox{IQR}_{j}^{\tau},\mbox{IQR}_{k}^{\tau})
    -\mbox{TailCoR}^{(jk)\xi} = o_p(1)$$
as $T\rightarrow\infty$. This follows from the asymptotic properties of sample quantiles under $S$--mixing (\cite{DominicyHormannOgataVeredas12}). Q.E.D.

\subsection*{Proof of Theorem 3}
By {\bf E1}, $\hat{\mbox{TailCoR}}^{(j\,k)\,\xi}_{T} =  2s_g(\xi,\tau)\hat{\mbox{Q}}^{(j\,k)\,\xi}_T$.
The term $2s_g(\xi,\tau)$ is a deterministic scale shift and the only source of randomness is $\hat{\mbox{Q}}^{(j\,k)\,\xi}_T$. Hence
\begin{eqnarray*}
E(\hat{\mbox{TailCoR}}^{(j\,k)\,\xi}_T) &=& 2s_g(\xi,\tau)E(\hat{\mbox{Q}}^{(j\,k)\,\xi}_T) \mbox{ and }\\
Var(\hat{\mbox{TailCoR}}^{(j\,k)\,\xi}_T) &=& 4 s_g(\xi,\tau)^2 Var(\hat{\mbox{Q}}^{(j\,k)\,\xi}_T).
\end{eqnarray*}
By the asymptotic properties of sample quantiles under $S$--mixing (\cite{DominicyHormannOgataVeredas12}) and the delta method the proof is completed. Q.E.D.

\subsection*{Proof of Theorem 4}
It follows the same lines as the proof of Theorem 3. By {\bf E1},
$vech \hat{\mbox{\bf TailCoR}}^{\xi}_{T}(\mbox{\bf Q}^{0.50}_j,\mbox{\bf Q}^{0.50}_k,\mbox{\bf IQR}^{\tau}_{j},\mbox{\bf IQR}^{\tau}_{k}) = 2s_g(\xi,\tau)\hat{\mathbf{Q}}^{\xi}_T$, where $\hat{\mathbf{Q}}^{\xi}_T$ is a $\tilde{N} \times 1$ vector of sample quantiles.
By the asymptotic properties of vectors of sample quantiles under $S$--mixing (\cite{DominicyHormannOgataVeredas12}) and the delta method the proof is completed. Q.E.D.


\subsection*{Semi moments}
We review standard definitions and properties for positive semi moments that we need. Results for negative semi moments are analogous. Let $I_{\{X_j>\mu_j\}}$ be an indicator function that takes value one if $X_j>\mu_j$. The positive semi variance of $X_j$ is $\sigma^{+\,2}_j=E[((X_j-\mu_j)I_{\{X_j>\mu_j\}})^2]$. It possesses the standard location and scale shift property: let $a$ and $b$ be a real and a positive real number, respectively, then the positive semi variance of $a + bX_j$ is $b^2\sigma^{+\,2}_j$. The positive semi covariance between $X_j$ and $X_k$ is $\sigma^{+}_{j\,k}=E[((X_j-\mu_j)I_{\{X_j>\mu_j\}})((X_k-\mu_k)I_{\{X_k>\mu_k\}})]$. It is invariant to location shifts but not to re-scaling: if $c$ is real and $d$ is real positive, the positive semi covariance
between $a + bX_j$ and $c + dX_j$ ($a$ and $b$ defined as above) is $bd\sigma^{+}_{j\,k}$. The positive semi correlation between $X_j$ and $X_k$ is $\rho^{+}_{j\,k} = \frac{\sigma^{+}_{j\,k}}{\sigma^{+}_j\sigma^{+}_k}$.
Last, the positive semi variance of a sum, say $X_j + X_k$, is $$\sigma^{+\,2}_{X_j + X_k} = E[((X_j-\mu_j)I_{\{X_j>\mu_j\}}+(X_k-\mu_k)I_{\{X_k>\mu_k\}})^2] = \sigma^{+\,2}_j + \sigma^{+\,2}_k + 2\sigma^{+}_{j\,k}.$$

\subsection*{Proof of Theorem 5}
Since the proof follows the same lines as the proof of Theorem 1, we only show the main steps for the upside $\mbox{TailCoR}^{(j\,k)\,\xi \,+}$. The proof for $\mbox{TailCoR}^{(j\,k)\,\xi \,-}$ is equivalent.

We first consider the case when $\rho_{j\,k}>0$ and hence $Z^{(j\,k)}_{t} = Y_{j\,t}\cos \phi + Y_{k\,t} \sin \phi$. The positive semi variance of $Y_{j\,t}$ is
\begin{equation*} 
\sigma^{+\,2}_{Y_{j}} = \frac{\sigma^{+\,2}_{X_j}}{({\rm IQR}^{\tau\,+}_j)^2},
\end{equation*} 
and likewise for $Y_{k\,t}$. The positive semi variance of $Z^{(j\,k)}_t$ is
\begin{equation*}
\sigma^{+\,2}_{(j\,k)} = \frac{\sigma^{+\,2}_{X_j} }{({\rm IQR}^{\tau\,+}_j)^2} \sin^2 \phi + \frac{ \sigma^{+\,2}_{X_k}}{({\rm IQR}^{\tau\,+}_k)^2}  \cos^2 \phi + 2 \sigma_{Y_j Y_k}^{+} \sin \phi \cos \phi , 
\end{equation*}
where $\sigma_{Y_j Y_k}^{+}$ is the positive semi covariance between $Y_{j\,t}$ and $Y_{k\,t}$. Since ${\rm IQR}^{\tau\,+}_j=  k(\tau, \alpha, \gamma)^{+}\sigma_{X_j}^{+}$, ${\rm IQR}^{\tau\,+}_k=  k(\tau, \alpha, \gamma)^{+}\sigma_{X_k}^{+}$, $\sin^2 \phi + \cos^2 \phi = 1$, $\sigma_{Y_j Y_k}^{+}=\sigma_{X_j X_k}^{+}/{\rm IQR}^{\tau\,+}_j {\rm IQR}^{\tau\,+}_k$, and by using the positive semi correlation, $\sigma^{+\,2}_{(j\,k)}$ simplifies to
\begin{equation}
\sigma^{+\,2}_{(j\,k)}  = \frac{1}{k(\tau, \alpha, \gamma)^2}\left(1 + 2 \rho_{j\,k}^{+} \sin \phi \cos \phi  \right).  \nonumber 
\end{equation}
Substituting $\sigma^{+}_{(j\,k)}$ for ${\rm IQR}^{(j\,k)\,\xi\,+}/k(\xi, \alpha, \gamma)^+$
\begin{equation*}
{\rm IQR}^{(j\,k)\,\xi\,+} = s(\xi,\tau,\alpha,\gamma)^{+}\sqrt{1+2 \rho_{j\,k}^{+} \sin \phi \cos \phi}, 
\end{equation*}
where $s(\xi,\tau,\alpha,\gamma)^+ = k(\xi, \alpha, \gamma)^{+}/k(\tau, \alpha, \gamma)^+$. Multiplying by the normalization $2s_g(\xi,\tau)$ yields
\begin{equation*}
 2s_g(\xi,\tau) {\rm IQR}^{(j\,k)\,\xi\,+} = 2s_g(\xi,\tau) s(\xi,\tau,\alpha,\gamma)^{+} \sqrt{1+2 \rho_{j\,k}^{+} \sin \phi \cos \phi}.
\end{equation*}
The same development follows for $\rho_{j\,k}^{+}<0$ and $Z^{(j\,k)}_{t} = Y_{j\,t}\cos \phi - Y_{k\,t} \sin \phi$, except that $\sqrt{1+2 \rho_{j\,k}^{+} \sin \phi \cos \phi}$ is replaced by $\sqrt{1-2 \rho_{j\,k}^{+} \sin \phi \cos \phi}$.

Now, $\sqrt{1+2 \rho_{j\,k}^{+} \sin \phi \cos \phi}$ and $\sqrt{1-2 \rho_{j\,k}^{+} \sin \phi \cos \phi}$ take the same value under the correct projection ($\cos \phi$ is negative when $\rho_{j\,k}<0$). Both expressions can be then written as
\begin{equation*}
\sqrt{1+2 |\rho_{j\,k}^{+}| |\sin \phi \cos \phi|}
\end{equation*}
Hence
\begin{equation*}
\mbox{TailCoR}^{(j\,k)\,\xi +} = 2s_g(\xi,\tau)s(\xi,\tau,\alpha,\gamma)^{+}\sqrt{1+2 |\rho_{j\,k}^{+}| |\sin \phi \cos \phi|}.
\end{equation*}
Q.E.D.

\newpage
 \newpage
\section*{Appendix T: Tables} \label{appendixT}

\begin{table}[h]
	\begin{center}
\caption{Tabulation of $s_g(\xi,\tau)$}
\label{tab:sg}
\begin{sideways}
\scriptsize
	\begin{tabular}{lcccccccccccccc}
	\hline
	\hline
	& \multicolumn{14}{c}{$\xi$} \\
$\tau$ & 0.700 & 0.725 & 0.750 & 0.775 & 0.800	 & 0.825 & 0.850 & 0.875 & 0.900 & 0.925 & 0.950 & 0.975 & 0.990 & 0.995\\
	\hline
0.600 & 0.483 & 0.424 & 0.375 & 0.335 & 0.301 & 0.271 & 0.244 & 0.220 & 0.198 & 0.176 & 0.154 & 0.129 & 0.109 & 0.098 \\
0.625
	&	0.607	&	0.533	&	0.472	&	0.422	&	0.379	&	0.341	&	0.307	&	0.277	&	0.249	&	0.221	&	0.194	&	0.163	&	0.137	&  0.124 \\
0.650	&	0.735	&	0.644	&	0.571	&	0.510	&	0.458	&	0.412	&	0.372	&	0.335	&	0.301	&	0.268	&	0.234	&	0.196	&	0.166	& 0.150 \\
0.675	&	0.865	&	0.759	&	0.673	&	0.601	&	0.539	&	0.486	&	0.438	&	0.394	&	0.354	&	0.315	&	0.276	&	0.231	&	0.195	& 0.176 \\
0.700	&	--	&	0.877	&	0.778	&	0.694	&	0.623	&	0.561	&	0.506	&	0.456	&	0.409	&	0.364	&	0.319	&	0.267	&	0.226	& 0.204 \\
0.725	&	--	&	--	&	0.886	&	0.791	&	0.711	&	0.640	&	0.577	&	0.520	&	0.466	&	0.415	&	0.363	&	0.305	&	0.257	& 0.232 \\
0.750	&	--	&	--	&	--	&	0.893	&	0.801	&	0.722	&	0.651	&	0.586	&	0.526	&	0.468	&	0.410	&	0.344	&	0.290	& 0.262 \\
0.775	&	--	&	--	&	--	&	--	&	0.898	&	0.808	&	0.729	&	0.657	&	0.589
	&	0.525	&	0.459	&	0.385	&	0.325	& 0.293 \\
0.800	&	--	&	--	&	--	&	--	&	--	&	0.900	&	0.812	&	0.731	&	0.657	&
	0.585	&	0.512	&	0.429	&	0.362	& 0.327 \\
0.825	&	--	&	--	&	--	&	--	&	--	&	--	&	0.902	&	0.812	&	0.729	&	0.649
	&	0.568	&	0.477	&	0.402	& 0.363 \\
0.850	&	--	&	--	&	--	&	--	&	--	&	--	&	--	&	0.901	&	0.809	&	0.720	&
	0.630	&	0.529	&	0.445	& 0.402 \\
0.875	&	--	&	--	&	--	&	--	&	--	&	--	&	--	&	--	&	0.898	&	0.799	&	0.699
	&	0.587	&	0.494	& 0.447 \\
0.900	&	--	&	--	&	--	&	--	&	--	&	--	&	--	&	--	&	--	&	0.890	&	0.779	&
	0.654	&	0.551	& 0.497 \\
	\hline
	\hline
\multicolumn{15}{p{16cm}}{\footnotesize \vspace{0.00cm} Interpolation can be used for values of $\xi$ and $\tau$ that are not in the table. Alternatively, $s_g(\xi,\tau)$ can be computed as $\frac{\Phi(\tau)^{-1}}{\Phi(\xi)^{-1}}$ where $\Phi(\cdot)$ is the cumulative distribution function of a standardized Gaussian distribution.} \label{tbl:sg}
%
	\end{tabular}
\end{sideways}
	\end{center}
\end{table}


\begin{landscape}
\begin{table}[ht]
\centering
\caption{Matrix of $\mbox{TailCoR}^{0.975}$}
\begin{adjustbox}{max width=1.3\textwidth}\label{tab:tailcor}
\begin{tabular}{rcccccccccccccccccccccc}
  \hline
   & \rot{90}{SP} & \rot{90}{NASDAQ} & \rot{90}{TSX} & \rot{90}{MERVAL} & \rot{90}{BOVESPA} & \rot{90}{IPC} & \rot{90}{AEX} & \rot{90}{ATX} & \rot{90}{FTSE} & \rot{90}{DAX} & \rot{90}{CAC} & \rot{90}{SMI} & \rot{90}{MIB} & \rot{90}{HGSG} & \rot{90}{NIKKEI} & \rot{90}{STRTIM} & \rot{90}{SSEC} & \rot{90}{BSE} & \rot{90}{KLSE} & \rot{90}{KOSPI} & \rot{90}{ALLORD} &\rot{90}{Average} \\ 
  \hline
SP & 2.33 & 2.33 & 2.12 & 1.87 & 1.78 & 2.03 & 2.00 & 1.83 & 1.95 & 1.99 & 1.97 & 1.91 & 1.93 & 1.67 & 1.59 & 1.77 & 1.65 & 1.71 & 1.59 & 1.80 & 1.62 & 1.86 \\ 
   & (13.69) & (12.89) & (13.53) & (8.77) & (7.87) & (10.59) & (10.43) & (10.22) & (10.11) & (10.17) & (10.49) & (9.93) & (8.98) & (8.28) & (7.68) & (9.78) & (7.92) & (9.21) & (8.38) & (8.91) & (9.70)& \\ 
NASDAQ &  & 2.36 & 2.10 & 1.83 & 1.78 & 1.97 & 2.00 & 1.74 & 1.93 & 1.95 & 1.92 & 1.86 & 1.91 & 1.71 & 1.61 & 1.77 & 1.68 & 1.66 & 1.64 & 1.81 & 1.63 &1.84 \\ 
   &  & (12.85) & (12.15) & (7.56) & (6.65) & (10.08) & (9.50) & (10.18) & (8.55) & (9.07) & (9.32) & (8.80) & (6.84) & (7.65) & (6.96) & (9.17) & (7.49) & (7.55) & (7.98) & (7.67) & (6.53)& \\ 
  TSX & & & 2.21 & 1.79 & 1.72 & 1.92 & 1.94 & 1.80 & 1.90 & 1.90 & 1.88 & 1.82 & 1.85 & 1.65 & 1.57 & 1.74 & 1.64 & 1.67 & 1.65 & 1.74 & 1.64 &1.80 \\ 
   & & & (15.55) & (7.31) & (8.76) & (10.22) & (10.89) & (9.91) & (10.36) & (9.66) & (9.86) & (9.84) & (8.59) & (9.48) & (7.52) & (9.61) & (8.97) & (10.34) & (8.92) & (9.22) & (9.01) &\\ 
  MERVAL & && & 2.11 & 1.61 & 1.72 & 1.75 & 1.68 & 1.72 & 1.72 & 1.73 & 1.66 & 1.70 & 1.55 & 1.52 & 1.63 & 1.56 & 1.59 & 1.54 & 1.65 & 1.59&1.67\\ 
   & && & (10.84) & (5.87) & (6.93) & (7.19) & (8.49) & (6.23) & (6.52) & (5.94) & (7.03) & (6.17) & (5.35) & (6.39) & (8.00) & (6.11) & (6.64) & (5.69) & (6.24) & (7.17) &\\ 
  BOVESPA & &&& & 1.71 & 1.64 & 1.64 & 1.55 & 1.61 & 1.61 & 1.60 & 1.54 & 1.56 & 1.45 & 1.37 & 1.53 & 1.46 & 1.47 & 1.44 & 1.57 & 1.42 &1.57\\ 
   & &&& & (6.26) & (7.37) & (6.26) & (7.35) & (6.44) & (6.40) & (6.84) & (6.28) & (4.57) & (5.87) & (4.65) & (6.05) & (7.19) & (8.89) & (5.70) & (7.32) & (6.52)& \\ 
  IPC & &&&& & 2.00 & 1.85 & 1.68 & 1.76 & 1.81 & 1.80 & 1.73 & 1.74 & 1.58 & 1.51 & 1.67 & 1.62 & 1.65 & 1.56 & 1.71 & 1.56 &1.73 \\ 
   & &&&& & (10.03) & (8.67) & (8.36) & (10.28) & (7.97) & (8.85) & (9.18) & (6.47) & (6.91) & (6.44) & (8.43) & (7.21) & (10.15) & (7.26) & (7.97) & (7.04)& \\ 
  AEX & &&&&& & 2.22 & 1.93 & 2.08 & 2.06 & 2.12 & 2.04 & 2.02 & 1.77 & 1.64 & 1.85 & 1.66 & 1.74 & 1.75 & 1.82 & 1.70 &1.87\\ 
   & &&&&& & (11.65) & (8.50) & (9.93) & (9.46) & (11.01) & (10.86) & (7.90) & (6.29) & (6.72) & (8.26) & (6.67) & (7.23) & (9.02) & (9.09) & (8.70)& \\ 
  ATX & &&&&&& & 2.06 & 1.89 & 1.90 & 1.86 & 1.84 & 1.86 & 1.64 & 1.54 & 1.76 & 1.61 & 1.65 & 1.61 & 1.74 & 1.64 &1.74 \\ 
   & &&&&&& & (12.88) & (9.75) & (9.27) & (9.43) & (8.52) & (8.89) & (7.85) & (7.97) & (10.79) & (6.11) & (10.15) & (8.59) & (7.90) & (9.28)& \\ 
  FTSE & &&&&&&& & 2.12 & 2.00 & 2.01 & 1.99 & 1.95 & 1.70 & 1.57 & 1.77 & 1.61 & 1.71 & 1.65 & 1.78 & 1.64 &1.81 \\ 
   & &&&&&&& & (8.07) & (7.66) & (9.22) & (9.38) & (7.42) & (6.66) & (5.86) & (7.97) & (6.83) & (7.96) & (6.89) & (8.13) & (7.40) &\\ 
  DAX & &&&&&&&& & 2.10 & 2.03 & 1.95 & 1.99 & 1.69 & 1.62 & 1.78 & 1.61 & 1.67 & 1.65 & 1.77 & 1.61 &1.82\\ 
   & &&&&&&&& & (9.71) & (9.23) & (9.05) & (7.58) & (6.60) & (6.48) & (8.04) & (5.42) & (6.13) & (6.86) & (7.48) & (7.23) &\\ 
  CAC & &&&&&&&&& & 2.08 & 1.99 & 2.02 & 1.70 & 1.59 & 1.78 & 1.59 & 1.70 & 1.64 & 1.76 & 1.64 &1.82\\ 
   & &&&&&&&&& & (10.31) & (9.62) & (7.53) & (6.25) & (6.03) & (8.38) & (6.22) & (7.58) & (6.87) & (7.60) & (6.68)& \\ 
  SMI & &&&&&&&&&& & 2.02 & 1.89 & 1.67 & 1.59 & 1.74 & 1.56 & 1.65 & 1.62 & 1.78 & 1.63 &1.77 \\ 
   & &&&&&&&&&& & (9.82) & (7.22) & (5.46) & (6.85) & (8.58) & (6.31) & (6.66) & (6.97) & (7.92) & (7.91) &\\ 
  MIB & &&&&&&&&&&& & 2.02 & 1.64 & 1.51 & 1.72 & 1.58 & 1.64 & 1.58 & 1.67 & 1.59 &1.77\\ 
   & &&&&&&&&&&& & (7.60) & (5.46) & (5.49) & (8.43) & (6.31) & (5.90) & (6.38) & (7.13) & (6.95) &\\ 
  HGSG & &&&&&&&&&&&& & 2.01 & 1.72 & 1.94 & 1.78 & 1.76 & 1.78 & 1.95 & 1.77 &1.71\\ 
   & &&&&&&&&&&&& & (9.41) & (7.31) & (12.40) & (8.06) & (9.48) & (7.78) & (8.32) & (9.28) &\\ 
  NIKKEI & &&&&&&&&&&&&& & 1.88 & 1.74 & 1.60 & 1.61 & 1.63 & 1.88 & 1.72 &1.61\\ 
   & &&&&&&&&&&&&& & (7.63) & (8.29) & (7.38) & (8.37) & (6.21) & (7.27) & (7.38)& \\ 
  STRTIM & &&&&&&&&&&&&&& & 2.15 & 1.74 & 1.86 & 1.84 & 2.01 & 1.82 &1.77\\ 
   & &&&&&&&&&&&&&& & (13.08) & (7.35) & (10.85) & (8.39) & (9.72) & (9.31)& \\ 
  SSEC & &&&&&&&&&&&&&&& & 2.19 & 1.68 & 1.73 & 1.79 & 1.72&1.64 \\ 
   & &&&&&&&&&&&&&&& & (9.10) & (6.96) & (6.63) & (7.15) & (7.51)& \\ 
  BSE & &&&&&&&&&&&&&&&& & 2.14 & 1.69 & 1.83 & 1.72 &1.68 \\ 
   & &&&&&&&&&&&&&&&& & (10.52) & (7.99) & (9.10) & (8.87)& \\ 
  KLSE & &&&&&&&&&&&&&&&&& & 2.10 & 1.86 & 1.72&1.66 \\ 
   & &&&&&&&&&&&&&&&&& & (8.35) & (7.25) & (7.20)& \\ 
  KOSPI & &&&&&&&&&&&&&&&&&& & 2.33 & 1.90 &1.79 \\ 
   & &&&&&&&&&&&&&&&&&& & (11.10) & (7.76)& \\ 
  ALLORD & &&&&&&&&&&&&&&&&&&& & 1.97 &1.66 \\ 
   & &&&&&&&&&&&&&&&&&&& & (9.45)& \\ 
   \hline
\end{tabular}
\end{adjustbox}
\end{table}
\end{landscape}

\begin{landscape}
\begin{table}[ht]
\centering
\caption{Matrix of linear contributions $\sqrt{1+|\rho|}$}\label{tab:linearComponent}
\begin{adjustbox}{max width=1.3\textwidth}
\begin{tabular}{rcccccccccccccccccccccc}
  \hline
   & \rot{90}{SP} & \rot{90}{NASDAQ} & \rot{90}{TSX} & \rot{90}{MERVAL} & \rot{90}{BOVESPA} & \rot{90}{IPC} & \rot{90}{AEX} & \rot{90}{ATX} & \rot{90}{FTSE} & \rot{90}{DAX} & \rot{90}{CAC} & \rot{90}{SMI} & \rot{90}{MIB} & \rot{90}{HGSG} & \rot{90}{NIKKEI} & \rot{90}{STRTIM} & \rot{90}{SSEC} & \rot{90}{BSE} & \rot{90}{KLSE} & \rot{90}{KOSPI} & \rot{90}{ALLORD} &\rot{90}{Average} \\ 
  \hline
  SP & 1.41 & 1.39 & 1.31 & 1.20 & 1.24 & 1.27 & 1.25 & 1.19 & 1.24 & 1.25 & 1.25 & 1.22 & 1.23 & 1.10 & 1.08 & 1.10 & 1.03 & 1.09 & 1.05 & 1.09 & 1.06 & 1.18\\ 
   & (0.00) & (0.18) & (0.50) & (1.01) & (0.76) & (0.62) & (0.59) & (0.88) & (0.61) & (0.67) & (0.64) & (0.71) & (0.63) & (0.74) & (0.71) & (0.61) & (0.80) & (0.73) & (0.76) & (0.65) & (0.84)& \\ 
  NASDAQ &  & 1.41 & 1.29 & 1.18 & 1.23 & 1.26 & 1.23 & 1.17 & 1.22 & 1.24 & 1.23 & 1.20 & 1.21 & 1.10 & 1.09 & 1.11 & 1.04 & 1.09 & 1.05 & 1.10 & 1.07 &1.17 \\ 
   &  & (0.00) & (0.51) & (1.03) & (0.86) & (0.60) & (0.58) & (0.93) & (0.68) & (0.58) & (0.62) & (0.78) & (0.59) & (0.66) & (0.69) & (0.63) & (0.81) & (0.73) & (0.84) & (0.64) & (0.69) &\\ 
  TSX & & & 1.41 & 1.20 & 1.24 & 1.24 & 1.24 & 1.19 & 1.24 & 1.24 & 1.24 & 1.20 & 1.22 & 1.13 & 1.10 & 1.13 & 1.06 & 1.11 & 1.07 & 1.12 & 1.10 &1.18\\ 
   & & & (0.00) & (1.08) & (0.83) & (0.75) & (0.60) & (0.87) & (0.60) & (0.62) & (0.62) & (0.75) & (0.69) & (0.72) & (0.74) & (0.71) & (0.79) & (0.83) & (0.84) & (0.70) & (0.77)& \\ 
  MERVAL & && & 1.41 & 1.22 & 1.19 & 1.16 & 1.15 & 1.16 & 1.16 & 1.16 & 1.13 & 1.15 & 1.10 & 1.07 & 1.09 & 1.04 & 1.08 & 1.05 & 1.08 & 1.06 &1.13\\ 
   & && & (0.00) & (0.96) & (0.93) & (1.07) & (1.04) & (1.07) & (1.05) & (1.03) & (0.98) & (1.04) & (0.83) & (0.89) & (0.86) & (0.94) & (0.88) & (0.86) & (0.78) & (0.82) &\\ 
  BOVESPA & &&& & 1.41 & 1.25 & 1.18 & 1.15 & 1.19 & 1.19 & 1.19 & 1.16 & 1.17 & 1.11 & 1.08 & 1.10 & 1.06 & 1.10 & 1.07 & 1.10 & 1.07 &1.15\\ 
   & &&& & (0.00) & (0.77) & (0.76) & (0.91) & (0.75) & (0.84) & (0.81) & (0.84) & (0.80) & (0.75) & (0.80) & (0.81) & (0.78) & (1.03) & (0.89) & (0.76) & (0.80) &\\ 
  IPC & &&&& & 1.41 & 1.21 & 1.18 & 1.21 & 1.21 & 1.22 & 1.19 & 1.20 & 1.13 & 1.09 & 1.13 & 1.05 & 1.12 & 1.09 & 1.12 & 1.09 &1.17\\ 
   & &&&& & (0.00) & (0.67) & (0.82) & (0.71) & (0.73) & (0.70) & (0.86) & (0.79) & (0.75) & (0.64) & (0.77) & (0.83) & (0.96) & (0.76) & (0.72) & (0.72) &\\ 
  AEX & &&&&& & 1.41 & 1.29 & 1.36 & 1.37 & 1.39 & 1.34 & 1.35 & 1.17 & 1.14 & 1.18 & 1.06 & 1.16 & 1.11 & 1.15 & 1.14 &1.22\\ 
   & &&&&& & (0.00) & (0.79) & (0.34) & (0.25) & (0.19) & (0.42) & (0.37) & (0.68) & (0.60) & (0.69) & (0.81) & (0.93) & (0.88) & (0.62) & (0.76)& \\ 
  ATX & &&&&&& & 1.41 & 1.28 & 1.29 & 1.30 & 1.26 & 1.29 & 1.17 & 1.15 & 1.18 & 1.07 & 1.15 & 1.12 & 1.14 & 1.15  &1.19\\ 
   & &&&&&& & (0.00) & (0.78) & (0.87) & (0.78) & (0.72) & (0.77) & (0.77) & (0.78) & (0.80) & (0.87) & (1.04) & (0.95) & (0.82) & (0.85) & \\ 
  FTSE & &&&&&&& & 1.41 & 1.34 & 1.36 & 1.33 & 1.33 & 1.17 & 1.14 & 1.18 & 1.06 & 1.16 & 1.10 & 1.14 & 1.14 &1.22 \\ 
   & &&&&&&& & (0.00) & (0.38) & (0.33) & (0.42) & (0.49) & (0.62) & (0.63) & (0.70) & (0.86) & (0.90) & (0.87) & (0.66) & (0.79) &\\ 
  DAX & &&&&&&&& & 1.41 & 1.38 & 1.34 & 1.35 & 1.16 & 1.13 & 1.17 & 1.05 & 1.15 & 1.10 & 1.14 & 1.12&1.22 \\ 
   & &&&&&&&& & (0.00) & (0.19) & (0.45) & (0.34) & (0.65) & (0.70) & (0.67) & (0.82) & (0.92) & (0.93) & (0.67) & (0.82)& \\ 
  CAC & &&&&&&&&& & 1.41 & 1.35 & 1.37 & 1.16 & 1.14 & 1.18 & 1.05 & 1.15 & 1.10 & 1.14 & 1.13&1.22 \\ 
   & &&&&&&&&& & (0.00) & (0.48) & (0.28) & (0.62) & (0.73) & (0.69) & (0.76) & (0.91) & (0.88) & (0.64) & (0.80)& \\ 
  SMI & &&&&&&&&&& & 1.41 & 1.31 & 1.16 & 1.14 & 1.17 & 1.04 & 1.15 & 1.10 & 1.14 & 1.13 &1.20\\ 
   & &&&&&&&&&& & (0.00) & (0.59) & (0.69) & (0.65) & (0.73) & (0.94) & (0.84) & (0.87) & (0.68) & (0.76)& \\ 
  MIB & &&&&&&&&&&& & 1.41 & 1.15 & 1.11 & 1.15 & 1.05 & 1.14 & 1.09 & 1.12 & 1.11 &1.21\\ 
   & &&&&&&&&&&& & (0.00) & (0.67) & (0.71) & (0.65) & (0.82) & (0.91) & (0.94) & (0.68) & (0.85) &\\ 
  HGSG & &&&&&&&&&&&& & 1.41 & 1.24 & 1.28 & 1.18 & 1.20 & 1.19 & 1.26 & 1.24 &1.17\\ 
   & &&&&&&&&&&&& & (0.00) & (0.61) & (0.66) & (0.93) & (0.87) & (0.88) & (0.60) & (0.72) &\\ 
  NIKKEI & &&&&&&&&&&&&& & 1.41 & 1.23 & 1.10 & 1.15 & 1.17 & 1.25 & 1.24 &1.14\\ 
   & &&&&&&&&&&&&& & (0.00) & (0.56) & (0.85) & (0.79) & (0.86) & (0.67) & (0.69)& \\ 
  STRTIM & &&&&&&&&&&&&&& & 1.41 & 1.12 & 1.20 & 1.21 & 1.24 & 1.22 &1.17\\ 
   & &&&&&&&&&&&&&& & (0.00) & (1.08) & (0.96) & (0.87) & (0.65) & (0.72)& \\ 
  SSEC & &&&&&&&&&&&&&&& & 1.41 & 1.08 & 1.09 & 1.11 & 1.10& 1.07 \\ 
   & &&&&&&&&&&&&&&& & (0.00) & (0.93) & (0.93) & (1.07) & (0.85) &\\ 
  BSE & &&&&&&&&&&&&&&&& & 1.41 & 1.14 & 1.18 & 1.15 &1.14 \\ 
   & &&&&&&&&&&&&&&&& & (0.00) & (1.02) & (0.78) & (0.81)& \\ 
  KLSE & &&&&&&&&&&&&&&&&& & 1.41 & 1.18 & 1.17 &1.11 \\ 
   & &&&&&&&&&&&&&&&&& & (0.00) & (0.93) & (0.95) &\\ 
  KOSPI & &&&&&&&&&&&&&&&&&& & 1.41 & 1.22&1.15 \\ 
   & &&&&&&&&&&&&&&&&&& & (0.00) & (0.73)& \\ 
  ALLORD & &&&&&&&&&&&&&&&&&&& & 1.41&1.14 \\ 
   & &&&&&&&&&&&&&&&&&&& & (0.00)& \\ 
  \hline
  \end{tabular}
\end{adjustbox}
\end{table}
\end{landscape}

\begin{landscape}
\begin{table}[ht]
\centering
\caption{Matrix of nonlinear contributions $s_g(0.975,0.75)s(0.975,0.75,\alpha)$}\label{tab:nonlinearComponent}
\begin{adjustbox}{max width=1.3\textwidth}
\begin{tabular}{rcccccccccccccccccccccc}
  \hline
   & \rot{90}{SP} & \rot{90}{NASDAQ} & \rot{90}{TSX} & \rot{90}{MERVAL} & \rot{90}{BOVESPA} & \rot{90}{IPC} & \rot{90}{AEX} & \rot{90}{ATX} & \rot{90}{FTSE} & \rot{90}{DAX} & \rot{90}{CAC} & \rot{90}{SMI} & \rot{90}{MIB} & \rot{90}{HGSG} & \rot{90}{NIKKEI} & \rot{90}{STRTIM} & \rot{90}{SSEC} & \rot{90}{BSE} & \rot{90}{KLSE} & \rot{90}{KOSPI} & \rot{90}{ALLORD} &\rot{90}{Average}\\ 
  \hline
  SP & 1.64 & 1.68 & 1.62 & 1.56 & 1.43 & 1.60 & 1.61 & 1.54 & 1.58 & 1.59 & 1.58 & 1.57 & 1.57 & 1.52 & 1.47 & 1.61 & 1.60 & 1.57 & 1.52 & 1.65 & 1.52 &1.57 \\ 
   & (9.68) & (9.27) & (10.11) & (7.04) & (6.04) & (8.07) & (8.35) & (8.37) & (8.03) & (7.88) & (8.23) & (7.88) & (7.13) & (7.31) & (7.04) & (8.70) & (7.44) & (7.99) & (7.83) & (8.01) & (8.54) &\\ 
  NASDAQ &  & 1.67 & 1.63 & 1.55 & 1.44 & 1.56 & 1.63 & 1.50 & 1.59 & 1.57 & 1.56 & 1.56 & 1.58 & 1.55 & 1.49 & 1.59 & 1.62 & 1.52 & 1.56 & 1.65 & 1.53 &1.57 \\ 
   &  & (9.09) & (9.24) & (6.08) & (5.15) & (7.86) & (7.76) & (8.33) & (6.91) & (7.18) & (7.54) & (7.16) & (5.57) & (6.82) & (6.49) & (8.18) & (7.06) & (6.53) & (7.44) & (6.90) & (5.82) &\\ 
  TSX & & & 1.56 & 1.49 & 1.39 & 1.54 & 1.57 & 1.51 & 1.53 & 1.53 & 1.52 & 1.51 & 1.52 & 1.46 & 1.42 & 1.54 & 1.55 & 1.50 & 1.54 & 1.56 & 1.49 &1.52\\ 
   & & & (11.00) & (5.97) & (6.63) & (7.88) & (8.67) & (7.96) & (8.14) & (7.62) & (7.70) & (7.84) & (6.83) & (8.14) & (6.60) & (8.32) & (8.28) & (8.87) & (8.25) & (8.15) & (7.86) &\\ 
  MERVAL & && & 1.49 & 1.32 & 1.45 & 1.51 & 1.46 & 1.49 & 1.48 & 1.50 & 1.47 & 1.47 & 1.42 & 1.42 & 1.49 & 1.50 & 1.47 & 1.47 & 1.53 & 1.49 &1.48 \\ 
   & && & (7.67) & (4.69) & (5.58) & (5.98) & (6.88) & (5.04) & (5.40) & (4.81) & (5.87) & (5.10) & (4.64) & (5.62) & (7.05) & (5.79) & (5.81) & (5.28) & (5.63) & (6.42)& \\ 
  BOVESPA & &&& & 1.21 & 1.31 & 1.39 & 1.34 & 1.35 & 1.36 & 1.35 & 1.33 & 1.34 & 1.30 & 1.27 & 1.38 & 1.38 & 1.35 & 1.35 & 1.43 & 1.33 &1.36\\ 
   & &&& & (4.43) & (5.56) & (5.13) & (5.88) & (5.15) & (5.15) & (5.46) & (4.96) & (3.73) & (4.98) & (4.08) & (5.25) & (6.38) & (7.47) & (5.08) & (6.50) & (5.65)& \\ 
  IPC & &&&& & 1.41 & 1.53 & 1.43 & 1.46 & 1.49 & 1.48 & 1.46 & 1.45 & 1.40 & 1.38 & 1.48 & 1.54 & 1.47 & 1.43 & 1.53 & 1.43 &1.47\\ 
   & &&&& & (7.09) & (6.91) & (6.71) & (8.14) & (6.21) & (7.00) & (7.28) & (5.12) & (5.76) & (5.59) & (7.10) & (6.54) & (8.43) & (6.40) & (6.95) & (6.10) &\\ 
  AEX & &&&&& & 1.57 & 1.50 & 1.53 & 1.51 & 1.53 & 1.52 & 1.49 & 1.51 & 1.43 & 1.56 & 1.56 & 1.50 & 1.57 & 1.58 & 1.49 &1.53\\ 
   & &&&&& & (8.24) & (6.41) & (7.19) & (6.89) & (7.87) & (7.90) & (5.76) & (5.32) & (5.81) & (6.90) & (6.01) & (5.92) & (7.87) & (7.79) & (7.41) &\\ 
  ATX & &&&&&& & 1.46 & 1.48 & 1.48 & 1.43 & 1.45 & 1.44 & 1.40 & 1.34 & 1.49 & 1.51 & 1.43 & 1.44 & 1.53 & 1.43 &1.46\\ 
   & &&&&&& & (9.11) & (7.36) & (7.04) & (6.99) & (6.57) & (6.63) & (6.26) & (6.50) & (8.70) & (5.43) & (8.08) & (7.15) & (6.67) & (7.54) &\\ 
  FTSE & &&&&&&& & 1.50 & 1.49 & 1.47 & 1.50 & 1.46 & 1.45 & 1.38 & 1.50 & 1.52 & 1.48 & 1.49 & 1.56 & 1.43 &1.49 \\ 
   & &&&&&&& & (5.71) & (5.64) & (6.65) & (6.88) & (5.50) & (5.55) & (4.93) & (6.69) & (6.15) & (6.57) & (5.94) & (6.88) & (6.05) &\\ 
  DAX & &&&&&&&& & 1.48 & 1.47 & 1.46 & 1.47 & 1.46 & 1.43 & 1.52 & 1.53 & 1.46 & 1.51 & 1.55 & 1.43 &1.49 \\ 
   & &&&&&&&& & (6.86) & (6.64) & (6.67) & (5.56) & (5.56) & (5.53) & (6.65) & (5.03) & (4.93) & (5.87) & (6.36) & (6.13) &\\ 
  CAC & &&&&&&&&& & 1.47 & 1.48 & 1.48 & 1.46 & 1.40 & 1.51 & 1.51 & 1.48 & 1.49 & 1.54 & 1.45 &1.48 \\ 
   & &&&&&&&&& & (7.29) & (6.92) & (5.46) & (5.23) & (5.00) & (6.90) & (5.62) & (6.08) & (5.84) & (6.51) & (5.52) \\ 
  SMI & &&&&&&&&&& & 1.43 & 1.44 & 1.45 & 1.39 & 1.49 & 1.50 & 1.43 & 1.47 & 1.56 & 1.44 &1.47 \\ 
   & &&&&&&&&&& & (6.94) & (5.31) & (4.67) & (5.85) & (7.08) & (5.82) & (5.47) & (6.15) & (6.93) & (6.70) &\\ 
  MIB & &&&&&&&&&&& & 1.43 & 1.44 & 1.36 & 1.49 & 1.51 & 1.43 & 1.45 & 1.49 & 1.44 &1.46\\ 
   & &&&&&&&&&&& & (5.38) & (4.58) & (4.57) & (7.02) & (5.76) & (4.68) & (5.50) & (6.16) & (5.75)& \\ 
  HGSG & &&&&&&&&&&&& & 1.42 & 1.39 & 1.51 & 1.51 & 1.46 & 1.49 & 1.55 & 1.43 &1.46 \\ 
   & &&&&&&&&&&&& & (6.65) & (5.71) & (9.34) & (6.36) & (7.43) & (6.26) & (6.52) & (7.25) &\\ 
  NIKKEI & &&&&&&&&&&&&& & 1.33 & 1.42 & 1.46 & 1.40 & 1.40 & 1.50 & 1.39 &1.41\\ 
   & &&&&&&&&&&&&& & (5.40) & (6.73) & (6.46) & (6.92) & (5.06) & (5.92) & (5.76)& \\ 
  STRTIM & &&&&&&&&&&&&&& & 1.52 & 1.56 & 1.55 & 1.52 & 1.62 & 1.49 &1.52\\ 
   & &&&&&&&&&&&&&& & (9.25) & (6.07) & (8.49) & (6.74) & (7.70) & (7.45) &\\ 
  SSEC & &&&&&&&&&&&&&&& & 1.55 & 1.55 & 1.58 & 1.61 & 1.56 &1.53 \\ 
   & &&&&&&&&&&&&&&& & (6.43) & (6.01) & (5.71) & (6.08) & (6.53) &\\ 
  BSE & &&&&&&&&&&&&&&&& & 1.51 & 1.49 & 1.56 & 1.49 &1.48 \\ 
   & &&&&&&&&&&&&&&&& & (7.44) & (6.49) & (7.37) & (7.21)& \\ 
  KLSE & &&&&&&&&&&&&&&&&& & 1.49 & 1.58 & 1.47 &1.49\\ 
   & &&&&&&&&&&&&&&&&& & (5.90) & (5.79) & (5.77) &\\ 
  KOSPI & &&&&&&&&&&&&&&&&&& & 1.65 & 1.56 &1.56 \\ 
   & &&&&&&&&&&&&&&&&&& & (7.85) & (6.36) &\\ 
  ALLORD & &&&&&&&&&&&&&&&&&&& & 1.39 &1.46\\ 
   & &&&&&&&&&&&&&&&&&&& & (6.68)& \\ 
  \hline
    \end{tabular}
\end{adjustbox}
\end{table}
\end{landscape}

\begin{landscape}
\begin{table}[ht]
\centering
\caption{Downside exceedance correlation: $\theta^{-}(<0.025)$}\label{tab:dexceedance}
\begin{adjustbox}{max width=1.35\textwidth}
\begin{tabular}{rcccccccccccccccccccccc}
  \hline
   & \rot{90}{SP} & \rot{90}{NASDAQ} & \rot{90}{TSX} & \rot{90}{MERVAL} & \rot{90}{BOVESPA} & \rot{90}{IPC} & \rot{90}{AEX} & \rot{90}{ATX} & \rot{90}{FTSE} & \rot{90}{DAX} & \rot{90}{CAC} & \rot{90}{SMI} & \rot{90}{MIB} & \rot{90}{HGSG} & \rot{90}{NIKKEI} & \rot{90}{STRTIM} & \rot{90}{SSEC} & \rot{90}{BSE} & \rot{90}{KLSE} & \rot{90}{KOSPI} & \rot{90}{ALLORD} &\rot{90}{Average} \\ 
  \hline
SP & 1.00 & 0.76 & 0.72 & 0.47 & 0.71 & 0.53 & 0.48 & 0.55 & 0.49 & 0.49 & 0.46 & 0.53 & 0.42 & -0.27 & 0.07 & -0.27 & -0.37 & 0.05 & 0.19 & -0.07 & 0.70 & 0.33 \\ 
  NASDAQ &  & 1.00 & 0.56 & 0.37 & 0.54 & 0.58 & 0.46 & 0.66 & 0.32 & 0.40 & 0.33 & 0.59 & 0.41 & -0.13 & 0.10 & 0.09 & -0.36 & 0.57 & 0.31 & -0.02 & 0.78 &0.36\\ 
  TSX & & & 1.00 & 0.46 & 0.71 & 0.45 & 0.46 & 0.77 & 0.62 & 0.60 & 0.56 & 0.58 & 0.65 & 0.11 & 0.30 & 0.31 & -0.25 & 0.29 & 0.26 & -0.11 & 0.76 & 0.44\\ 
  MERVAL & && & 1.00 & 0.77 & 0.22 & 0.36 & 0.35 & 0.25 & 0.34 & 0.42 & 0.37 & 0.49 & -0.12 & 0.00 & 0.13 & -0.02 & 0.12 & 0.06 & -0.14 & 0.49 &0.27 \\ 
  BOVESPA & &&& & 1.00 & 0.45 & 0.37 & 0.64 & 0.46 & 0.53 & 0.49 & 0.49 & 0.58 & -0.01 & -0.10 & 0.20 & -0.28 & 0.24 & 0.26 & 0.01 & 0.59 & 0.38\\ 
  IPC & &&&& & 1.00 & 0.47 & 0.36 & 0.34 & 0.36 & 0.31 & 0.48 & 0.31 & 0.30 & -0.05 & 0.43 & 0.33 & 0.12 & 0.15 & -0.01 & 0.18 &0.32\\ 
  AEX & &&&&& & 1.00 & 0.60 & 0.79 & 0.79 & 0.82 & 0.75 & 0.62 & 0.11 & 0.31 & 0.33 & 0.09 & 0.29 & -0.06 & -0.04 & 0.44 & 0.42\\ 
  ATX & &&&&&& & 1.00 & 0.69 & 0.70 & 0.73 & 0.57 & 0.62 & 0.21 & 0.36 & 0.28 & -0.00 & 0.34 & 0.27 & -0.04 & 0.78 &0.47 \\ 
  FTSE & &&&&&&& & 1.00 & 0.72 & 0.77 & 0.76 & 0.62 & 0.11 & 0.29 & 0.29 & 0.13 & 0.28 & -0.12 & -0.07 & 0.50 &0.41\\ 
  DAX & &&&&&&&& & 1.00 & 0.88 & 0.72 & 0.81 & 0.14 & 0.20 & 0.48 & 0.09 & 0.45 & -0.01 & 0.11 & 0.52 &0.47 \\ 
  CAC & &&&&&&&&& & 1.00 & 0.71 & 0.80 & -0.04 & 0.21 & 0.34 & 0.03 & 0.19 & -0.05 & -0.08 & 0.51 &0.42\\ 
  SMI & &&&&&&&&&& & 1.00 & 0.57 & 0.07 & 0.11 & 0.45 & -0.24 & 0.42 & 0.13 & -0.07 & 0.64 &0.43\\ 
  MIB & &&&&&&&&&&& & 1.00 & -0.06 & 0.21 & 0.33 & 0.17 & 0.39 & -0.03 & -0.10 & 0.51 &0.41 \\ 
  HGSG & &&&&&&&&&&&& & 1.00 & 0.52 & 0.71 & 0.10 & 0.24 & 0.25 & 0.52 & 0.30 &0.15\\ 
  NIKKEI & &&&&&&&&&&&&& & 1.00 & 0.38 & -0.24 & 0.29 & 0.27 & 0.57 & 0.44 &0.21\\ 
  STRTIM & &&&&&&&&&&&&&& & 1.00 & 0.17 & 0.42 & 0.49 & 0.53 & 0.55 &0.33\\ 
  SSEC & &&&&&&&&&&&&&&& & 1.00 & -0.11 & -0.22 & -0.07 & -0.20& -0.06\\ 
  BSE & &&&&&&&&&&&&&&&& & 1.00 & 0.20 & 0.15 & 0.43 &0.27\\ 
  KLSE & &&&&&&&&&&&&&&&&& & 1.00 & 0.40 & 0.49& 0.16 \\ 
  KOSPI & &&&&&&&&&&&&&&&&&& & 1.00 & 0.20 &0.08 \\ 
  ALLORD & &&&&&&&&&&&&&&&&&&& & 1.00&0.48 \\ 
 \hline
    \end{tabular}
\end{adjustbox}
\end{table}
\end{landscape}

\begin{landscape}
\begin{table}[ht]
\centering
\caption{Upside exceedance correlation: $\theta^{+}(>0.975)$}\label{tab:uexceedance}
\begin{adjustbox}{max width=1.35\textwidth}
\begin{tabular}{rcccccccccccccccccccccc}
  \hline
   & \rot{90}{SP} & \rot{90}{NASDAQ} & \rot{90}{TSX} & \rot{90}{MERVAL} & \rot{90}{BOVESPA} & \rot{90}{IPC} & \rot{90}{AEX} & \rot{90}{ATX} & \rot{90}{FTSE} & \rot{90}{DAX} & \rot{90}{CAC} & \rot{90}{SMI} & \rot{90}{MIB} & \rot{90}{HGSG} & \rot{90}{NIKKEI} & \rot{90}{STRTIM} & \rot{90}{SSEC} & \rot{90}{BSE} & \rot{90}{KLSE} & \rot{90}{KOSPI} & \rot{90}{ALLORD} &\rot{90}{Average} \\ 
  \hline
  SP & 1.00 & 0.56 & 0.77 & 0.53 & 0.73 & 0.63 & 0.61 & 0.47 & 0.52 & 0.70 & 0.60 & 0.57 & 0.66 & 0.41 & 0.56 & 0.29 & -0.06 & 0.05 & -0.12 & 0.02 & 0.45 &0.45 \\ 
  NASDAQ &  & 1.00 & 0.47 & 0.37 & 0.44 & 0.51 & 0.45 & 0.36 & 0.35 & 0.38 & 0.33 & 0.45 & 0.61 & 0.36 & 0.78 & 0.26 & -0.21 & 0.40 & -0.27 & 0.23 & 0.32 &0.36\\ 
  TSX & & & 1.00 & 0.64 & 0.78 & 0.41 & 0.60 & 0.52 & 0.44 & 0.68 & 0.52 & 0.50 & 0.45 & 0.35 & 0.58 & 0.39 & 0.07 & 0.17 & 0.15 & 0.19 & 0.35 &0.45\\ 
  MERVAL & && & 1.00 & 0.49 & 0.22 & 0.54 & 0.60 & 0.64 & 0.56 & 0.69 & 0.67 & 0.51 & 0.22 & 0.47 & 0.33 & 0.43 & 0.13 & -0.05 & -0.12 & 0.83& 0.44 \\ 
  BOVESPA & &&& & 1.00 & 0.60 & 0.66 & 0.49 & 0.50 & 0.77 & 0.58 & 0.76 & 0.59 & 0.72 & 0.53 & 0.54 & 0.14 & 0.11 & 0.04 & 0.14 & 0.62 & 0.51 \\ 
  IPC & &&&& & 1.00 & 0.57 & 0.50 & 0.42 & 0.65 & 0.60 & 0.61 & 0.56 & 0.47 & 0.54 & 0.20 & -0.10 & 0.58 & -0.02 & 0.11 & 0.15 & 0.41\\ 
  AEX & &&&&& & 1.00 & 0.72 & 0.83 & 0.78 & 0.84 & 0.71 & 0.70 & 0.69 & 0.42 & 0.50 & 0.09 & 0.31 & 0.33 & 0.22 & 0.44 & 0.55\\ 
  ATX & &&&&&& & 1.00 & 0.78 & 0.57 & 0.79 & 0.73 & 0.75 & 0.34 & 0.40 & 0.48 & 0.30 & 0.25 & 0.48 & 0.12 & 0.56 &0.51\\ 
  FTSE & &&&&&&& & 1.00 & 0.75 & 0.82 & 0.71 & 0.72 & 0.63 & 0.31 & 0.61 & 0.05 & 0.62 & 0.50 & 0.18 & 0.41 & 0.54\\ 
  DAX & &&&&&&&& & 1.00 & 0.77 & 0.72 & 0.59 & 0.63 & 0.66 & 0.53 & -0.05 & 0.60 & -0.17 & 0.29 & 0.53  & 0.55\\ 
  CAC & &&&&&&&&& & 1.00 & 0.72 & 0.80 & 0.69 & 0.50 & 0.56 & 0.33 & 0.40 & 0.13 & 0.14 & 0.21 &0.55\\ 
  SMI & &&&&&&&&&& & 1.00 & 0.64 & 0.56 & 0.54 & 0.54 & 0.02 & 0.56 & -0.26 & 0.29 & 0.42 &0.52\\ 
  MIB & &&&&&&&&&&& & 1.00 & 0.55 & 0.43 & 0.55 & -0.03 & 0.28 & 0.16 & 0.19 & 0.37 &0.50 \\ 
  HGSG & &&&&&&&&&&&& & 1.00 & 0.33 & 0.53 & -0.03 & 0.54 & -0.02 & 0.31 & 0.47 &0.44 \\ 
  NIKKEI & &&&&&&&&&&&&& & 1.00 & 0.36 & -0.33 & 0.59 & -0.33 & 0.47 & 0.44  & 0.41\\ 
  STRTIM & &&&&&&&&&&&&&& & 1.00 & 0.02 & 0.51 & 0.37 & 0.50 & 0.58 & 0.43 \\ 
  SSEC & &&&&&&&&&&&&&&& & 1.00 & 0.22 & 0.33 & -0.32 & 0.41 & 0.06 \\ 
  BSE & &&&&&&&&&&&&&&&& & 1.00 & 0.12 & 0.39 & 0.53 & 0.37\\ 
  KLSE & &&&&&&&&&&&&&&&&& & 1.00 & 0.33 & 0.35 &0.10\\ 
  KOSPI & &&&&&&&&&&&&&&&&&& & 1.00 & 0.41 &0.20 \\ 
  ALLORD & &&&&&&&&&&&&&&&&&&& & 1.00 &0.44\\ 
  \hline
    \end{tabular}
\end{adjustbox}
\end{table}
\end{landscape}

\begin{landscape}
\begin{table}[ht]
\centering
\caption{$t$-copula tail dependence: $\tau_p$}\label{tab:param}
\begin{adjustbox}{max width=1.35\textwidth}
\begin{tabular}{rcccccccccccccccccccccc}
  \hline
   & \rot{90}{SP} & \rot{90}{NASDAQ} & \rot{90}{TSX} & \rot{90}{MERVAL} & \rot{90}{BOVESPA} & \rot{90}{IPC} & \rot{90}{AEX} & \rot{90}{ATX} & \rot{90}{FTSE} & \rot{90}{DAX} & \rot{90}{CAC} & \rot{90}{SMI} & \rot{90}{MIB} & \rot{90}{HGSG} & \rot{90}{NIKKEI} & \rot{90}{STRTIM} & \rot{90}{SSEC} & \rot{90}{BSE} & \rot{90}{KLSE} & \rot{90}{KOSPI} & \rot{90}{ALLORD} &\rot{90}{Average} \\ 
  \hline
SP & 1.00 & 0.69 & 0.44 & 0.18 & 0.28 & 0.32 & 0.38 & 0.22 & 0.36 & 0.38 & 0.37 & 0.30 & 0.29 & 0.13 & 0.10 & 0.16 & 0.01 & 0.12 & 0.08 & 0.12 & 0.10 & 0.25\\
  NASDAQ &  & 1.00 & 0.40 & 0.17 & 0.25 & 0.30 & 0.32 & 0.14 & 0.30 & 0.33 & 0.32 & 0.26 & 0.22 & 0.12 & 0.09 & 0.15 & 0.00 & 0.13 & 0.10 & 0.16 & 0.07  & 0.23\\
  TSX & & & 1.00 & 0.16 & 0.26 & 0.29 & 0.29 & 0.24 & 0.30 & 0.26 & 0.29 & 0.23 & 0.23 & 0.17 & 0.12 & 0.20 & 0.03 & 0.17 & 0.12 & 0.16 & 0.16 & 0.23 \\
  MERVAL & && & 1.00 & 0.21 & 0.11 & 0.12 & 0.07 & 0.11 & 0.12 & 0.11 & 0.08 & 0.11 & 0.04 & 0.02 & 0.06 & 0.00 & 0.03 & 0.02 & 0.03 & 0.02 & 0.09 \\
  BOVESPA & &&& & 1.00 & 0.29 & 0.18 & 0.12 & 0.17 & 0.17 & 0.17 & 0.15 & 0.11 & 0.10 & 0.06 & 0.12 & 0.01 & 0.11 & 0.07 & 0.11 & 0.08 & 0.15 \\
  IPC & &&&& & 1.00 & 0.19 & 0.12 & 0.23 & 0.19 & 0.20 & 0.15 & 0.11 & 0.14 & 0.09 & 0.16 & 0.02 & 0.14 & 0.12 & 0.15 & 0.10  & 0.17\\
  AEX & &&&&& & 1.00 & 0.38 & 0.58 & 0.66 & 0.70 & 0.55 & 0.46 & 0.19 & 0.14 & 0.22 & 0.03 & 0.14 & 0.10 & 0.15 & 0.14 & 0.30\\
  ATX & &&&&&& & 1.00 & 0.35 & 0.36 & 0.39 & 0.28 & 0.35 & 0.18 & 0.09 & 0.18 & 0.05 & 0.14 & 0.07 & 0.07 & 0.19 & 0.20 \\
  FTSE & &&&&&&& & 1.00 & 0.52 & 0.59 & 0.50 & 0.40 & 0.20 & 0.13 & 0.22 & 0.02 & 0.16 & 0.11 & 0.14 & 0.17 &0.28 \\
  DAX & &&&&&&&& & 1.00 & 0.70 & 0.50 & 0.51 & 0.16 & 0.12 & 0.19 & 0.01 & 0.11 & 0.10 & 0.14 & 0.10 &0.28 \\
  CAC & &&&&&&&&& & 1.00 & 0.54 & 0.55 & 0.16 & 0.12 & 0.19 & 0.01 & 0.13 & 0.09 & 0.13 & 0.13 & 0.30 \\
  SMI & &&&&&&&&&& & 1.00 & 0.37 & 0.16 & 0.16 & 0.20 & 0.02 & 0.12 & 0.10 & 0.14 & 0.15 & 0.25\\
  MIB & &&&&&&&&&&& & 1.00 & 0.14 & 0.09 & 0.14 & 0.01 & 0.07 & 0.05 & 0.06 & 0.11 &0.22 \\
  HGSG & &&&&&&&&&&&& & 1.00 & 0.22 & 0.38 & 0.12 & 0.24 & 0.20 & 0.29 & 0.26 &0.18 \\
  NIKKEI & &&&&&&&&&&&&& & 1.00 & 0.23 & 0.01 & 0.12 & 0.14 & 0.24 & 0.25 &0.13 \\
  STRTIM & &&&&&&&&&&&&&& & 1.00 & 0.06 & 0.25 & 0.26 & 0.29 & 0.26 & 0.20 \\
  SSEC & &&&&&&&&&&&&&&& & 1.00 & 0.03 & 0.04 & 0.02 & 0.04 & 0.03 \\
  BSE & &&&&&&&&&&&&&&&& & 1.00 & 0.16 & 0.19 & 0.17 & 0.14 \\
  KLSE & &&&&&&&&&&&&&&&&& & 1.00 & 0.21 & 0.12 & 0.11 \\
  KOSPI & &&&&&&&&&&&&&&&&&& & 1.00 & 0.18 &0.15 \\
  ALLORD & &&&&&&&&&&&&&&&&&&& & 1.00 &0.14 \\
     \hline
    \end{tabular}
\end{adjustbox}
\end{table}
\end{landscape}

\begin{landscape}
\begin{table}[ht]
\centering
\caption{Non-parametric tail dependence: $\tau_{np}$ }\label{tab:non-param}
\begin{adjustbox}{max width=1.35\textwidth}
\begin{tabular}{rcccccccccccccccccccccc}
  \hline
   & \rot{90}{SP} & \rot{90}{NASDAQ} & \rot{90}{TSX} & \rot{90}{MERVAL} & \rot{90}{BOVESPA} & \rot{90}{IPC} & \rot{90}{AEX} & \rot{90}{ATX} & \rot{90}{FTSE} & \rot{90}{DAX} & \rot{90}{CAC} & \rot{90}{SMI} & \rot{90}{MIB} & \rot{90}{HGSG} & \rot{90}{NIKKEI} & \rot{90}{STRTIM} & \rot{90}{SSEC} & \rot{90}{BSE} & \rot{90}{KLSE} & \rot{90}{KOSPI} & \rot{90}{ALLORD} &\rot{90}{Average} \\ 
  \hline
  SP & 1.00 & 0.81 & 0.47 & 0.18 & 0.34 & 0.50 & 0.38 & 0.25 & 0.35 & 0.44 & 0.38 & 0.31 & 0.34 & 0.19 & 0.15 & 0.22 & 0.07 & 0.17 & 0.09 & 0.15 & 0.18 &0.30\\ 
  NASDAQ &  & 1.00 & 0.46 & 0.15 & 0.32 & 0.41 & 0.31 & 0.17 & 0.29 & 0.37 & 0.31 & 0.26 & 0.28 & 0.16 & 0.14 & 0.21 & 0.06 & 0.16 & 0.09 & 0.17 & 0.15 &0.26 \\ 
  TSX & & & 1.00 & 0.20 & 0.35 & 0.39 & 0.31 & 0.27 & 0.33 & 0.30 & 0.30 & 0.28 & 0.28 & 0.24 & 0.20 & 0.26 & 0.09 & 0.19 & 0.14 & 0.19 & 0.22 &0.28\\ 
  MERVAL & && & 1.00 & 0.20 & 0.19 & 0.19 & 0.19 & 0.17 & 0.17 & 0.17 & 0.16 & 0.16 & 0.15 & 0.14 & 0.16 & 0.07 & 0.12 & 0.11 & 0.12 & 0.11& 0.16 \\ 
  BOVESPA & &&& & 1.00 & 0.38 & 0.27 & 0.23 & 0.24 & 0.25 & 0.25 & 0.22 & 0.22 & 0.20 & 0.16 & 0.22 & 0.09 & 0.18 & 0.13 & 0.15 & 0.17 & 0.23\\ 
  IPC & &&&& & 1.00 & 0.29 & 0.26 & 0.30 & 0.29 & 0.27 & 0.22 & 0.24 & 0.21 & 0.16 & 0.22 & 0.11 & 0.19 & 0.15 & 0.19 & 0.15 & 0.26\\ 
  AEX & &&&&& & 1.00 & 0.35 & 0.60 & 0.52 & 0.73 & 0.55 & 0.45 & 0.23 & 0.19 & 0.28 & 0.09 & 0.17 & 0.14 & 0.15 & 0.19 & 0.32\\ 
  ATX & &&&&&& & 1.00 & 0.37 & 0.36 & 0.40 & 0.31 & 0.36 & 0.25 & 0.18 & 0.30 & 0.10 & 0.21 & 0.17 & 0.16 & 0.24 & 0.26 \\ 
  FTSE & &&&&&&& & 1.00 & 0.51 & 0.68 & 0.58 & 0.47 & 0.23 & 0.18 & 0.26 & 0.06 & 0.18 & 0.13 & 0.18 & 0.24 & 0.32\\ 
  DAX & &&&&&&&& & 1.00 & 0.64 & 0.45 & 0.51 & 0.20 & 0.16 & 0.24 & 0.07 & 0.13 & 0.13 & 0.18 & 0.17 & 0.30\\ 
  CAC & &&&&&&&&& & 1.00 & 0.61 & 0.54 & 0.23 & 0.18 & 0.27 & 0.07 & 0.15 & 0.13 & 0.17 & 0.20 & 0.33\\ 
  SMI & &&&&&&&&&& & 1.00 & 0.44 & 0.19 & 0.18 & 0.23 & 0.05 & 0.13 & 0.12 & 0.14 & 0.21 & 0.28\\ 
  MIB & &&&&&&&&&&& & 1.00 & 0.19 & 0.17 & 0.21 & 0.07 & 0.12 & 0.10 & 0.12 & 0.18 & 0.27\\ 
  HGSG & &&&&&&&&&&&& & 1.00 & 0.30 & 0.52 & 0.17 & 0.32 & 0.23 & 0.35 & 0.32 &0.24 \\ 
  NIKKEI & &&&&&&&&&&&&& & 1.00 & 0.30 & 0.08 & 0.20 & 0.15 & 0.27 & 0.29 & 0.19\\ 
  STRTIM & &&&&&&&&&&&&&& & 1.00 & 0.11 & 0.34 & 0.26 & 0.35 & 0.34 & 0.27\\ 
  SSEC & &&&&&&&&&&&&&&& & 1.00 & 0.09 & 0.12 & 0.06 & 0.09 &0.08 \\ 
  BSE & &&&&&&&&&&&&&&&& & 1.00 & 0.19 & 0.21 & 0.24 &0.19 \\ 
  KLSE & &&&&&&&&&&&&&&&&& & 1.00 & 0.16 & 0.17 & 0.15 \\ 
  KOSPI & &&&&&&&&&&&&&&&&&& & 1.00 & 0.24 & 0.19\\ 
  ALLORD & &&&&&&&&&&&&&&&&&&& & 1.00 & 0.21 \\ 
    \hline
    \end{tabular}
\end{adjustbox}
\end{table}
\end{landscape}

\end{document}